\documentclass[twocolumn,aps,prc,showpacs,superscriptaddress]{revtex4-2}

\usepackage{color,amsmath,amssymb,epsfig,graphicx}
\usepackage{dcolumn}
\usepackage{bm}
\usepackage{mathptmx, courier, pifont}
\usepackage[scaled=0.92]{helvet}
\usepackage[T1]{fontenc}
\usepackage{textcomp}
\usepackage{xspace}
\usepackage[colorlinks=true,linkcolor=blue,filecolor=blue,urlcolor=blue,citecolor=blue]{hyperref}

\usepackage{multirow}
\usepackage{longtable}
\usepackage[english]{babel}
\usepackage[autostyle, english = american]{csquotes}
\MakeOuterQuote{"}

\begin{document}
\author{Sameer Ahmad Mir}
\email{sameerphst@gmail.com}
\affiliation{Department of Physics, Jamia Millia Islamia, New Delhi, 110 025, India}
\author{Iqbal Mohi Ud Din}
\affiliation{Department of Physics, Jamia Millia Islamia, New Delhi, 110 025, India}
\author{Nasir Ahmad Rather}
\affiliation{Department of Physics, Jamia Millia Islamia, New Delhi, 110 025, India}
\author{Saeed Uddin}
\email{suddin@jmi.ac.in}
\affiliation{Department of Physics, Jamia Millia Islamia, New Delhi, 110 025, India}
\author{M. Farooq Mir}
\affiliation{Department of Physics, University of Kashmir, Jammu and Kashmir, 190 006, India}
\title{Particle production in HRG with thermodynamically consistent EoS and partially deformable hadrons}
\begin{abstract}
\noindent
In the present work, we analyze several strange as well as non-strange relative hadronic yields obtained in the ultra-relativistic heavy-ion collisions (URHIC) experiments over a wide range of center-of-mass collision energy ($\sqrt{s_{NN}}$). We invoke the formation of a hot and dense hadronic resonance gas (HRG) in the final stage following the URHIC. We use an earlier proposed thermodynamically consistent approach for obtaining the equation of state (EoS) of a HRG. It takes into account an important aspect of the hadronic interaction, viz., the hadronic hard-core repulsion, by assigning hard-core volumes to the hadrons, leading to an excluded volume (EV) type effect. We have invoked the bag model approach to assign hard-core volumes to baryons (antibaryons) while treating mesons to be point particles. We employ ansatz to obtain the dependence of the temperature (\textit{T}) and baryon chemical potential (BCP) of HRG system on the center-of-mass energy in URHIC. We also find strong evidence of a double freeze-out scenario, corresponding to baryons (antibaryons) and mesons, respectively. Strangeness (anti-strangeness) imbalance factor is also seen to play an important role in explaining the ratio of strange hadrons to the non-strange ones. The HRG model can explain the experimental data on various relative hadronic multiplicities quite satisfactorily over a wide range of $\sqrt{s_{NN}}$, ranging from the lowest RHIC energies to the highest LHC energies using one set of model parameters by obtaining the best theoretical fits to the experimental data using the minimum $\chi^{2}$/dof method.
\end{abstract}

\pacs{05.70.Ce, 12.38.Mh, 12.39.Ba, 13.30.Eg, 13.75.Cs, 24.10.Pa, 24.85.+p, 25.75.-q, 25.75.Dw}

\date{\today}

\maketitle
\section{Introduction}
\label{Introduction}
The study of hadronic and nuclear collisions at ultra-relativistic energies constitutes a highly focused subfield of high-energy physics. The ultra-relativistic heavy-ion collisions (URHIC) provide an immense potential to investigate the fundamental constituents of matter i.e., quarks and gluons, and the complex forces governing their interactions. The energy densities obtained by colliding high-energy nuclear beams in a laboratory environment are significant enough to enable the transformation of strongly interacting matter (SIM), where hadrons are the fundamental degrees of freedom, with quarks and gluons confined inside various hadronic species, to a state where quarks and gluons themselves become the fundamental degrees of freedom due to the melting of hadrons in an environment of high temperature and density. Existence of such a state is predicted by quantum chromodynamics (QCD), which is the fundamental SU(3) gauge theory of strong interactions~\cite{IEoS1}. In this altered state, also known as quark-gluon plasma (QGP)~\cite{IEoS2,Shuryak:1977ut,Bazavov,Alice2022,Alice2023}, the quarks and gluons are de-confined over an extended region, spreading over almost the entire physical volume of the system initially formed. As particles are continuously produced within QGP, which continuously expands in the outward direction, the density and temperature drops to a critical level where quarks and gluons can no longer remain free due to the highly non-perturbative nature of the strong force. Consequently, the system undergoes a process called hadronization, where quarks and gluons again get confined and hadrons are produced. These hadrons continue to interact with one another producing more hadrons until a certain stage is reached where a chemical freeze-out (CFO) occurs. 
The term "chemical freeze-out" refers to the stage of the evolving SIM after the collision, where particle “composition” becomes fixed or frozen in time and the system is no longer undergoing any significant changes in terms of particle creation and annihilation.
Therefore, URHIC presents a unique and crucial opportunity for investigating the properties and behaviour of SIM under extreme conditions of temperature, pressure and density. The abundant production of various hadronic species in these collisions has been extensively studied in the framework of statistical thermal models. Though the dynamics governing the evolution of system is quite complex however, the system is expected to reach a reasonably high degree of chemical and thermal equilibrium~\cite{Muller1992,Bass2000}.
Under this condition, the properties leading to the production of particles in the final state of system before its breakup can be described using statistical models with temperature $(T)$ and baryonic chemical potential $(\mu_{B})$ as two important independent free model parameters~\cite{IEoS7,IEoS8,IEoS9,IEoS10,IEoS11,IEoS12,IEoS13,IEoS14,IEoS15}. It is worthwhile to note that baryon chemical potential (BCP) is also an indicator of abundance of baryons over antibaryons in the system.
In other words, this description assumes the emission of particles from a source that is not only thermally but also chemically  equilibrated, a state that is created by the elastic and inelastic particle reactions within the system. The statistical model parameters defining the CFO stage of the colliding systems are determined by the actual yield of produced hadrons \cite{IEoS20,IEoS22,IEoS23}.

The ideal hadron resonance gas (Id-HRG) model is a phenomenological framework providing an equation of state (EoS) that has been used extensively to analyse many experimental hadron yield data from URHIC. This straightforward approach has produced fairly good descriptions of experimental data over a broad range of energies, ranging from SchwerIonen-Synchroton (SIS) to the Relativistic Heavy-Ion Collider (RHIC) at BNL and further to the highest energies of the Large Hadron Collider (LHC). These studies have enriched our broad understanding of such states to a great extent~\cite{IEoS24,IEoS25,IEoS26}.
This model, within the temperature range of approximately 100–150 MeV and at both negligible and finite BCP $(\mu_{B})$ at CFO, offers a useful and straightforward phenomenological framework. It can effectively reproduce various hadronic yields, which are otherwise a complex phenomenon observed in lattice QCD calculations~\cite{IEoS27,IEoS28,IEoS29,IEoS30}.
However, some deviations from the Id-HRG models are anticipated. The exploration of extensions beyond Id-HRG framework has been the centre of attention. In particular, the excluded volume-based hadronic resonance gas, i.e., the EV-HRG model, which has been studied in depth~\cite{IEoS31,IEoS32,IEoS33}. The EV-HRG model mimics the hard-core repulsive interactions among hadrons at short distances~\cite{IEoS33,IEoS34,cleymanssatz,cleymansredlich,MunzingerXu,IEoS39,IEoS40}.
It has also been shown that the absence of repulsive interactions in the hadronic EoS restricts the construction of a first-order quark-hadron phase transition within the Bag model approach \cite{Bag} according to the Gibbs criteria across the entire $T- \mu_B$ plane. The short-range repulsive hard-core interactions between a pair of baryons or antibaryons is incorporated using a phenomenological approach in the excluded volume inspired HRG models \cite{cleymans-Suhonen,IEoS52,hagedorn,kuonotakagi,rischke,IEoS55,IEoS56,IEoS58,IEoS59,uddinsingh,uddinplb}.
The baryons in the system are typically assumed to be incompressible. Furthermore, in previous works they are considered either as completely non-deformable spherical objects or as fully deformable objects under extreme high-density conditions, where they can change shape but maintain the same volume due to their non-compressibility. The result of these two seemingly extreme and opposing assumptions is that for baryons (or antibaryons) with radius $r$, the excluded volume per baryon (or antibaryon) is $(16/3) \pi r^{3}$ in the former case \cite{Sam:2025,Ref4/3} and $(4/3) \pi r^{3}$ in the latter case \cite{SKT}.

It is important and interesting to understand that how systems consisting of hot and dense hadronic gas can behave when formed in the most central heavy-ion collisions over a wide range of centre of mass collision energy, $\sqrt{s_{NN}}$. 
There is a strong experimental evidence that towards smaller collision energies, the colliding nuclei are able to stop each other completely. However, as the centre of mass frame energy of the colliding beams of heavy nuclei increases beyond SPS energy, i.e., $\sqrt{s_{NN}}\:\sim$ 17.3 GeV per nucleon, the nuclear transparency effect sets in.
This effect turns out to be partial in both SPS aswell as RHIC energies. When the energy is 
further increased, the nuclear transparency effect tends to become more prominent. The main outcome of this effect is that at low and intermediate energies, the system that is formed due to the high degree of stopping in nuclear collisions contains almost all the colliding (participating) nucleons in the most central collisions, along with other secondary hadrons produced in the bulk system~\cite{Baryonstopping}. Hence, a baryon-rich system is formed, which maintains a large BCP. On the other hand, at sufficiently large collision energies, as a result of the nuclear transparency effect setting in, the bulk of the system is formed in the region between the two receding nuclei due to highly excited vacuum. Consequently, the system is almost baryon-symmetric, which maintains a small chemical potential. Hence, with increasing collision energy, this effect becomes more prominent and the systems formed in URHIC tend to maintain smaller chemical potentials~\cite{Baryonstopping,Nucleartransparency}. It is found in experiments that though there is a rapid increase in the thermal temperature of the system as the energy is increased from a reasonably low value of $\sqrt{s_{NN}}$, it however tends to almost saturate towards very high collision energies at RHIC and LHC~\cite{Transparency,Transparency1,Buzzatti}.

The final state relative hadronic yields obtained from these experiments are known to depend very sensitively on the energy of the colliding beam nuclei. Hence, there are strong indications that relative hadronic yields might serve as an important tool to explore the properties at CFO of hot and dense system formed at various collision energies. One can therefore expect that within the framework of the EV-HRG model a correlation between the thermal parameters of the system and the relative yields of various hadronic species can be established.
Therefore, in this work, our aim is to use a realistic phenomenological EoS such that it can describe the relative hadronic yields over a wide range of collision energy and hence extract reasonable values of various model parameters. We will also attempt to understand the impact of finite baryon sizes which lead to excluded volume type effect in the system on the hadronic yields, as already discussed above. In addition, we will explore how the consideration of incompressible but $partly$ deformable baryonic (antibaryonic) states can affect the relative particle yields. We shall also apply various conservation criteria, such as strangeness and electric charge conservation, in the system during its evolution till the CFO. The study of energy dependence of these quantities, along with other model parameters, can shed light on the behaviour of the systems formed at different collision energies as they approach the CFO stage.

This paper is structured as follows: In section \ref{Statisticalapproach}, we present a brief overview of the excluded volume hadron resonance gas (EV-HRG) model. In section \ref{ResultsDiscussion}, obtained results are discussed in depth. Finally, we summarise and conclude our results in Section \ref{summaryconclusion}.

\section{The Statistical Approach}
\label{Statisticalapproach}

The Ideal Hadron Resonance Gas (Id-HRG) model is a statistical-thermal framework utilized to describe the properties of a system regarded as a hot and dense gas of various hadronic species that are considered to be in a state of thermo-chemical equilibrium \cite{Andronic}. It aims to help us understand the macroscopic properties of the system by incorporating the statistical behaviour of microscopic constituents i.e., hadrons. The hadrons within the system continue to interact inelastically until freeze-out, hence their numbers are not conserved. Under this condition, a grand canonical partition function-based EoS assuming thermal distribution functions for the constituent hadrons of the system is a suitable way to compute various thermodynamical quantities for a HRG system. This can also provide hadronic yields seen in URHIC.

The hadronic EoS within the framework of statistical models has also been employed through various computational tools like THERMUS \cite{thermus}, Thermal-FIST \cite{fist}, and THERMINATOR \cite{therminator}. These authors have calculated several thermodynamic parameters of the system.

As discussed in Section \ref{Introduction}, the hard-core repulsion is an essential feature of the hadronic interaction at short distances, which is assumed to act between a pair of baryons or a pair of antibaryons \cite{cleymans-Suhonen,kuonotakagi}. Hence, in a system at sufficiently high densities, the mean separation between its constituents becomes small enough, and the effect of the short range repulsive interaction becomes important. In a simplistic approach, it is assumed that a given baryon in the system cannot move freely over the entire volume of the system but only in the $available$ volume, which is free from other baryons. Hence, an effective proper volume $\nu_{0}$ is assigned to every baryonic species. Following van der Waals’ excluded volume (EV) type treatment and for a system with $N$ number of particles, the volume $V$ is replaced by $V -\nu_{0} N$ \cite{GorensteinPLB}. However, in a strict sense, the procedure is thermodynamically inconsistent. For example, the net baryon density $(n_{B})$ of the system calculated from the grand partition function or thermodynamic potential $(\Omega)$ cannot be derived as $n_{B} \neq \partial\Omega/\partial\mu_{B}$ \cite{hagedorn-rafelski,hagedorn,cleymans-Suhonen,kuonotakagi}.

Several models have emerged to rectify these inconsistencies by offering varied perspectives and solutions. In this context, a thermodynamically consistent EoS formulation of the EV-HRG was developed by Rischke et al. \cite{rischke}. In this approach, strong repulsive interactions exist between all pairs of baryons and all pairs of antibaryons \cite{vovchenkoprl,vovchenkoprc,vovchenkoprc1,samanthamohanty} while the interactions between baryon-antibaryon pairs are only attractive in nature and can lead to annihilation processes \cite{vovchenkoprl,andronicplb,granddon}. Similarly, the strong meson-meson and meson-(anti)baryon interactions are also predominantly of an attractive nature.

The EV-HRG model proposed by Rischke et. al. can be viewed as an extension of the Id-HRG model \cite{rischke}. It is expected that thermodynamically consistent EoS can provide a more reasonable understanding of the behavior of nuclear matter under extreme conditions of temperature and densities. In the following, we first briefly review the grand canonical ensemble (GCE) based formulation for an Id-HRG consisting of point particles.


The expression for the partition function of the $i^{th}$ particle specie can be written as \cite{Andronic,andronicplb,kkpradhan}

\begin{equation}\label{eqn:spectralfunc}
    ln\mathcal{Z}_{i}^{id}(T,\mu_i,V)=\frac{Vg_{i}}{2\pi^{2}}\int_{0}^{\infty}p^{2}dp \;ln\{1\pm e^{-(E_{i}-\mu_{i})/T}\}
\end{equation}
where + sign corresponds to fermions and anti-fermions, while the - sign corresponds to
bosons. The quantities $g_{i}$ and $E_{i}=\sqrt{p^{2}+m_{i}^{2}}$ are respectively the spin-isospin degeneracy and energy of the $i^{th}$ hadronic specie. $V$ is the physical volume of the system and $\mu_{i}$ denotes the chemical potential of $i^{th}$ specie which is defined as

\begin{equation}\label{eqn:chempot}
    \mu_{i}=B_{i}\mu_{B}+S_{i}\mu_{S}+Q_{i}\mu_{Q}
\end{equation}
where $B_{i}$, $S_{i}$, and $Q_{i}$ are the baryon number, strangeness, and electric charge of the $i^{th}$ hadronic specie. The quantities $\mu_{B}$, $\mu_{S}$, $\mu_{Q}$ represent the BCP, strange chemical potential and electric chemical potential, which control the net baryon, net strangeness and net electric charge content of the system, respectively.

Using the grand canonical partition function $\mathcal{Z}_{i}^{id}(T,\mu_{i},V)$, one can get 
pressure $p_{i}^{id}(T,\mu_{i})$ and number density $n^{id}_{i}(T,\mu_{i})$ as

\begin{multline}
    p_{i}^{id}(T,\mu_{i})=\lim_{V\to\infty}T\frac{ln\mathcal{Z}_{i}^{id}(T,\mu_{i},V)}{V} \\ 
    =\frac{Tg_{i}}{2\pi^{2}}\int_{0}^{\infty}p^{2}dp\;ln\{1\pm e^{-(E_{i}-\mu_{i})/T}\}
    \label{eqn:pressideal}
\end{multline}

\begin{equation}\label{eqn:numdenideal}
    n^{id}_{i}(T,\mu_{i})=\left(\frac{\partial p^{id}_{i}}{\partial \mu_{i}}\right)=\frac{g_{i}}{2\pi^{2}}\int_{0}^{\infty}\frac{p^{2}dp}{e^{(E_{i}-\mu_{i})/T}\pm 1}
\end{equation}

The entropy density, represented as $s_{i}^{id}$, is determined as $s_{i}^{id}=(\partial p_{i}^{id}/\partial T)_{\mu_{i}}$. Meanwhile, the energy density $\epsilon_{i}^{id}$, which can be derived from another thermodynamical relation $\epsilon_{i}^{id}(T,\mu_{i})=Ts_{i}^{id}(T,\mu_{i})-p_{i}^{id}(T,\mu_{i})-\mu_{i} n_{i}^{id}(T,\mu_{i})$, is written as \cite{GorensteinSingle}

\begin{equation}\label{eqn:edensityid}
    \epsilon_{i}^{id}(T,\mu_{i})=\frac{g_{i}}{2\pi^{2}}\int_{0}^{\infty}\frac{p^{2}dp}{e^{(E_{i}-\mu_{i})/T}\pm 1}\:E_{i}
\end{equation}

One can also write the ideal grand canonical partition function $\mathcal{Z}_{GC}^{id}(T,\mu_{i},V)$ for any hadronic specie in terms of the corresponding canonical partition function as
\begin{equation}\label{eqn:partfunc}
    \mathcal{Z}_{GC}^{id}(T,\mu_{i},V)=\sum_{N=0}^{\infty}e^{\mu_{i}N/T}\mathcal{Z}_{C}^{id}(T,N,V)
\end{equation}
Here $\mathcal{Z}_{C}^{id}(T,N,V)$ denotes the canonical partition function for a system consisting of $N$ particles at temperature $T$ and volume $V$. As discussed above, the grand partition function, which incorporates baryonic repulsive interactions, can be obtained by assigning a finite size to all baryons (antibaryons). This, as discussed in section \ref{Introduction}, results in an excluded volume effect where a hard-core volume $\nu_{0}$ is assigned to each baryon. Thus, for a single component system ($i^{th}$) having $N$ number of finite size particles, the volume $V$ in the ideal canonical partition function is replaced by the volume available to any given baryon in the system, i.e., $V-\nu_{0}N$. This leads to a modified grand canonical partition function incorporating the effect of repulsive interaction as follows
\begin{equation}\label{eqn:exclpf}
    \mathcal{Z}_{GC}^{excl}(T,\mu_{i},V)=\sum_{N=0}^{\infty}e^{\mu_{i}N/T}\mathcal{Z}_{C}^{id}(T,N,V-\nu_{0}N)\theta(V-\nu_{0}N)
\end{equation}

The canonical partition function $\mathcal{Z}_{C}^{id}(T,N,V-\nu_{0}N)$ thus incorporates the excluded volume effect resulting from the hard-core repulsion among hadrons.

Here the total excluded volume arising due to $N$ number of particles having hard core interactions is $\nu_{0}N$ \cite{kuonotakagi,rischke,uddinsingh,uddinplb}.

The evaluation of the sum over $N$ in Eq. \eqref{eqn:exclpf} poses a challenge due to the dependence of the available volume, i.e., $V-\nu_{0}N$ on $N$. To address this, one can take the Laplace transform on Eq. \eqref{eqn:exclpf} \cite{rischke} to get
\begin{equation}\label{eqn:exclpf1half}
    \hat{\mathcal{Z}}_{GC}^{excl}(T,\mu_{i},x)=\int_{0}^{\infty}dV\,exp(-xV)\mathcal{Z}_{GC}^{excl}(T,\mu_{i},V)
\end{equation}
For the finiteness of the above integral and avoid the extreme right singularity in the limit $V\xrightarrow{}\infty$, it is required that we should have \cite{rischke}
\begin{equation}
    x=\lim_{V\to\infty}\frac{ln\mathcal{Z}_{GC}^{excl}(T,\mu_{i},V)}{V}
\end{equation}
This gives
\begin{equation}\label{eqn:xT}
    xT=T\lim_{V\to\infty}\frac{ln\mathcal{Z}_{GC}^{excl}(T,\mu_{i},V)}{V}=p_{i}^{excl}(T,\mu_{i})
\end{equation}
Using Eq. \eqref{eqn:exclpf}, one can rewrite Eq. \eqref{eqn:exclpf1half} as \cite{rischke}
\begin{multline}
    \hat{\mathcal{Z}}_{GC}^{excl}(T,\mu_{i},x)= \int_{0}^{\infty}dV\:exp(-xV)\:\sum_{N=0}^{\infty}\:exp(\mu_i N/T)\\
    \times \mathcal{Z}_{C}^{id}(T,N,V-\nu_{0}N)
    \label{eqn:11}
\end{multline}
Interchanging the order of summation and integration, we can rewrite the above Eq. \eqref{eqn:11} as
\begin{multline}
    \hat{\mathcal{Z}}_{GC}^{excl}(T,\mu_{i},x)= \sum_{N=0}^{\infty} \int_{0}^{\infty}dV\:exp(-xV)\:\:exp(\mu_{i} N/T)\\
    \times \mathcal{Z}_{C}^{id}(T,N,V-\nu_{0}N)
\end{multline}

Writing $V^{*}=V-\nu_{0} N$ i.e., $V=V^{*}+\nu_{0} N$, we get
\begin{multline}
    \hat{\mathcal{Z}}^{excl}_{GC}(T,\mu_{i},x)=\sum_{N=0}^{\infty}\:\int_{0}^{\infty}\: dV\:exp(-x(V^{*}\:+\:\nu_{0}N))\\
    \times \: exp(\mu_i N/T)\: \mathcal{Z}_{C}^{id}(T,N,V^{*})
    \label{eqn:13}
\end{multline}

As $N$ is constant in each integral, hence $dV=dV^{*}$, we can rewrite Eq. \eqref{eqn:13} as
\begin{multline}
    \hat{\mathcal{Z}}_{GC}^{excl}(T,\mu_{i},x)= \sum_{N=0}^{\infty} \int_{0}^{\infty}dV^{*}\:exp(-xV^{*})\\
    \times exp((\mu_{i}-\nu_{0}Tx)(N/T))\mathcal{Z}_{C}^{id}(T,N,V^{*})
\end{multline}

Further substituting $\mu_{i}^{*}=\mu_{i}-\nu_{0}Tx$ and again interchanging the order of the integration and summation (replacing the dummy integration variable $V^{*}$ by $V$)
\begin{multline}
    \hat{\mathcal{Z}}_{GC}^{excl}(T,\mu_{i},x)=\int_{0}^{\infty}dV\:exp(-xV)\sum_{N=0}^{\infty}exp(\mu_{i}^{*}N/T)\\
    \times \mathcal{Z}_{C}^{id}(T,N,V)
\end{multline}

The above result is as that for the case of point-like particles but with $\mu_{i}$ replaced by $\mu_{i}^{*}$, hence we can finally write
\begin{equation}\label{eqn:samepoint}
    \hat{\mathcal{Z}}_{GC}^{excl}(T,\mu_{i},x)=\int_{0}^{\infty}dV\:exp(-xV)\:\mathcal{Z}_{GC}^{id}(T,\mu_{i}^{*},V)
\end{equation}
Comparing Eq. \eqref{eqn:exclpf1half} and \eqref{eqn:samepoint}, we get
\begin{equation}
    \mathcal{Z}_{GC}^{excl}(T,\mu_{i},V)=\mathcal{Z}_{GC}^{id}(T,\mu_{i}^{*},V)
\end{equation}

This gives $p_{i}^{excl}(T,\mu_{i})=p_{i}^{id}(T,\mu_{i}^{*})$ with $\mu_{i}^{*}=\mu_{i}-\nu_{0}Tx$. Combining with Eq. \eqref{eqn:xT}, it becomes $\mu_{i}^{*}=\mu_{i}-\nu_{0}p_{i}^{excl}(T,\mu_{i})=\mu_{i}-\nu_{0}p_{i}^{id}(T,\mu_{i}^{*})$. Hence, one can obtain the pressure of the system in the thermodynamic limit $V\xrightarrow{}\infty$ by solving the transcendental equations.



The number density of the finite size particles can now be obtained in a thermodynamically consistent manner by taking the derivative of the pressure term and after rearranging the terms, we get

\begin{equation}\label{eqn:numdenexcl}
    n_{i}^{excl}(T,\mu_{i})=\left(\frac{\partial p_{i}^{excl}(T,\mu_{i})}{\partial\mu_{i}}\right)_{T}=\frac{n_{i}^{id}(T,\mu_{i}^{*})}{1+\nu_{0}\,n_{i}^{id}(T,\mu_{i}^{*})}
\end{equation}

Using $p_{i}^{excl}(T,\mu_{i})$ and $n_{i}^{excl}(T,\mu_{i})$, one could derive the $s_{i}^{excl}(T,\mu_{i})$ and by using the thermodynamical relation $\epsilon_{i}^{excl}(T,\mu_{i})=Ts_{i}^{excl}(T,\mu_{i})-p_{i}^{excl}(T,\mu_{i})-\mu_{i} n_{i}^{excl}(T,\mu_{i})$, we write for the $i^{th}$ hadronic specie

\begin{equation}\label{eqn:entropyexcl}
    s_{i}^{excl}(T,\mu_{i})=\left(\frac{\partial p_{i}^{excl}(T,\mu_{i})}{\partial T}\right)_{T}=\frac{s_{i}^{id}(T,\mu_{i}^{*})}{1+\nu_{0}\,n_{i}^{id}(T,\mu_{i}^{*})}
\end{equation}

\begin{equation}\label{eqn:enegydensityexcl}
    \epsilon_{i}^{excl}(T,\mu_{i})=Ts_{i}^{excl}-p_{i}^{excl}+\mu_{i}n_{i}^{excl}=\frac{\epsilon_{i}^{id}(T,\mu_{i}^{*})}{1+\nu_{0}\,n_{i}^{id}(T,\mu_{i}^{*})}
\end{equation}

The above equations reveals two suppression effect because of the van der Waals (vdW) repulsion, as

\begin{enumerate}
    \item The transformation of chemical potential $\mu_{i}\xrightarrow{}\mu_{i}^{*}$.
    \item A suppression factor $\left[1+\nu_{0}\,n_{i}^{id}(T,\mu_{i}^{*})\right]^{-1}<1$.
\end{enumerate}

Eqs. \eqref{eqn:numdenexcl}, \eqref{eqn:entropyexcl}, \eqref{eqn:enegydensityexcl} describe the  number density, entropy density, energy density for a single component of hadronic matter, taking into account finite size effects in a thermodynamically consistent manner. We can generalize this for "$n$" number of hadronic species. 



The excluded volume approach can be extended to accommodate several particle species. 
The excluded volume grand canonical partition function $\mathcal{Z}_{GC}^{excl}(T,\mu_{1},...,\mu_{n},V)$ for several particle species equals to the product of the partition functions, denoted as $\mathcal{Z}^{excl}_{GC}(T,\mu_{i},V)$ for each individual particle species "$i$" having proper volumes $\nu_{1},...,\nu_{n}$ can be defined as

\begin{multline}
    \mathcal{Z}_{GC}^{excl}(T,\mu_{1},...,\mu_{n},V)=\sum_{N_{1}=0}^{\infty}\:\cdots\:\sum_{N_{n}=0}^{\infty}\: \prod_{i=1}^{n}\: exp{\left(\frac{\mu_{i}N_{i}}{T}\right)}\\
    \times\mathcal{Z}^{id}_{GC}(T,N_{i},V^{*})\theta(V^{*})
    \label{eqn:exclpffs}
\end{multline}

where the available volume is $V^{*}=V-\sum_{i=1}^{n}\nu_{i}N_{i}$. The Laplace transform of Eq. \eqref{eqn:exclpffs} gives the total pressure
\begin{multline}
    p^{excl}(T,\mu_{1},...,\mu_{n})=T\lim_{V\to\infty}\: \frac{ln\mathcal{Z}_{GC}^{excl}(T,\mu_{1},...,\mu_{n},V)}{V}\\
    =p^{id}(T,\mu_{1}^{*},...\mu_{n}^{*})=\sum_{i=1}^{n}p_{i}^{id}(T,\mu_{i}^{*})
    \label{eqn:pressexclmultcom}
\end{multline}

The effective baryonic chemical potential $\mu_{i}^{*}$ for $i^{th}$ baryonic component can then be written as 

\begin{equation}\label{eqn:effcpotbar}
    \mu_{i}^{*}=\mu_{i}-\nu_{i}\:p^{excl}(T,\mu_{1},...,\mu_{n})
\end{equation}

The modified number density of any hadronic species for the multi-component EV-HRG system in the baryonic sector can be obtained in a thermodynamically consistent manner and is thus given as

\begin{equation}\label{eqn:numdenexclbarhalf}
    n_{i}^{excl}(T,\mu_{1},...,\mu_{n})=\left(\frac{\partial p^{excl}}{\partial \mu_{i}}\right)_{T,\mu_{1},...,\mu_{n}}\nonumber
\end{equation}
which can be written as 
\begin{equation}\label{eqn:numdenexclbar}
    \hspace{35mm}=\frac{n_{i}^{id}(T,\mu_{i}^{*})}{1+\sum_{j=1}^{n}\nu_{j}n_{j}^{id}(T,\mu_{j}^{*})}
\end{equation}

The effective chemical potential for the $i^{th}$ antibaryonic component i.e., $\overline{\mu_{i}^{*}}$ can be obtained in a similar way as given by  Eq. \eqref{eqn:effcpotbar}, by replacing number density of particles $n^{id}_{i}$ with antiparticles $\overline{{n}^{id}_{i}}$ and is written as:

\begin{equation}\label{eqn:effcpotantibar}
    \overline{\mu_{i}^{*}}=\overline{\mu}_{i}-\nu_{i}\:p^{excl}(T,\overline{\mu}_{1},...,\overline{\mu}_{n});\;\;\;\;\; \overline{\mu_{i}}=-\mu_{i}
\end{equation}

The modified number density of the hadronic species for the multi-component EV-HRG system in the antibaryonic sector is thus given as

\begin{multline}
    \overline{n}_{i}^{excl}(T,\overline{\mu}_{1},...,\overline{\mu}_{n})=\left(\frac{\partial \overline{p}^{excl}}{\partial \overline{\mu}_{i}}\right)_{T,\overline{\mu}_{1},...,\overline{\mu}_{n}}\\
    =\frac{\overline{n}_{i}^{id}(T,\overline{\mu_{i}^{*}})}{1+\sum_{j=1}^{n}\nu_{j}\overline{n}_{j}^{id}(T,\overline{\mu_{j}^{*}})}
    \label{eqn:numdenexclantibar}
\end{multline}

In the above, we have presented a comprehensive thermodynamically consistent formulation of the vdW repulsion within the grand canonical ensemble given by Rischke et. al. \cite{rischke}.
The formulation also addresses a non-trivial problem: the necessity of thermodynamical self-consistency. Other thermal model formulations with ad hoc corrections fall short of this essential condition. Unlike other earlier approaches described in the Ref.’s \cite{cleymanssatz,cleymansredlich,redlich,MunzingerXu}, this approach differs significantly in a way that leads to a modified baryon (antibaryon) chemical potential, as shown in Eqs. \eqref{eqn:effcpotbar} and \eqref{eqn:effcpotantibar}. 

\section{Results and Discussion}
\label{ResultsDiscussion}

In the preceding sections, we have presented a comprehensive description of the EoS of an EV-HRG, with the main emphasis being on incorporating an essential feature of strong interaction i.e., the hard-core type repulsion among pairs of two baryons or two antibaryons. This effect resembling van der Waals (VDW) type repulsion is incorporated within a grand canonical ensemble based formulation in a thermodynamically consistent manner. In the present approach, the mesons are treated as point-like particles as they do not exhibit repulsive interaction.

In Eq. \eqref{eqn:numdenexclbar}, the summation over the index $j$ in the denominator involves all baryonic degrees of freedom, including the $i^{th}$ species of baryons (antibaryons). 
It is noteworthy that the equation of state reflects the influence of repulsive hard-core interactions through the effective BCP $(\mu^{*})$ as given in Eqs. \eqref{eqn:effcpotbar} and \eqref{eqn:effcpotantibar}. In order to fix the hard-core volumes of various baryons (antibaryons), we invoke the basic feature of the Bag model, where quarks and gluons are essentially supposed to exist inside a Bag (i.e. hadron), which is assumed to be incompressible \cite{ChodosBag,DeGrandBag,CleymansBag}.
The energy density of such a bag is given by $4B$, where $B$ is the bag constant. Using this Bag model approach, the hard-core volume of a baryon with a given mass $M$ can be written as $\nu_{0}=M_{i}/4B$, where $M_{i}$ is the mass of the $i^{th}$ hadron \cite{Guru1}.
The value of the Bag constant, \textit{B}, has been obtained by choosing the widely used value of the hard-core radius of proton, i.e., \textit{r\:=} 0.59 fm~\cite{vovchenkoprl} which is fixed by obtaining a reasonable description of the properties of the ground state nuclear matter~\cite{samanthamohanty}. This yields \textit{B\:=} 272 MeV/fm$^{3}$. The hard-core volumes of the remaining baryons are fixed by using this value of Bag constant. 
Further, in our approach, we have considered hadrons as incompressible objects but allowed them to get deformed under extreme conditions of temperature and pressure, thus leading to different effective values of the excluded volumes in the system arising due to each baryon (antibaryon). For the case of completely non-deformable spherical baryons (antibaryons), each baryon gives rise to an excluded volume, which is 4$\nu_{0}$ while for a fully deformable case, it turns out to be $\nu_{0}$ only.
In our present analysis, we have therefore considered a more realistic situation by allowing the baryons and antibaryons to become partly deformable. Hence we have multiplied the baryonic (antibaryonic) hard-core volume $\nu_{0}$ by a deformation factor $\zeta$ which can vary between 1 and 4. As discussed in Sec.~\ref{Introduction} also, the values 1 and 4 respectively correspond to two extreme ideal situations when baryons (antibaryons) are fully deformable and when they are not deformable at all and are able to retain spherical shape. While using the Eqs.~\eqref{eqn:numdenexclbar} and ~\eqref{eqn:numdenexclantibar} to describe the relative particle-antiparticle yields, we have therefore treated the quantity $\zeta$ as a free model parameter along with the other unknown free parameters of the equation. These are fixed by using the best fit method of the experimental data through minimizing the $\chi^{2}$/dof.

As discussed earlier in Sec.~\ref{Statisticalapproach}, the chemical potential $\mu_{i}$ of a given $i^{th}$ hadronic species is defined in Eq.~\eqref{eqn:chempot}. Since baryons are made up of three valence quarks, the light quark $(u,d)$ baryonic chemical potential is defined as $\mu_{q}=\mu_{B}/3$. 
The corresponding baryonic fugacity of the light quarks will be $\lambda_{q}=e^{\mu_{q}/T}$ while the light antiquark fugacity will be $\lambda_{\overline{q}}=\lambda^{-1}_{q}=e^{-\mu_{q}/T}$. 
Here we have neglected the isospin asymmetry and hence used $\lambda_{u}=\lambda_{d}=\lambda_{s}$. This is required to define the fugacities of all the hadrons. 
In addition, charged hadrons will also acquire electric chemical potential, $\mu_{Q}$ in order to control the net electric charge of the system. The corresponding electric fugacity for a positively charged hadron ($\pi^{+},p, \Sigma^{+},\overline{\Sigma^{-}}$ etc) will be $\lambda_{Q}=e^{\mu_{Q}/T}$. Similarly for a negatively charged hadron ($\pi^{-},\overline{p},\overline{\Sigma^{+}},\Sigma^{-}$ etc), we will have $\lambda_{\overline{Q}}=\lambda_{Q}^{-1}=e^{-\mu_{Q}/T}$, where
$\mu_{Q}$ is treated as a free parameter and is fixed by conserving the charge-to-baryon ratio of the system.
For neutral hadrons, $Q_{i}=0$, which means their chemical potential is independent of $\mu_{Q}$.
The kaon however, contains a light quark (antiquark) and an antistrange (strange) quark therefore, we define kaon fugacities as $\lambda_{K^{+}}=\lambda_{q}\lambda^{-1}_{s}\lambda_{Q}$, $\lambda_{K^{0}}=\lambda_{q}\lambda^{-1}_{s}$ etc. The fugacity associated with strange quarks is defined here as $\lambda_{s}=e^{\mu_{s}/T}$. For strange antiquarks the associated fugacity will be $\lambda_{\overline{s}}=\lambda_{s}^{-1}=e^{-\mu_{s}/T}$, where $\mu_{s}$ is the strange chemical potential and is fixed by applying the criteria of the overall strangeness conservation. 
As proton and neutron are composed of three light quarks, their fugacities are given as $\lambda_{p}=\lambda^{3}_{q}\lambda_{Q}$ and $\lambda_{n}=\lambda^{3}_{q}$. In a similar fashion for the neutral singly strange hyperon $(\Lambda^{0}, \Sigma^{0})$ the fugacities becomes $\lambda_{\Lambda^{0},\Sigma^{0}}=\lambda^{2}_{q}\lambda_{s}$ and for charged singly strange hyperon $(\Sigma^{+},\Sigma^{-})$ fugacity are defined as $\lambda_{\Sigma^{+}}=\lambda^{2}_{q}\lambda_{s}\lambda_{Q}$, $\lambda_{\Sigma^{-}}=\lambda^{2}_{q}\lambda_{s}\lambda_{Q}^{-1}$. For the case of neutral doubly strange hyperon $(\Xi^{0})$ fugacity will be $\lambda_{\Xi^{0}}=\lambda_{q}\lambda^{2}_{s}$ while for the charged doubly strange hyperon i.e., $({\Xi^{-}})$, the fugacity will be $\lambda_{\Xi^{-}}=\lambda_{q}\lambda^{2}_{s}\lambda^{-1}_{Q}$. In a similar fashion, we can define the $\Omega^{-}$ baryon fugacity as $\lambda_{\Omega^{-}}=\lambda^{3}_{s}\lambda^{-1}_{Q}$. The chemical potential of all antiparticles is always taken to be the negative of their corresponding particle’s chemical potential. Consequently, the fugacities of all antiparticles will be the reciprocal of their corresponding particle’s fugacities \cite{IEoS15,IEoS16,uddinplb,Sam:2025,CleymansBag}.

There is a strong relationship between these chemical potentials and the temperature of the
system i.e., $\mu_{B}$, $\mu_{Q}$, $\mu_{S}$ and $T$ inside the hot and dense matter created in the URHIC \cite{Albactob,Sumana}. As mentioned, in our present approach we have used the net charge to the net baryon number ratio to fix the value of the electric chemical potential, $\mu_{Q}$ as

\begin{equation}\label{eqn:chargeconservation}
    \frac{\sum_{i}n_{i}(T,\mu_{B},\mu_{S},\mu_{Q},\zeta)\:\:Q_{i}}{\sum_{i}n_{i}(T,\mu_{B},\mu_{S},\mu_{Q},\zeta)\:\:B_{i}}\:\:=\:\: \text{constant}
\end{equation}

The summation in the numerator is over all charged particle densities, where $Q_{i}=$+1(+2) and -1(-2) respectively for positive singly (doubly) and negative singly (doubly) charged particle. In the denominator the summation is over all baryon and antibaryon densities where $B_{i}=$ +1 for baryons and -1 for antibaryons.
The physical systems formed in the ultra-relativistic nucleus-nucleus collisions determine the ratio of the net charge to the net baryon number. It is seen that the ratio of the number of protons $(N_{p})$ to the mass number $(N_p+N_n)$ of the colliding nuclei in the URHIC experiments is nearly constant, i.e., $\sim$ 0.4. 

The strange chemical potential $\mu_{S}(T,\mu_{B},\mu_{Q},\zeta)$ of the system is determined by the overall $net\: zero$ strangeness

\begin{multline}
    n_{S}(T,\mu_{B},\mu_{S},\mu_{Q},\zeta)=\sum_{i} s_{i}n_{i}(T,\mu_{B},\mu_{S},\mu_{Q},\zeta)\:\:-\\
    \sum_{i}\overline{s}_{i}\overline{n}_{i}(T,\mu_{B},\mu_{S},\mu_{Q},\zeta)=0
    \label{eqn:strangenessconsvation}
\end{multline}

In the above equation, the $s_{i}$ and $n_{i}$ are the strangeness content and number density of the $i^{th}$ strange hadron. Similarly, the $\overline{s_{i}}$ and $\overline{n_{i}}$ are for the cases of the corresponding antiparticles.

Now the two remaining thermodynamic parameters, i.e., temperature $(T)$ and BCP $(\mu_{B})$ at various collision energies can be fixed by fitting the experimental data on antibaryon to baryon ratios available over a wide range of collision energy ($\sqrt{s_{NN}}$). 
As already discussed the BCP reflects the excess of baryons over antibaryons. 
On the other hand, an increasing temperature leads to significant excitation within the system, causing the densities of all particles to increase and hence strongly affecting the particle ratios.

Hence, $T$ and $\mu_{B}$ of the systems formed in the URHIC at different collision energies ($\sqrt{s_{NN}}$) at CFO must be precisely determined in order to perform a useful comparison between theoretical and experimental data. An effective way to achieve this goal is to create a relationship between these two freeze-out parameters and $\sqrt{s_{NN}}$. The following type of ansatz has been widely used in the literature to fit the extracted values of $T$ and $\mu_{B}$ from the experimental data \cite{ansatztempchepot1,ansatztempchepot2,ansatztempchepot3,ansatztempchepot4,ansatztempchepot5,ansatztempchepot6,Guru}.

\begin{equation}\label{eqn:cpot}
    \mu_{B}\:=\:\frac{a}{1\:+\:b\:\sqrt{s_{NN}}}
\end{equation}
\begin{equation}\label{eqn:Temp}
    T\:=c\:-d\:\mu_{B}^{2}-\:e\:\mu_{B}^{4}
\end{equation}

This kind of ansatz has shown promising results in examining the characteristics of hot and dense matter in URHIC using thermal models across various colliding energies. Analysis over a broad collision energy range \cite{IEoS14,IEoS24,Becattini} have contributed significantly in advancing our comprehension of the criteria governing thermo-chemical freeze-out.

It is customary to include all baryonic and mesonic resonances in the system and the contribution of their subsequent decays to the observed final state hadrons in the experiments. It is also important to note that the final number density of hadronic species is modified due to the feed-down decay contribution of the heavier resonances. We have included hadronic resonances upto 2 GeV mass \cite{hp2GeV1,hp2GeV2,pdg}. The final state particle multiplicities are then represented by the following

\begin{equation}\label{eqn:decay}
    n_{i}^{f}=n_{i}^{excl}+\sum_{j}\:\Gamma_{j\xrightarrow{}i}\:n_{j}^{excl} 
\end{equation}
\begin{equation}\label{eqn:decaybar}
    \overline{n}_{i}^{f}=\overline{n}_{i}^{excl}+\sum_{j}\:\Gamma_{j\xrightarrow{}i}\:\overline{n}_{j}^{excl} 
\end{equation}

where $n_{i}^{excl}$ is the thermal particle number density which is calculated using Eq. \eqref{eqn:numdenexclbar} and $\overline{n}_{i}^{excl}$ is the thermal antiparticle number density, which is calculated using Eq. \eqref{eqn:numdenexclantibar}. 
In the above equations,
$\Gamma_{j\xrightarrow{}i}$ is the probability of particle $j$ decay into particle $i$ \cite{pdg}.

In our present analysis, the values of the model parameters in the above Eqs.~\eqref{eqn:cpot} and~\eqref{eqn:Temp} are obtained by the least $\chi^{2}$- fitting procedure.
This approach involves simultaneously fitting the available experimental data on all antibaryon-to-baryon ratios by the corresponding theoretical results. We have, for the sake of comparison and better understanding of the finite-size effect, done this for three different values of hard-core baryonic (antibaryonic) radius viz, \textit{r} = 0.40 fm, 0.59 fm and 0.80 fm over a wide range of collision energy. The $\chi^{2}$ is defined as
\begin{equation}\label{eqn:chisquare}
    \chi^{2}=\sum_{i=1}^{h}\left(\dfrac{R_{i}^{theo}-R_{i}^{expt}}{\sigma_{i}}\right)^{2}
\end{equation}

The $R_{i}^{theo}$ and $R_{i}^{expt}$ are the theoretical results and experimental values of the particle ratios for $i^{th}$ collision energy $\sqrt{s_{NN}}$ and $\sigma_{i}$ represent the experimental (statistical + systematic) errors for corresponding energies. 
The freeze-out parameter values obtained for the radius \textit{r} = 0.40 fm, 0.59 fm and 0.80 fm can be seen in table~\ref{t1}. 
\begin{table*}[htb]
    \centering
    \caption{Freeze-out parameters of Eqs.~\eqref{eqn:cpot} and~\eqref{eqn:Temp} are extracted by fitting the experimental results of anti-baryon to baryon ratios using our theoretical calculations for finite size particles having radius \textit{r} = 0.40 fm, 0.59 fm and 0.80 fm.}
    \vspace{0.2cm}
    \label{t1}
    \begin{tabular}{c c c c c c c c}
    \hline
    \hline
        $r$ (fm) \hspace{0.7cm} $a$ (GeV)\hspace{0.7cm} & $b$ (GeV)$^{-1}$\hspace{0.7cm} & c (MeV)\hspace{0.7cm} &
        $d$ (GeV)$^{-1}$\hspace{0.7cm} & $e$ (GeV)$^{-3}$\hspace{0.7cm} & 
        $\zeta$ \hspace{0.7cm}\\
        \hline
       \hspace{0.1cm} 0.40 \hspace{0.7cm} 1.050$\pm$0.05\hspace{0.7cm} & 0.30$\pm$0.009\hspace{0.7cm} & 153$\pm$02\hspace{0.7cm} & 0.28$\pm$0.009\hspace{0.7cm} & 0.015$\pm$0.03\hspace{0.7cm} & 
       3.5 $\pm$0.09\\ 
       \hline
       \hspace{0.1cm} 0.59 \hspace{0.7cm} 1.150$\pm$0.05\hspace{0.7cm} & 0.30$\pm$0.009\hspace{0.7cm} & 155$\pm$02\hspace{0.7cm} & 0.25$\pm$0.002\hspace{0.7cm} & 0.015$\pm$0.03\hspace{0.7cm} & 
       3.1 $\pm$0.09\\ 
      \hline
      \hspace{0.1cm} 0.80 \hspace{0.7cm} 1.200$\pm$0.04\hspace{0.7cm} & 0.29$\pm$0.009\hspace{0.7cm} & 155$\pm$02\hspace{0.7cm} & 0.26$\pm$0.005\hspace{0.7cm} & 0.015$\pm$0.03\hspace{0.7cm} & 
      3.0 $\pm$ 0.09\\
      \hline
      \hline
    \end{tabular}
\end{table*}

\begin{figure}[htb]
    \includegraphics[scale=0.70]{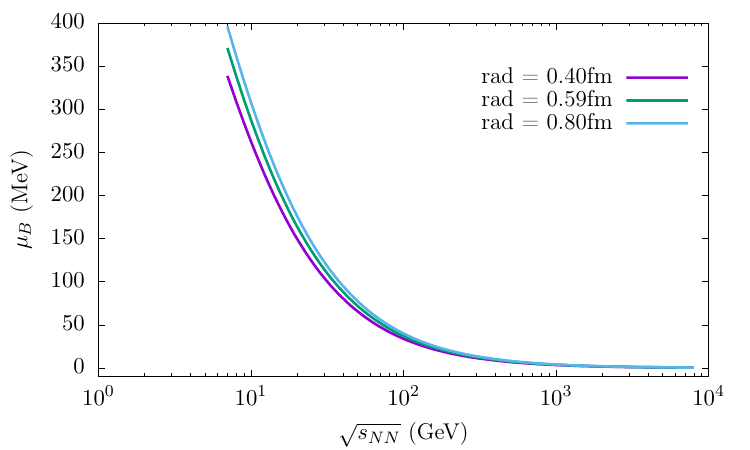}
    \includegraphics[scale=0.70]{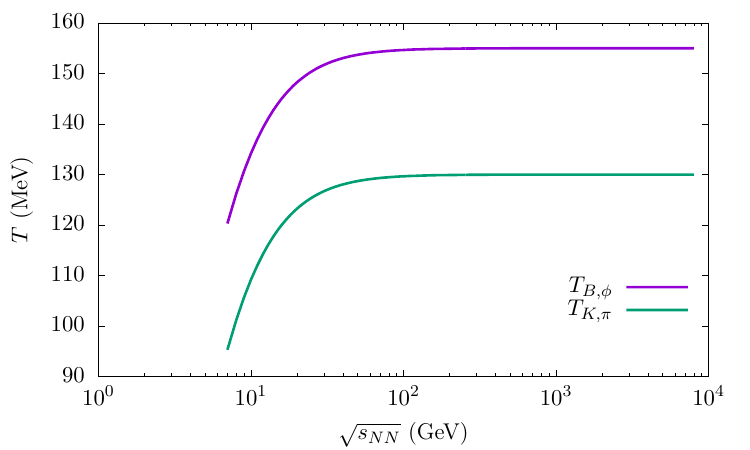}
    \renewcommand{\figurename}{FIG.}
    \caption{(Color online) Dependence of BCP ($\mu_{B}$) and temperature $(T)$ on center-of-mass energy ($\sqrt{s_{NN}}$) for the two freeze-out scenarios. The violet line represents the temperature of baryons including $\phi$-meson while the green one represents the temperature of pions and kaons. The temperature curves showing variations with $\sqrt{s_{NN}}$ almost merge for different radii ($r=0.4\:\text{fm},\; 0.59\:\text{fm}\; \text{and}\; 0.80\:\text{fm}$). However, there are significant deviations in chemical potential parameters for the three cases.}
    \label{fig:cpottemp}
\end{figure}

Using these extracted values of the system parameters, we first show the variation of temperature $(T)$ and BCP $(\mu_{B})$ of the baryons in the system with the centre of mass collision energy $\sqrt{s_{NN}}$, in Fig.~\ref{fig:cpottemp}, represented by Eqs. \eqref{eqn:cpot} and \eqref{eqn:Temp}, respectively. The extracted values of $\mu_{B}$ falls monotonically from SIS and AGS energy range to Relativistic Heavy-Ion Collider (RHIC) at BNL and further towards the LHC energies while extracted temperature $T$ is seen to rise asymptotically almost saturating at around 155 MeV for baryons (antibaryons) for all cases discussed. In other words, the central collisions at higher energies can be characterized by a unique (almost) energy independent CFO temperature. This baryonic freeze-out temperature value is in close proximity to the phase-transition temperature predicted by the lattice QCD calculations. This indicates that in very high energy nuclear collisions, the system undergoes a chemical decoupling close to the phase boundary.
\begin{figure}[htb]
    \includegraphics[scale=0.70]{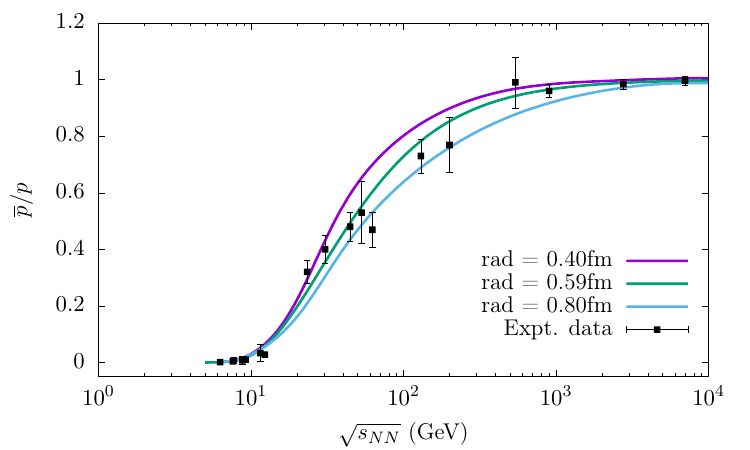}
    \renewcommand{\figurename}{FIG.}
    \caption{(Color online) $\overline{p}/p$ dependence on center-of-mass collision energy, $\sqrt{s_{NN}}$. Experimental data is taken from the Refs.~\cite{Adamczyk2017,Rossi1975,Aggarwal2023,Adams2003,Abbas2013,Ahle1999} represented with error bars. The theoretical results which are represented in violet, green and blue represents for $r$ = 0.40 fm, 0.59 fm and 0.80 fm respectively. }
    \label{fig:protons}
\end{figure}
\begin{figure}[htb]
    \includegraphics[scale=0.70]{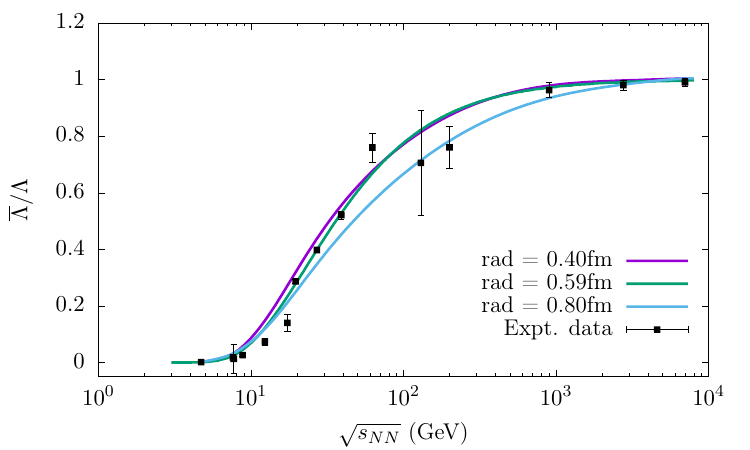}
    \renewcommand{\figurename}{FIG.}
    \caption{(Color online) $\overline{\Lambda}/\Lambda$ dependence on center-of-mass collision energy, $\sqrt{s_{NN}}$. Experimental data is taken from the Refs.~\cite{Aggarwal2023,Abbas2013,Abelev2010,Adam2020,Mischke2002} represented with error bars. The theoretical results which are represented in violet, green and blue represents for $r$ = 0.40 fm, 0.59 fm and 0.80 fm respectively.}
    \label{fig:lambda}
\end{figure}
\begin{figure}[htb]
    \includegraphics[scale=0.70]{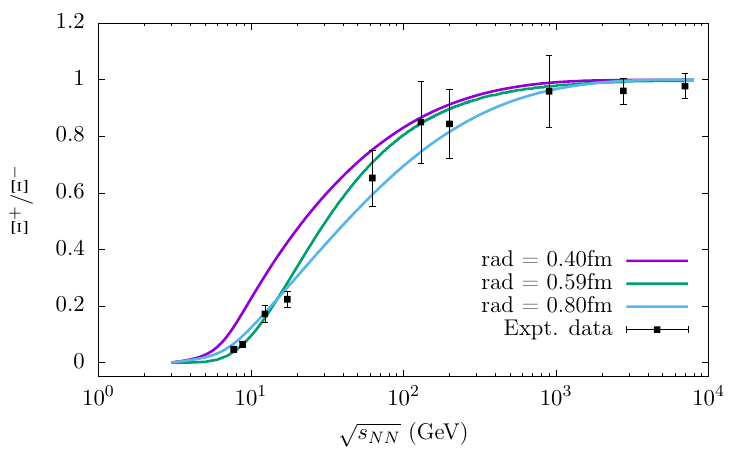}
    \renewcommand{\figurename}{FIG.}
    \caption{(Color online) $\Xi^{+}/\Xi^{-}$ dependence on center-of-mass collision energy, $\sqrt{s_{NN}}$. Experimental data is taken from~\cite{Aggarwal2023,Abbas2013,Abelev2010,Adam2020,Mischke2002} represented with error bars. The theoretical results which are represented in violet, green and blue represents for $r$ = 0.40 fm, 0.59 fm and 0.80 fm respectively.}
    \label{fig:cascade}
\end{figure}
\begin{figure}[htb]
    \includegraphics[scale=0.70]{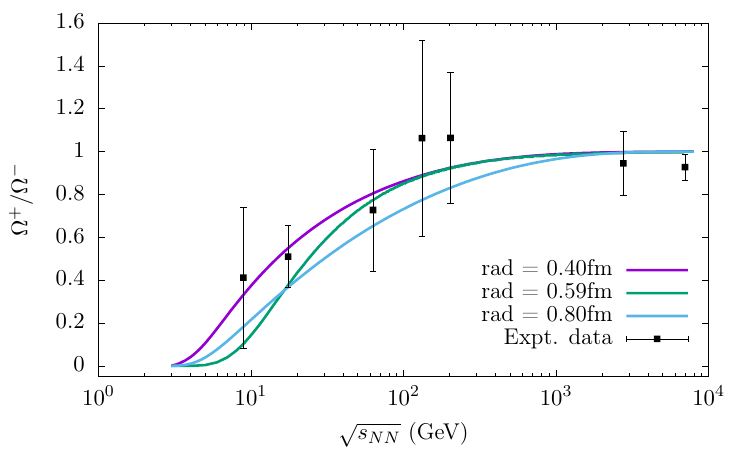}
    \renewcommand{\figurename}{FIG.}
    \caption{(Color online) $\Omega^{+}/\Omega^{-}$ dependence on center-of-mass collision energy, $\sqrt{s_{NN}}$. Experimental data is taken from~\cite{Aggarwal2023,Abbas2013,Abelev2010,Adam2020,Mischke2002} represented with error bars. The theoretical results which are represented in violet, green and blue represents for $r$ = 0.40 fm, 0.59 fm and 0.80 fm respectively.}
    \label{fig:omega}
\end{figure}

The proton (antiproton) is the lightest baryon (antibaryon). As most of the high-mass baryons (antibaryons) decay into protons (antiprotons), it is worthwhile to note that protons and antiprotons are of particular importance because of their large abundance in the final state among all baryons in URHIC. 
The inclusive yield of protons (as well as all other hadrons) is reported as the sum of their primordial yields and the decay contributions after CFO. We have taken into account the contributions of single weak decay processes and where a weak decay is followed by a strong decay. The final state hadron multiplicities can only be understood in a better way by taking these decay modes into account~\cite{decayprotonsprl}.

The decreasing value of $\overline{p}/p$ towards lower $\sqrt{s_{NN}}$ reflects an increase in the net baryon density due to higher baryon stopping at lower collision energies, as seen in Fig.~\ref{fig:protons}. The antiproton yield however increases with increasing collision energies more rapidly than the proton. Towards sufficiently higher values of $\sqrt{s_{NN}}$, the production of protons and antiprotons becomes almost equal due to increasing temperature, as there is relatively more thermal production of antiprotons at higher temperatures. 
Further, due to the increasing transparency effect in nuclear collisions at higher energies, the bulk of the secondary hadronic matter is formed between the two receding nuclei. Consequently, the system tends to become symmetric between baryons and antibaryons and hence maintains a small BCP as their is very little excess of baryons over antibaryons. In Fig.~\ref{fig:protons} we have shown the best possible theoretical fits for the three choices of the proton hard-core radius.

Similarly, other antibaryon-to-baryon ratios, like $\overline{\Lambda}/\Lambda$, $\Xi^{+}/\Xi^{-}$ and $\Omega^{+}/\Omega^{-}$ are shown in Figs.~\ref{fig:lambda},~\ref{fig:cascade},~\ref{fig:omega}, respectively. We have independently fitted the above data sets to extract their respective freeze-out parameters for the three different choices of the proton radius i.e., \textit{r} = 0.4 fm, 0.59 fm and 0.80 fm. We have found that the values of the model parameters for the three cases barely change when we consider different antibaryon to baryon ratios and are well within the limits of the calculated statistical errors. The parameters thus obtained are listed in table~\ref{t1}. This clearly indicates a near simultaneous freeze-out of all baryonic (antibaryonic) species in the system~\cite{vovchenkoprl}.
Like the antiproton to proton ratio case the antihyperon to hyperon ratios also show an increasing trend with increasing collision energy. Beyond $\sqrt{s_{NN}}\sim$ 200 GeV, the ratios tend to saturate towards unity, exhibiting clear evidence of nearly symmetric production of baryons and antibaryons in the URHIC. In all the above four cases, we find that the theoretically fitted curves are in better agreement with the trend exhibited by the experimental data for the choice of the proton radius 0.59 fm.
The minimum $\chi^{2}$/dof for the best fitted curves in all the antibaryon to baryon ratio cases for \textit{r} = 0.40 fm are 4.8, 3.0, 1.52, 1.92. While for the two other radii i.e., \textit{r} = 0.59 fm and 0.80 fm, it turns out to be 2.0, 1.8, 1.42, 0.59 and 3.34, 4.0, 1.54, 0.94, respectively.

One important observation in our study is that a lower CFO temperature than that of baryons is required to explain the $\sqrt{s_{NN}}$ dependence of the $K^{-}/K^{+}$ ratio as well as the charged pion ratio, i.e., $\pi^{-}/\pi^{+}$.

\begin{figure}[htb]
    \includegraphics[scale=0.70]{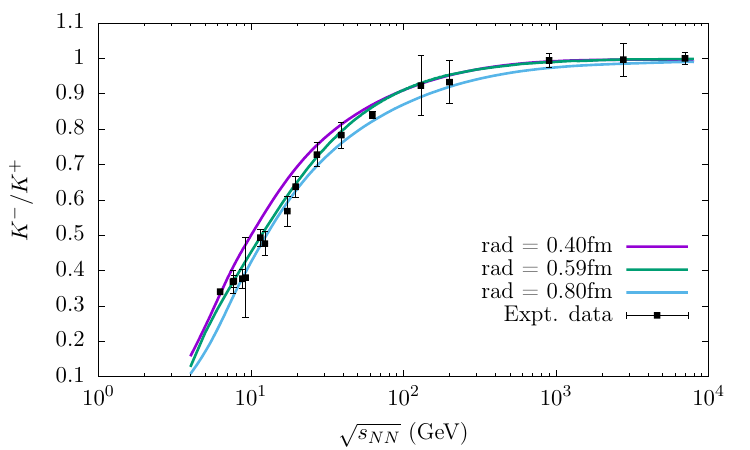}
    \renewcommand{\figurename}{FIG.}
    \caption{(Color online) $K^{-}/K^{+}$ dependence on center-of-mass collision energy, $\sqrt{s_{NN}}$. Experimental data is taken from~\cite{Adams2003,Abelev2010,Abelev2013} represented with error bars. The theoretical results which are represented in violet, green and blue represents for $r$ = 0.40 fm, 0.59 fm and 0.80 fm respectively.}
    \label{fig:kaons}
\end{figure}

We find clear evidence that the baryons have a higher CFO temperature than that of mesons.
In earlier studies, two different freeze-out temperatures for baryon and meson sectors in thermal models have been discussed in relation to Hagedorn’s states \cite{Broniowski2,Broniowski1}. 
The best-fit curve shows the $K^{-}/K^{+}$ dependence on $\sqrt{s_{NN}}$ which is obtained only by using a lower value of the temperature ansatz parameter, i.e., $c$ = 130 $\pm$ 3.0 MeV in Eq. \eqref{eqn:Temp} for \textit{r} = 0.59 fm. The values of the other parameters remain unchanged. Again, as done for the four cases of antibaryon to baryon ratios, we have fitted the $K^{-}/K^{+}$ data using the other two choices of the proton radius. The best fitted curves for the three choices are shown in Fig.~\ref{fig:kaons} with $\chi^2$/dof for \textit{r} = 0.40 fm, 0.59 fm and 0.80 fm being 2.08, 1.91 and 4.37, respectively. The temperature fixing model parameters thus obtained for the three choices of the proton radius are only slightly affected and are found to be within the limits of the statistical errors.

We find that the $K^{-}/K^{+}$ ratio increases with increasing collision energy, $\sqrt{s_{NN}}$ and then saturates as it approaches unity beyond $\sqrt{s_{NN}}\sim$ 200 GeV. This may again be attributed to almost symmetric particle and antiparticle production at high energies,

The increase in the $K^{-}/K^{+}$ ratio with $\sqrt{s_{NN}}$ is an indicator that the production channels of $K^{+}$ and $K^{-}$ contribute almost equally to their abundances as the matter becomes baryon symmetric. 
Consequently, baryons and antibaryons contribute to the abundance of $K^{+}$ and $K^{-}$ almost equally. However, at lower energies, the production of $K^{+}$ dominates due to an excess of baryons over antibaryons.

The reactions for the production of $K^{+}$ involving the above-mentioned interactions are mainly:
\begin{equation}
    \pi^{+}\:+\:n\longrightarrow\:K^{+}\:+\: \Lambda \nonumber
\end{equation}
\begin{equation}
     \pi^{0}\:+\:p\longrightarrow\:K^{+}\:+\: \Lambda \nonumber
\end{equation}
\begin{equation}
    p\:+\:p\longrightarrow\:p\:+\:\Lambda\:+\:K^{+} \nonumber
\end{equation}
\begin{equation}
    n\:+\:p\longrightarrow\:n\:+\:\Lambda\:+\:K^{+} \nonumber
\end{equation}

While the reactions for the production of $K^{-}$ (involving antinucleons) are mainly:
\begin{equation}
    \pi^{0}\:+\:\overline{p}\longrightarrow\:K^{-}\:+\: \overline{\Lambda} \nonumber
\end{equation}
\begin{equation}
    \pi^{-}\:+\:\overline{n}\longrightarrow\:K^{-}\:+\: \overline{\Lambda} \nonumber
\end{equation}
\begin{equation}
    \overline{p}\:+\:\overline{p}\longrightarrow\:\overline{p}\:+\:\overline{\Lambda}\:+\:K^{-} \nonumber
\end{equation}
\begin{equation}
    \overline{n}\:+\:\overline{p}\longrightarrow\:\overline{n}\:+\:\overline{\Lambda}\:+\:K^{-} \nonumber
\end{equation}

The $\pi\:\pi\:\longrightarrow\:K^{+}\:K^{-}$ reaction channel however produces $K^{-}$ and $K^{+}$ in a symmetric manner. As illustrated above, at lower collision energies, there is more production of $K^{+}$ in the system due to the dominance of baryons over antibaryons states.
Our thermal model-based predictions are in accordance with this scenario. The experimental data on $K^{-}$ and $K^{+}$ abundance ratios closely matches the predictions of the thermal model. This agreement strengthens our notion of that thermal models may adequately represent the dynamics of particle production at different collision energies.

\begin{figure}[htb]
    \includegraphics[scale=0.70]{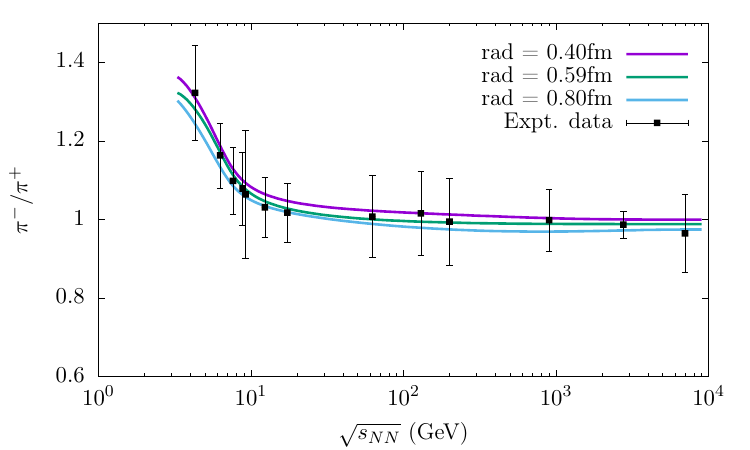}
    \renewcommand{\figurename}{FIG.}
    \caption{(Color online) $\pi^{-}/\pi^{+}$ dependence on center-of-mass collision energy, $\sqrt{s_{NN}}$. Experimental data is taken from~\cite{Adamczyk2017,Abelev2010} represented with error bars. The theoretical results which are represented in violet, green and blue represents for $r$ = 0.40 fm, 0.59 fm and 0.80 fm respectively.}
    \label{fig:pi}
\end{figure}

Charged hadrons are produced at different stages in the system formed in URHIC during its evolution \cite{IEoS46,Pratt,ChargePratt4}. Pion is the lightest among all charged hadrons and is thus abundantly produced during the evolution of the system until freeze-out and carries only electric charge (i.e. neither strangeness nor baryon number). The thermal production ratio $\pi^{-}/\pi^{+}$ is, however, not only governed by the electric chemical potential, i.e., $\mu_{Q}$, but also by the CFO temperature $T$. Besides, several heavier hadronic resonances, including baryons and kaons, contribute to its final stage abundance through their various decay channels. Hence, the ratio $\pi^{-}/\pi^{+}$, besides $\mu_{Q}$ and $T$, also exhibits its dependence on BCP $\mu_{B}$.

At low energies, the secondary heavier resonances are much less produced and the positive charges are mostly trapped inside the protons. Hence any extra positive charge of the
secondary hadrons is compensated by the production of $\pi^{-}$. This causes an increase in the $\pi^{-}$ production as compared to the $\pi^{+}$ production. However, at sufficiently higher energies, there is almost symmetric production of positive and negative pions in the system as the system maintains an overall low net baryon density (i.e., small $\mu_{B}$). The ratio thus tends to saturate to unity for the energies beyond $\sqrt{s_{NN}}\sim$ 10 GeV, as shown in Fig. \ref{fig:pi}.

\begin{figure}[htb]
    \includegraphics[scale=0.70]{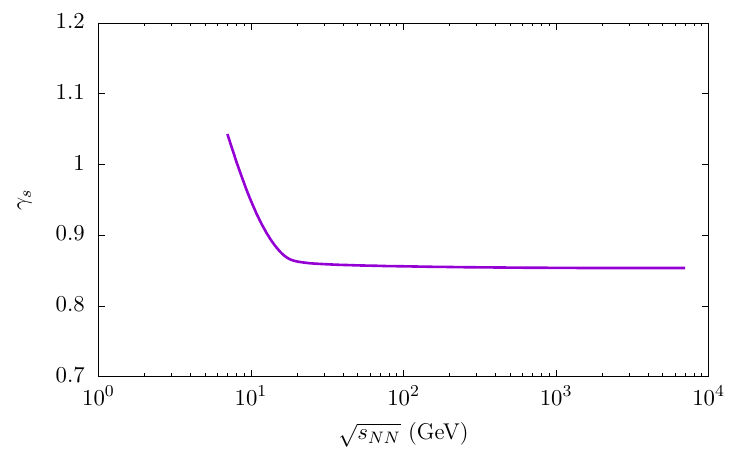}
    \renewcommand{\figurename}{FIG.}
    \caption{(Color online) Dependence on strangeness imbalance factor on increasing $\sqrt{s_{NN}}$}
    \label{fig:gammas}
\end{figure}

Kaon is the most abundantly produced strange hadron in heavy-ion collisions as it is the lightest strange particle having mass much less than the lightest baryon i.e., proton $(m_{K}<<m_{p})$.
The strangeness (antistrangeness) content of the system is thus mostly carried by the
kaons. Similarly, as discussed above, pions $(\pi)$ being the most abundantly produced particles in URHIC carry bulk of the entropy content of the system.
The $K/\pi$ ratio is therefore of particular interest, as it reflects the strangeness content relative to the entropy content of the system formed in ultra-relativistic heavy-ion collisions. This ratio is also an indicator of the enhanced strangeness production in the highly excited thermalized system formed in the URHIC~\cite{RafelskiMuller,Pal,Bass}.
Initially, it was suggested that the enhanced $K/\pi$ ratio could potentially serve as a possible signature of QGP~\cite{Adamczyk64} formation, but their precise connection to the properties of phase transition remains skeptical~\cite{WangLiu}.
Therefore, there still remains an ambiguity as to how $K/\pi$ could be regarded an unambiguous indicator of the transition between QGP and hadronic matter in heavy-ion collisions. It is also been suggested that a chemically equilibrated hadronic resonance gas (HRG) might exhibit strangeness comparable to or greater than that of QGP~\cite{KapustaMekjian,LeeRhodesHeinz,Baym,McLarren}.
The present calculation clearly shows that in order to explain all strange (antistrange) to non-strange hadronic ratios obtained within the framework of thermal hadron resonance gas (HRG) model, it requires a strangeness (antistrangeness) imbalance correction.
We have thus introduced a strangeness imbalance factor, $\gamma_{s}$, which is defined at different $i^{th}$ collision energies as mentioned in the works~\cite{Tawfik:2021txc,Biswas:2020dsc} as
\begin{equation}
    \gamma_s^i=\frac{(K^{-}/\pi^{-})^i_{exp}}{(K^{-}/\pi^{-})^i_{theo}}
    \label{eq35}
\end{equation}
In order to make this factor convenient to apply to other particle ratios also, and at any given $\sqrt{s_{NN}}$, we have chosen an energy dependent ansatz of $\gamma_s$ as
\begin{equation}
    \gamma_s=\gamma_s^0+x_1e^{-x_2(\sqrt{s_{NN}}-x_3)}
    \label{eq36}
\end{equation}
where the values of the constants are $\gamma_s^0$ = 0.85$\pm$0.02, $x_1$ = 0.38$\pm$0.003, $x_2 =\,\text{0.25$\pm$0.03 GeV}^{-1}$ and $x_3$ = 4.29 GeV. These are extracted by the best spline fit to the discrete values of $\gamma_s^i$ obtained through the application of Eq.~\eqref{eq35}. The curve showing $\gamma_s$ with increasing collision energy has a decreasing trend as shown in Fig.~\ref{fig:gammas}. 
Using the above ansatz from Eq.~\eqref{eq36}, we have calculated the modified particle number densities of strange (\textit{s}) as well as antistrange ($\overline{s}$) particles at any given collision energy by redefining their fugacities as 
\begin{equation}
    \lambda_{s,\overline{s}}^{mod}=\gamma_s\lambda_{s,\overline{s}}
    \label{eq37}
\end{equation}
where $\lambda_{s,\overline{s}}^{mod}$ is the modified (or corrected) fugacity of strange (antistrange) hadrons. 
Above $\sqrt{s_{NN}}\:\sim$ 10 GeV, the factor $\gamma_{s}$ turns out to be less than unity (i.e., $\gamma_{s}<1$) indicating a suppression in the $K^{-}/\pi^{-}$ yield in the experiments. While below this energy, there is evidence of excess experimental yield compared to the theoretically obtained values, thus providing $\gamma_{s}>1$.
This phenomenon may be interpreted as evidence of a non-equilibrium effect in the production of strangeness \cite{Gazdzicki,GazdzickiJ.Ph}.
Our results are also consistent with works of Bugaev et al. as well as V.V. Sagun~\cite{Bugaev:2013sfa,Sagun:2014mka} where this factor follows a similar decreasing trend and tends to saturate at a value lower than unity at higher $\sqrt{s_{NN}}$. Thus in this case the strange sector doesn't represent a fully chemically equilibrated phase.
It has also been suggested that an observably low $K^{-}/\pi^{-}$ yield above $\sqrt{s_{NN}}\sim$ 10 GeV may be a signature of QGP presence followed by a non-equilibrium effect during the QGP to HRG transition, especially in the kaonic sector \cite{Torrieri,Blume}.
While at lower energies, where the experimental $K^{-}/\pi^{-}$ ratios are greater than the theoretical ones, it can be safely assumed that no QGP is formed and the system remains in the HRG phase. In such a system, a certain fraction of pions may get absorbed through the strangeness producing reactions $(\pi\pi\xrightarrow{}K\overline{K},\:\pi N\xrightarrow{}K\Lambda,\:\pi\overline{N}\xrightarrow{}\overline{K}\:\overline{\Lambda})$, while due to possible non-equilibrium effects, the backward reaction rates remain slower than the forward ones, leading to a depletion of pions and increase in the kaon population, consequently leading to an enhanced $K^{-}/\pi^{-}$ ratio.
Thus, this ratio in URHIC may offer a more insightful perspective, especially when compared to proton-proton collisions \cite{B421,ansatztempchepot2,Jajati}. The present work thus also attempts to study the $K/\pi$ ratio and the effect of strangeness imbalance over a wide range of collision energy by taking into account the hard-core hadronic repulsions. We find that near $\sqrt{s_{NN}}$ = 10 GeV, our theoretical value of the $K^{-}/\pi^{-}$ ratio is $\sim$ 0.1, which matches quite well with experimental data.
Hence, in this case, the strangeness imbalance factor $\gamma_{s}$ turn out to be $\sim$ 1. This may be due to the competing effect of a predominantly HRG type system and the beginning of the onset of deconfinement.

\begin{figure}[htb]
    \includegraphics[scale=0.70]{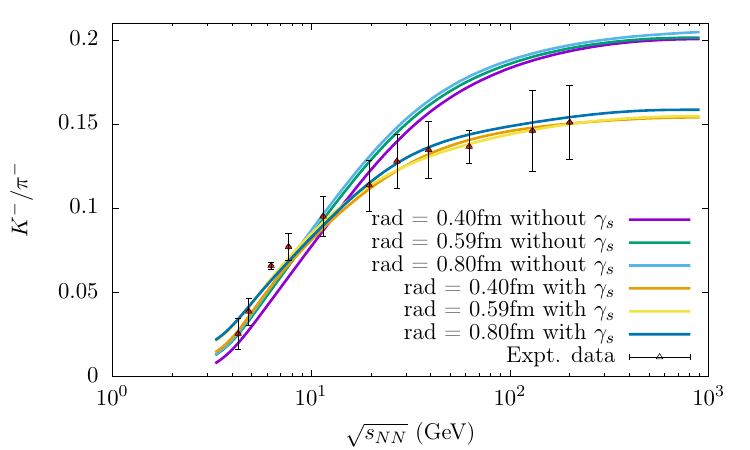}
    \renewcommand{\figurename}{FIG.}
    \caption{(Color online) $K^{-}/\pi^{-}$ dependence on center-of-mass energy, $\sqrt{s_{NN}}$. Experimental data is taken from~\cite{Adams2003,Abelev2010,Abelev2013,Adam2020} represented with error bars. The theoretical results which are represented in violet, green and blue represents for $r$ = 0.40 fm, 0.59 fm and 0.80 fm without the inclusion of strangeness imbalance factor respectively where as orange, yellow and dark-blue represents the theoretical curves for the radius $r$ = 0.40 fm, 0.59 fm and 0.80 fm respectively where strangeness imbalance factor is taken into account.}
    \label{fig:kpiminus}
\end{figure}

In Fig. \ref{fig:kpiminus}, we have shown the variation of $K^{-}/\pi^{-}$ ratio with increasing $\sqrt{s_{NN}}$. The theoretical results in violet, green and blue colors are for \textit{r} = 0.40 fm, 0.59 fm and 0.80 fm without the inclusion of a strangeness imbalance factor while with the strangeness imbalance factor taken into consideration, the resulting theoretical curves are shown in orange, yellow and dark-blue colors. It can be clearly seen that the theoretical results align with the experimental data only after incorporating the strangeness imbalance factor.
The implementation of $\gamma_{s}$ leads to a more accurate 
representation of the $K^{-}/\pi^{-}$ experimental data ratio across a wide range of center of mass energy, $\sqrt{s_{NN}}$.
The agreement between the corrected curve and the experimental data highlights the need of strangeness imbalance factor in describing the ratio of strange to non-strange particles.
The calculations provide the $\chi^{2}$/dof to be 10.89, 7.83 and 7.24 for $K^{-}/\pi^{-}$ ratio for the cases having radii \textit{r} = 0.40 fm, 0.59 fm and 0.80 fm without the inclusion of a strangeness imbalance factor. However, the $\chi^{2}$/dof improves when we introduce the strangeness imbalance factor and it turns out to be 0.62, 0.24 and 0.30, respectively. We find that the choice of the proton radius 0.59 fm again yields better result. The ratio increases steadily with increasing $\sqrt{s_{NN}}$ up to $\sim$ 50 GeV in the URHIC and saturates towards further higher energies at $\sim$ 0.16 when strangeness imbalance factor is taken into account, and is in good agreement with the experimental data. Without strangeness imbalance factor taken into account the theoretical curves saturate at 0.20. 

\begin{figure}[htb]
    \includegraphics[scale=0.70]{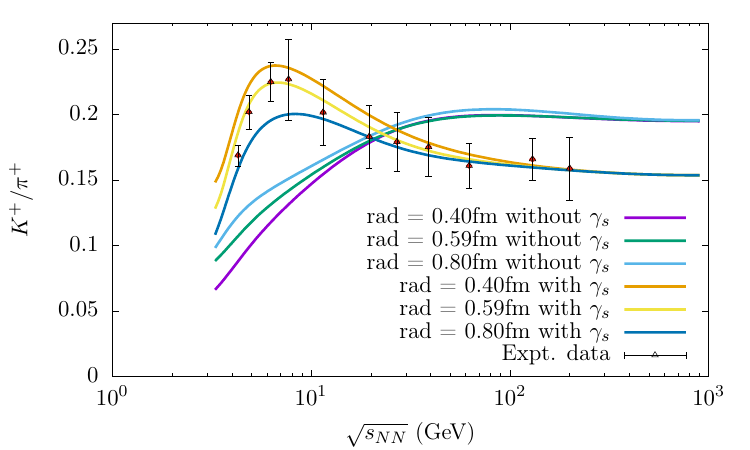}
    \renewcommand{\figurename}{FIG.}
    \caption{(Color online) $K^{+}/\pi^{+}$ dependence on center-of-mass collision energy, $\sqrt{s_{NN}}$. Experimental data is taken from the Refs.~\cite{Adams2003,Abelev2010,Abelev2013,Adam2020} along with error bars. The theoretical results which are represented in violet, green and blue represents for $r$ = 0.40 fm, 0.59 fm and 0.80 fm, respectively but without the inclusion of strangeness imbalance factor. The orange, yellow and dark-blue curves represent the corresponding theoretical calculations when strangeness imbalance factor is taken into account.}
    \label{fig:kpiplus}
\end{figure}

It is very interesting to note here that the values of the strangeness imbalance factor obtained from the analysis of the $K^{-}/\pi^{-}$ ratio can be used to obtain reasonably correct values of the $K^{+}/\pi^{+}$ ratio over a wide range of collision energy. The resulting (corrected) theoretical values are found to fit the experimental data quite well, including the region of energy (5–7) GeV \cite{Cleymans1,Busza}, where a horn structure is seen in Fig. \ref{fig:kpiplus}.

For a nearly baryon symmetric matter formed in URHIC having $\mu_{B}\sim$ 0, the $K^{-}/\pi^{-}$ and $K^{+}/\pi^{+}$ would become nearly identical. This is clearly seen in the two Figs. \ref{fig:kpiminus} and \ref{fig:kpiplus} where both ratios tend to saturate at $\sim$ 0.16 for large values of $\sqrt{s_{NN}}$.
The existence of the “horn” structure in the $K^{+}/\pi^{+}$ ratio may be explained in terms of higher kaon production rates compared to that of the antikaons \cite{WangPLB,WangXu} as already discussed above.
Here it is worthwhile to mention that the previously shown antihyperon to hyperon ratios
in Figs. \ref{fig:lambda}, \ref{fig:cascade} and \ref{fig:omega} are almost independent of $\gamma_{s}$ as their strangeness-antistrangeness contents are equal and are hence equally affected by this factor. 

\begin{figure}[htb]
    \includegraphics[scale=0.70]{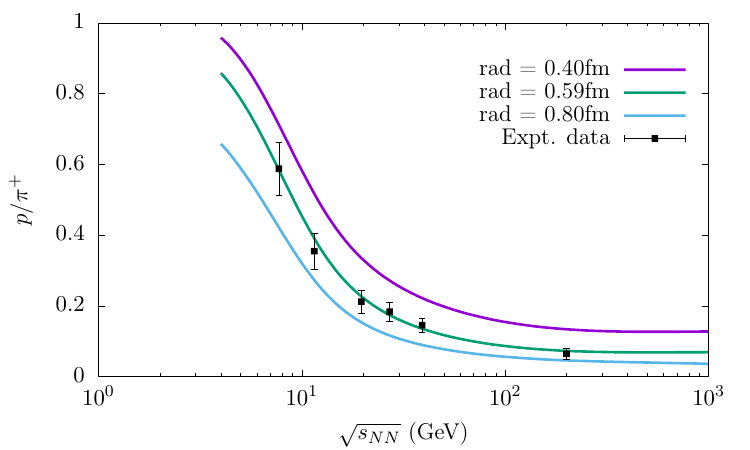}
    \renewcommand{\figurename}{FIG.}
    \caption{(Color online) $p/\pi^{+}$ dependence on center-of-mass collision energy, $\sqrt{s_{NN}}$. Experimental data is taken from the Refs.~\cite{Adamczyk2017,Aggarwal2023,Abelev2010} represented with error bars. The theoretical results which are represented in violet, green and blue represents for $r$ = 0.40 fm, 0.59 fm and 0.80 fm, respectively.}
    \label{fig:protonpiplus}
\end{figure}

\begin{figure}[htb]
    \includegraphics[scale=0.70]{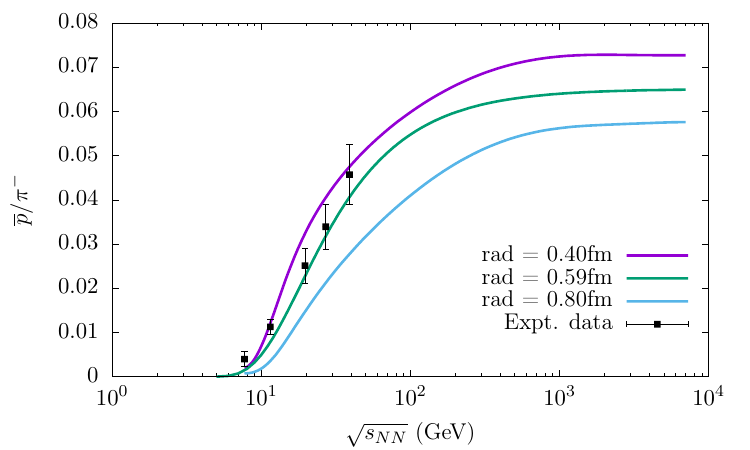}
    \renewcommand{\figurename}{FIG.}
    \caption{(Color online) $\overline{p}/\pi^{-}$ dependence on center-of-mass collision energy, $\sqrt{s_{NN}}$. Experimental data is taken from the Refs.~\cite{Adamczyk2017,Aggarwal2023,Abelev2010} represented with error bars. The theoretical results which are represented in violet, green and blue represents for $r$ = 0.40 fm, 0.59 fm and 0.80 fm respectively.}
    \label{fig:pbarpim}
\end{figure}

In Figs.~\ref{fig:protonpiplus} and ~\ref{fig:pbarpim}, we have shown the dependence of $p/\pi^{+}$ and $\overline{p}/\pi^{-}$ ratios on $\sqrt{s_{NN}}$ for three different radii i.e., \textit{r} = 0.40 fm, 0.59 fm and 0.80 fm, respectively. These ratios serve as indicators of the relative yields of the lightest baryons (antibaryons) to mesons. The apparently good agreement between the theoretical and experimental value can characterize the thermal nature of the hadron production within hot and dense fireball consisting of various hadronic resonances.
In Fig.~\ref{fig:protonpiplus}, the value of $p/\pi^{+}$ ratio decreases monotonically.
This could be due to the fact that at lower collision energies, BCP of the system is high ($\sim$ 500 MeV) which tends to enhance the baryon abundance while suppressing the antibaryon abundance. 
Hence at lower $\sqrt{s_{NN}}$ we have a proton rich system. However, with increasing collision energy the temperature of the HRG system formed in URHIC increases (asymptotically) but the BCP simultaneously decreases. This results in a relatively slower increase in the fermionic, i.e., protonic, abundance than that of the light bosonic i.e., pionic abundance in the system thus leading to a decrease in the $p/\pi^{+}$ ratio with increasing $\sqrt{s_{NN}}$.

\begin{figure}[htb]
    \includegraphics[scale=0.70]{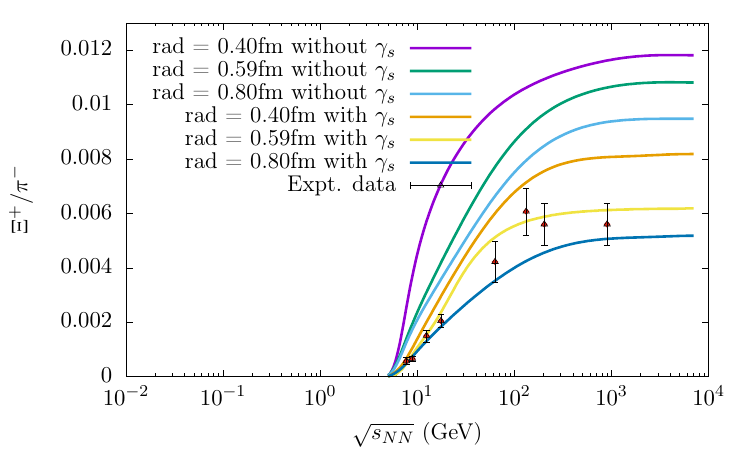}
    \renewcommand{\figurename}{FIG.}
    \caption{(Color online) $\Xi^{+}/\pi^{-}$ dependence on $\sqrt{s_{NN}}$. Experimental data is taken from the Refs.~\cite{Adamczyk2017,Aggarwal2023,Adam2020,Abelev2010} represented with error bars. The theoretical results which are represented in violet, green and blue represents for $r$ = 0.40 fm, 0.59 fm and 0.80 fm without the inclusion of strangeness imbalance factor respectively where as orange, yellow and dark-blue represents the theoretical curves for the radius $r$ = 0.40 fm, 0.59 fm and 0.80 fm respectively where strangeness imbalance factor is taken into account.}
    \label{fig:Xipi}
\end{figure}
\begin{figure}[htb]
    \includegraphics[scale=0.70]{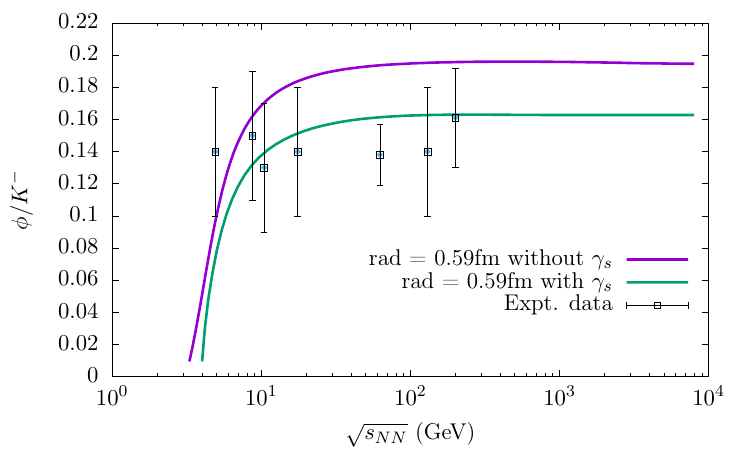}
    \renewcommand{\figurename}{FIG.}
    \caption{(Color online) $\phi/K^{-}$ dependence on center-of-mass collision energy, $\sqrt{s_{NN}}$. Experimental data is taken from the Refs.~\cite{STAR:2008bgi} represented with error bars. The theoretical results which are represented in violet and green colors represents for $r$ = 0.59 fm without and with the inclusion of strangeness imbalance factor respectively.}
    \label{fig:phik}
\end{figure}

Since towards higher collision energies the decreasing BCP and the rising temperature leads to a rapid increase in the antiproton abundance than that of the proton hence the ratio $\overline{p}/\pi^{-}$ is seen to rise with $\sqrt{s_{NN}}$ in Fig. \ref{fig:pbarpim}. The $\chi^2$/dof of these two ratios i.e., $p/\pi^{+}$ and $\overline{p}/\pi^{-}$ for the radius \textit{r} = 0.40 fm, 0.59 fm and 0.80 fm are 4.72, 0.82 and 3.16 and 2.73, 2.01 and 3.89, respectively.

In Fig.~\ref{fig:Xipi}, we have shown the dependence of the cascade antibaryon to pion ratio i.e., $\Xi^{+}/\pi^{-}$ ratio on the increasing center-of-mass collision energy, $\sqrt{s_{NN}}$. The results indicate that the abundance of the doubly strange antihyperon $\Xi^{+}$ increases more rapidly with $\sqrt{s_{NN}}$ compared to $\pi^{-}$. This trend can be explained by the competing effects of the decreasing strangeness suppression factor $\gamma_{s}$, which te nds to reduce the $\Xi^{+}$ yield,combined  with the influence of the decreasing baryon chemical potential (BCP) and increasing temperature, both of which tend to enhance its production.
Furthermore, it also highlights the fact that when different hard-core radius of protons are considered i.e., \textit{r} = 0.40 fm, 0.59 fm and 0.80 fm, the theoretical predictions again appear to align more closely with the experimental data when \textit{r} = 0.59 fm compared to \textit{r} = 0.40 fm and 0.80 fm. For the case of \textit{r} = 0.40 fm, the theoretical values systematically overestimate the experimental data over the entire energy regime. The reason behind this overestimation is that the calculated number densities of baryons and antibaryons are higher for the smaller radii. This discrepancy is not apparent when antiparticle-to-particle ratios with similar masses are analyzed.
The $\chi^{2}$/dof of the above-mentioned ratio i.e., $\Xi^{+}/\pi^{-}$ for the cases of radius \textit{r} = 0.40 fm, 0.59 fm and 0.80 fm when strangeness imbalance factor is not taken into account are 34.67, 25.83 and 15.87, respectively while the $\chi^{2}$/dof when this factor is taken into account are 8.76, 0.83 and 4.87, respectively.
The ratio $\phi/K^{-}$ is also analyzed and its variation with the increasing center-of-mass collision energy ($\sqrt{s_{NN}}$) can be seen in Fig.~\ref{fig:phik}. The experimental data is taken from Ref.~\cite{STAR:2008bgi}. A reasonably good theoretical fit is obtained for \textit{r} = 0.59 fm when the strangeness imbalance factor is implemented. The $\chi^2$/dof for the above mentioned ratio for the case of \textit{r} = 0.59 fm when the strangeness imbalance factor is not considered is 3.42 while it is 1.02 when this factor is taken into account.

\section{Summary and Conclusion}
\label{summaryconclusion}
An improved thermodynamically consistent hadron resonance gas (HRG) model is used in this study by taking excluded volume (EV) effects into consideration in order to explain several relative hadronic yields in URHIC over a wide range of collision $\sqrt{s_{NN}}$ energy using three different set of model parameters for the three cases of the proton hard-core radii i.e., \textit{r} = 0.40 fm, 0.59 fm, 0.80 fm. We have used the bag model approach to fix the hard-core volume of different baryons (antibaryons) according to their masses. The mesons however have been treated as point particles.
We find that though the baryons and antibaryons may be treated as incompressible however they appear to be partly deformable, where the deformation factor $\zeta$ $\sim$ 3.5, 3.1 and 3.0 for the above mentioned three cases.
Our approach is seen to be quiet valid even at values of thermodynamic variables, i.e., temperature ($T$) and BCP ($\mu_{B}$) of the HRG system under extreme conditions. The centre-of-mass energy $\sqrt{s_{NN}}$ dependent ansatz is used for fixing the values of these two thermodynamic variables ($T$) and ($\mu_{B}$).
Clear evidence of a double CFO scenario, one corresponding to baryons (antibaryons) and the other to mesons, is seen. The baryons are found to freeze-out nearly simultaneously but earlier than the mesons. We also take into account the final state contributions of the heavier hadronic resonances with masses up to 2 GeV through their weak decays. We have included single weak decays as well as the double decays where weak decay is followed by strong decay.
We use minimum $\chi^{2}$/dof fits to theoretically explain the variation of the experimental particle ratios with centre-of-mass collision energy. When attempting to explain the unlike mass ratios of strange and non-strange hadrons, e.g., $K/\pi$ and $\Xi^{+}/\pi^{-}$, we find that the theoretical results overestimate the experimental data significantly. In order to achieve a reasonably good agreement between the theoretical results with the experimental data, we have introduced a strangeness (anti)strangeness imbalance factor, $\gamma_{s}$, which can explain the abundance of strange hadrons relative to non-strange ones, especially pions, which are the most abundantly produced non-strange hadrons. After applying the same strangeness imbalance correction factor $\gamma_{s}$ to each case, we find that the theoretical curves match quite well with the trend of the experimental data, with reasonable values of the minimum $\chi^{2}$/dof for each case. The theoretical results obtained for the case when proton hard-core radius is taken to be 0.59 fm are found to yield better results. This value is also in agreement with that obtained through a satisfactory description of the ground state nuclear matter~\cite{vovchenkoprl}. 
\section{Data Availability Statement}
The data supporting the findings of this study are included in the article and are referenced appropriately throughout the text. Additional results from this study will be made available upon request to the corresponding author.

\section{Future Work}
We intend to incorporate the effect of attractive interaction and also consider the medium effect on the hadron masses and see how these can modify our current results on various particle ratios in the system.

\bibliographystyle{apsrev4-2}
\bibliography{part_prod_EoS}

\begin{thebibliography}{129}%
\makeatletter
\providecommand \@ifxundefined [1]{%
 \@ifx{#1\undefined}
}%
\providecommand \@ifnum [1]{%
 \ifnum #1\expandafter \@firstoftwo
 \else \expandafter \@secondoftwo
 \fi
}%
\providecommand \@ifx [1]{%
 \ifx #1\expandafter \@firstoftwo
 \else \expandafter \@secondoftwo
 \fi
}%
\providecommand \natexlab [1]{#1}%
\providecommand \enquote  [1]{``#1''}%
\providecommand \bibnamefont  [1]{#1}%
\providecommand \bibfnamefont [1]{#1}%
\providecommand \citenamefont [1]{#1}%
\providecommand \href@noop [0]{\@secondoftwo}%
\providecommand \href [0]{\begingroup \@sanitize@url \@href}%
\providecommand \@href[1]{\@@startlink{#1}\@@href}%
\providecommand \@@href[1]{\endgroup#1\@@endlink}%
\providecommand \@sanitize@url [0]{\catcode `\\12\catcode `\$12\catcode
  `\&12\catcode `\#12\catcode `\^12\catcode `\_12\catcode `\%12\relax}%
\providecommand \@@startlink[1]{}%
\providecommand \@@endlink[0]{}%
\providecommand \url  [0]{\begingroup\@sanitize@url \@url }%
\providecommand \@url [1]{\endgroup\@href {#1}{\urlprefix }}%
\providecommand \urlprefix  [0]{URL }%
\providecommand \Eprint [0]{\href }%
\providecommand \doibase [0]{https://doi.org/}%
\providecommand \selectlanguage [0]{\@gobble}%
\providecommand \bibinfo  [0]{\@secondoftwo}%
\providecommand \bibfield  [0]{\@secondoftwo}%
\providecommand \translation [1]{[#1]}%
\providecommand \BibitemOpen [0]{}%
\providecommand \bibitemStop [0]{}%
\providecommand \bibitemNoStop [0]{.\EOS\space}%
\providecommand \EOS [0]{\spacefactor3000\relax}%
\providecommand \BibitemShut  [1]{\csname bibitem#1\endcsname}%
\let\auto@bib@innerbib\@empty
\bibitem [{\citenamefont {Wilczek}(1999)}]{IEoS1}%
  \BibitemOpen
  \bibfield  {author} {\bibinfo {author} {\bibfnamefont {F.}~\bibnamefont
  {Wilczek}},\ }\href {https://doi.org/10.1103/RevModPhys.71.S85} {\bibfield
  {journal} {\bibinfo  {journal} {Rev. Mod. Phys.}\ }\textbf {\bibinfo {volume}
  {71}},\ \bibinfo {pages} {S85} (\bibinfo {year} {1999})}\BibitemShut
  {NoStop}%
\bibitem [{\citenamefont {Cabibbo}\ and\ \citenamefont {Parisi}(1975)}]{IEoS2}%
  \BibitemOpen
  \bibfield  {author} {\bibinfo {author} {\bibfnamefont {N.}~\bibnamefont
  {Cabibbo}}\ and\ \bibinfo {author} {\bibfnamefont {G.}~\bibnamefont
  {Parisi}},\ }\href {https://doi.org/10.1016/0370-2693(75)90158-6} {\bibfield
  {journal} {\bibinfo  {journal} {Phys. Lett. B}\ }\textbf {\bibinfo {volume}
  {59}},\ \bibinfo {pages} {67} (\bibinfo {year} {1975})}\BibitemShut {NoStop}%
\bibitem [{\citenamefont {Shuryak}(1978)}]{Shuryak:1977ut}%
  \BibitemOpen
  \bibfield  {author} {\bibinfo {author} {\bibfnamefont {E.~V.}\ \bibnamefont
  {Shuryak}},\ }\href@noop {} {\bibfield  {journal} {\bibinfo  {journal} {Sov.
  Phys. JETP}\ }\textbf {\bibinfo {volume} {47}},\ \bibinfo {pages} {212}
  (\bibinfo {year} {1978})}\BibitemShut {NoStop}%
\bibitem [{\citenamefont {Bazavov}\ \emph {et~al.}(2014)\citenamefont
  {Bazavov}, \citenamefont {Bhattacharya}, \citenamefont {DeTar}, \citenamefont
  {Ding}, \citenamefont {Gottlieb}, \citenamefont {Gupta}, \citenamefont
  {Hegde}, \citenamefont {Heller}, \citenamefont {Karsch}, \citenamefont
  {Laermann}, \citenamefont {Levkova}, \citenamefont {Mukherjee}, \citenamefont
  {Petreczky}, \citenamefont {Schmidt}, \citenamefont {Schroeder},
  \citenamefont {Soltz}, \citenamefont {Soeldner}, \citenamefont {Sugar},
  \citenamefont {Wagner},\ and\ \citenamefont {Vranas}}]{Bazavov}%
  \BibitemOpen
  \bibfield  {author} {\bibinfo {author} {\bibfnamefont {A.}~\bibnamefont
  {Bazavov}}, \bibinfo {author} {\bibfnamefont {T.}~\bibnamefont
  {Bhattacharya}}, \bibinfo {author} {\bibfnamefont {C.}~\bibnamefont {DeTar}},
  \bibinfo {author} {\bibfnamefont {H.-T.}\ \bibnamefont {Ding}}, \bibinfo
  {author} {\bibfnamefont {S.}~\bibnamefont {Gottlieb}}, \bibinfo {author}
  {\bibfnamefont {R.}~\bibnamefont {Gupta}}, \bibinfo {author} {\bibfnamefont
  {P.}~\bibnamefont {Hegde}}, \bibinfo {author} {\bibfnamefont {U.~M.}\
  \bibnamefont {Heller}}, \bibinfo {author} {\bibfnamefont {F.}~\bibnamefont
  {Karsch}}, \bibinfo {author} {\bibfnamefont {E.}~\bibnamefont {Laermann}},
  \bibinfo {author} {\bibfnamefont {L.}~\bibnamefont {Levkova}}, \bibinfo
  {author} {\bibfnamefont {S.}~\bibnamefont {Mukherjee}}, \bibinfo {author}
  {\bibfnamefont {P.}~\bibnamefont {Petreczky}}, \bibinfo {author}
  {\bibfnamefont {C.}~\bibnamefont {Schmidt}}, \bibinfo {author} {\bibfnamefont
  {C.}~\bibnamefont {Schroeder}}, \bibinfo {author} {\bibfnamefont {R.~A.}\
  \bibnamefont {Soltz}}, \bibinfo {author} {\bibfnamefont {W.}~\bibnamefont
  {Soeldner}}, \bibinfo {author} {\bibfnamefont {R.}~\bibnamefont {Sugar}},
  \bibinfo {author} {\bibfnamefont {M.}~\bibnamefont {Wagner}},\ and\ \bibinfo
  {author} {\bibfnamefont {P.}~\bibnamefont {Vranas}} (\bibinfo {collaboration}
  {HotQCD Collaboration}),\ }\href {https://doi.org/10.1103/PhysRevD.90.094503}
  {\bibfield  {journal} {\bibinfo  {journal} {Phys. Rev. D}\ }\textbf {\bibinfo
  {volume} {90}},\ \bibinfo {pages} {094503} (\bibinfo {year}
  {2014})}\BibitemShut {NoStop}%
\bibitem [{\citenamefont {Collaboration}(2022)}]{Alice2022}%
  \BibitemOpen
  \bibfield  {author} {\bibinfo {author} {\bibfnamefont {A.}~\bibnamefont
  {Collaboration}},\ }\href@noop {} {\bibinfo {title} {The alice experiment : A
  journey through qcd}} (\bibinfo {year} {2022}),\ \Eprint
  {https://arxiv.org/abs/2211.04384} {arXiv:2211.04384 [nucl-ex]} \BibitemShut
  {NoStop}%
\bibitem [{\citenamefont {Collaboration}(2023)}]{Alice2023}%
  \BibitemOpen
  \bibfield  {author} {\bibinfo {author} {\bibfnamefont {A.}~\bibnamefont
  {Collaboration}},\ }\href@noop {} {\bibinfo {title} {Multiplicity dependence
  of charged-particle intra-jet properties in pp collisions at $\sqrt{s}$ = 13
  tev}} (\bibinfo {year} {2023}),\ \Eprint {https://arxiv.org/abs/2311.13322}
  {arXiv:2311.13322 [hep-ex]} \BibitemShut {NoStop}%
\bibitem [{\citenamefont {Müller}(1992)}]{Muller1992}%
  \BibitemOpen
  \bibfield  {author} {\bibinfo {author} {\bibfnamefont {B.}~\bibnamefont
  {Müller}},\ }\href
  {https://doi.org/https://doi.org/10.1016/0375-9474(92)90567-4} {\bibfield
  {journal} {\bibinfo  {journal} {Nucl. Phys. A}\ }\textbf {\bibinfo {volume}
  {544}},\ \bibinfo {pages} {95} (\bibinfo {year} {1992})}\BibitemShut
  {NoStop}%
\bibitem [{\citenamefont {Bass}\ and\ \citenamefont
  {Dumitru}(2000)}]{Bass2000}%
  \BibitemOpen
  \bibfield  {author} {\bibinfo {author} {\bibfnamefont {S.~A.}\ \bibnamefont
  {Bass}}\ and\ \bibinfo {author} {\bibfnamefont {A.}~\bibnamefont {Dumitru}},\
  }\href {https://doi.org/10.1103/PhysRevC.61.064909} {\bibfield  {journal}
  {\bibinfo  {journal} {Phys. Rev. C}\ }\textbf {\bibinfo {volume} {61}},\
  \bibinfo {pages} {064909} (\bibinfo {year} {2000})}\BibitemShut {NoStop}%
\bibitem [{\citenamefont {Mekjian}(1977)}]{IEoS7}%
  \BibitemOpen
  \bibfield  {author} {\bibinfo {author} {\bibfnamefont {A.}~\bibnamefont
  {Mekjian}},\ }\href {https://doi.org/10.1103/PhysRevLett.38.640} {\bibfield
  {journal} {\bibinfo  {journal} {Phys. Rev. Lett.}\ }\textbf {\bibinfo
  {volume} {38}},\ \bibinfo {pages} {640} (\bibinfo {year} {1977})}\BibitemShut
  {NoStop}%
\bibitem [{\citenamefont {Gosset}\ \emph {et~al.}(1978)\citenamefont {Gosset},
  \citenamefont {Kapusta},\ and\ \citenamefont {Westfall}}]{IEoS8}%
  \BibitemOpen
  \bibfield  {author} {\bibinfo {author} {\bibfnamefont {J.}~\bibnamefont
  {Gosset}}, \bibinfo {author} {\bibfnamefont {J.~I.}\ \bibnamefont
  {Kapusta}},\ and\ \bibinfo {author} {\bibfnamefont {G.~D.}\ \bibnamefont
  {Westfall}},\ }\href {https://doi.org/10.1103/PhysRevC.18.844} {\bibfield
  {journal} {\bibinfo  {journal} {Phys. Rev. C}\ }\textbf {\bibinfo {volume}
  {18}},\ \bibinfo {pages} {844} (\bibinfo {year} {1978})}\BibitemShut
  {NoStop}%
\bibitem [{\citenamefont {Mekjian}(1978)}]{IEoS9}%
  \BibitemOpen
  \bibfield  {author} {\bibinfo {author} {\bibfnamefont {A.~Z.}\ \bibnamefont
  {Mekjian}},\ }\href
  {https://doi.org/https://doi.org/10.1016/0375-9474(78)90604-8} {\bibfield
  {journal} {\bibinfo  {journal} {Nucl. Phys. A}\ }\textbf {\bibinfo {volume}
  {312}},\ \bibinfo {pages} {491} (\bibinfo {year} {1978})}\BibitemShut
  {NoStop}%
\bibitem [{\citenamefont {Csernai}\ and\ \citenamefont
  {Kapusta}(1986)}]{IEoS10}%
  \BibitemOpen
  \bibfield  {author} {\bibinfo {author} {\bibfnamefont {L.}~\bibnamefont
  {Csernai}}\ and\ \bibinfo {author} {\bibfnamefont {J.~I.}\ \bibnamefont
  {Kapusta}},\ }\href
  {https://doi.org/https://doi.org/10.1016/0370-1573(86)90031-1} {\bibfield
  {journal} {\bibinfo  {journal} {Phys. Rep.}\ }\textbf {\bibinfo {volume}
  {131}},\ \bibinfo {pages} {223} (\bibinfo {year} {1986})}\BibitemShut
  {NoStop}%
\bibitem [{\citenamefont {Hahn}\ and\ \citenamefont
  {St\"ocker}(1987)}]{IEoS11}%
  \BibitemOpen
  \bibfield  {author} {\bibinfo {author} {\bibfnamefont {D.}~\bibnamefont
  {Hahn}}\ and\ \bibinfo {author} {\bibfnamefont {H.}~\bibnamefont
  {St\"ocker}},\ }\href {https://doi.org/10.1103/PhysRevC.35.1311} {\bibfield
  {journal} {\bibinfo  {journal} {Phys. Rev. C}\ }\textbf {\bibinfo {volume}
  {35}},\ \bibinfo {pages} {1311} (\bibinfo {year} {1987})}\BibitemShut
  {NoStop}%
\bibitem [{\citenamefont {Hahn}\ and\ \citenamefont {Stöcker}(1988)}]{IEoS12}%
  \BibitemOpen
  \bibfield  {author} {\bibinfo {author} {\bibfnamefont {D.}~\bibnamefont
  {Hahn}}\ and\ \bibinfo {author} {\bibfnamefont {H.}~\bibnamefont
  {Stöcker}},\ }\href
  {https://doi.org/https://doi.org/10.1016/0375-9474(88)90332-6} {\bibfield
  {journal} {\bibinfo  {journal} {Nucl. Phys. A}\ }\textbf {\bibinfo {volume}
  {476}},\ \bibinfo {pages} {718} (\bibinfo {year} {1988})}\BibitemShut
  {NoStop}%
\bibitem [{\citenamefont {Braun-Munzinger}\ and\ \citenamefont
  {Stachel}(1996)}]{IEoS13}%
  \BibitemOpen
  \bibfield  {author} {\bibinfo {author} {\bibfnamefont {P.}~\bibnamefont
  {Braun-Munzinger}}\ and\ \bibinfo {author} {\bibfnamefont {J.}~\bibnamefont
  {Stachel}},\ }\href
  {https://doi.org/https://doi.org/10.1016/0375-9474(96)00198-4} {\bibfield
  {journal} {\bibinfo  {journal} {Nucl. Phys. A}\ }\textbf {\bibinfo {volume}
  {606}},\ \bibinfo {pages} {320} (\bibinfo {year} {1996})}\BibitemShut
  {NoStop}%
\bibitem [{\citenamefont {Becattini}\ \emph {et~al.}(2001)\citenamefont
  {Becattini}, \citenamefont {Cleymans}, \citenamefont {Ker\"anen},
  \citenamefont {Suhonen},\ and\ \citenamefont {Redlich}}]{IEoS14}%
  \BibitemOpen
  \bibfield  {author} {\bibinfo {author} {\bibfnamefont {F.}~\bibnamefont
  {Becattini}}, \bibinfo {author} {\bibfnamefont {J.}~\bibnamefont {Cleymans}},
  \bibinfo {author} {\bibfnamefont {A.}~\bibnamefont {Ker\"anen}}, \bibinfo
  {author} {\bibfnamefont {E.}~\bibnamefont {Suhonen}},\ and\ \bibinfo {author}
  {\bibfnamefont {K.}~\bibnamefont {Redlich}},\ }\href
  {https://doi.org/10.1103/PhysRevC.64.024901} {\bibfield  {journal} {\bibinfo
  {journal} {Phys. Rev. C}\ }\textbf {\bibinfo {volume} {64}},\ \bibinfo
  {pages} {024901} (\bibinfo {year} {2001})}\BibitemShut {NoStop}%
\bibitem [{\citenamefont {Uddin}(1998)}]{IEoS15}%
  \BibitemOpen
  \bibfield  {author} {\bibinfo {author} {\bibfnamefont {S.}~\bibnamefont
  {Uddin}},\ }\href {https://doi.org/10.1007/s100520050347} {\bibfield
  {journal} {\bibinfo  {journal} {Eur. Phys. J. C}\ }\textbf {\bibinfo {volume}
  {6}},\ \bibinfo {pages} {355} (\bibinfo {year} {1998})}\BibitemShut {NoStop}%
\bibitem [{\citenamefont {Heinz}(2004)}]{IEoS20}%
  \BibitemOpen
  \bibfield  {author} {\bibinfo {author} {\bibfnamefont {U.~W.}\ \bibnamefont
  {Heinz}},\ }\href@noop {} {} (\bibinfo {year} {2004}),\ \Eprint
  {https://arxiv.org/abs/hep-ph/0407360} {arXiv:hep-ph/0407360 [hep-ph]}
  \BibitemShut {NoStop}%
\bibitem [{\citenamefont {Bhat}\ \emph {et~al.}(2015)\citenamefont {Bhat},
  \citenamefont {Uddin},\ and\ \citenamefont {ul~Bashir}}]{IEoS22}%
  \BibitemOpen
  \bibfield  {author} {\bibinfo {author} {\bibfnamefont {R.~A.}\ \bibnamefont
  {Bhat}}, \bibinfo {author} {\bibfnamefont {S.}~\bibnamefont {Uddin}},\ and\
  \bibinfo {author} {\bibfnamefont {I.}~\bibnamefont {ul~Bashir}},\ }\href
  {https://doi.org/https://doi.org/10.1016/j.nuclphysa.2015.01.002} {\bibfield
  {journal} {\bibinfo  {journal} {Nucl. Phys. A}\ }\textbf {\bibinfo {volume}
  {935}},\ \bibinfo {pages} {43} (\bibinfo {year} {2015})}\BibitemShut
  {NoStop}%
\bibitem [{\citenamefont {Cleymans}\ \emph {et~al.}(2005)\citenamefont
  {Cleymans}, \citenamefont {K\"ampfer}, \citenamefont {Kaneta}, \citenamefont
  {Wheaton},\ and\ \citenamefont {Xu}}]{IEoS23}%
  \BibitemOpen
  \bibfield  {author} {\bibinfo {author} {\bibfnamefont {J.}~\bibnamefont
  {Cleymans}}, \bibinfo {author} {\bibfnamefont {B.}~\bibnamefont {K\"ampfer}},
  \bibinfo {author} {\bibfnamefont {M.}~\bibnamefont {Kaneta}}, \bibinfo
  {author} {\bibfnamefont {S.}~\bibnamefont {Wheaton}},\ and\ \bibinfo {author}
  {\bibfnamefont {N.}~\bibnamefont {Xu}},\ }\href
  {https://doi.org/10.1103/PhysRevC.71.054901} {\bibfield  {journal} {\bibinfo
  {journal} {Phys. Rev. C}\ }\textbf {\bibinfo {volume} {71}},\ \bibinfo
  {pages} {054901} (\bibinfo {year} {2005})}\BibitemShut {NoStop}%
\bibitem [{\citenamefont {Letessier}\ and\ \citenamefont
  {Rafelski}(2008)}]{IEoS24}%
  \BibitemOpen
  \bibfield  {author} {\bibinfo {author} {\bibfnamefont {J.}~\bibnamefont
  {Letessier}}\ and\ \bibinfo {author} {\bibfnamefont {J.}~\bibnamefont
  {Rafelski}},\ }\href {https://doi.org/10.1140/epja/i2007-10546-7} {\bibfield
  {journal} {\bibinfo  {journal} {Eur. Phys. J. A}\ }\textbf {\bibinfo {volume}
  {35}},\ \bibinfo {pages} {221} (\bibinfo {year} {2008})}\BibitemShut
  {NoStop}%
\bibitem [{\citenamefont {Becattini}(2009)}]{IEoS25}%
  \BibitemOpen
  \bibfield  {author} {\bibinfo {author} {\bibfnamefont {F.}~\bibnamefont
  {Becattini}},\ }\href@noop {} {\bibinfo {title} {An introduction to the
  statistical hadronization model}} (\bibinfo {year} {2009}),\ \Eprint
  {https://arxiv.org/abs/0901.3643} {arXiv:0901.3643 [hep-ph]} \BibitemShut
  {NoStop}%
\bibitem [{\citenamefont {Andronic}\ \emph {et~al.}(2018)\citenamefont
  {Andronic}, \citenamefont {Braun-Munzinger}, \citenamefont {Redlich},\ and\
  \citenamefont {Stachel}}]{IEoS26}%
  \BibitemOpen
  \bibfield  {author} {\bibinfo {author} {\bibfnamefont {A.}~\bibnamefont
  {Andronic}}, \bibinfo {author} {\bibfnamefont {P.}~\bibnamefont
  {Braun-Munzinger}}, \bibinfo {author} {\bibfnamefont {K.}~\bibnamefont
  {Redlich}},\ and\ \bibinfo {author} {\bibfnamefont {J.}~\bibnamefont
  {Stachel}},\ }\href {https://doi.org/10.1038/s41586-018-0491-6} {\bibfield
  {journal} {\bibinfo  {journal} {Nature}\ }\textbf {\bibinfo {volume} {561}},\
  \bibinfo {pages} {321} (\bibinfo {year} {2018})}\BibitemShut {NoStop}%
\bibitem [{\citenamefont {Borsányi}\ \emph {et~al.}(2012)\citenamefont
  {Borsányi}, \citenamefont {Fodor}, \citenamefont {Katz}, \citenamefont
  {Krieg}, \citenamefont {Ratti},\ and\ \citenamefont {Szabó}}]{IEoS27}%
  \BibitemOpen
  \bibfield  {author} {\bibinfo {author} {\bibfnamefont {S.}~\bibnamefont
  {Borsányi}}, \bibinfo {author} {\bibfnamefont {Z.}~\bibnamefont {Fodor}},
  \bibinfo {author} {\bibfnamefont {S.}~\bibnamefont {Katz}}, \bibinfo {author}
  {\bibfnamefont {S.}~\bibnamefont {Krieg}}, \bibinfo {author} {\bibfnamefont
  {C.}~\bibnamefont {Ratti}},\ and\ \bibinfo {author} {\bibfnamefont
  {K.}~\bibnamefont {Szabó}},\ }\href
  {https://doi.org/10.1007/JHEP01(2012)138} {\bibfield  {journal} {\bibinfo
  {journal} {J. High Energ. Phys.}\ }\textbf {\bibinfo {volume} {2012}},\
  \bibinfo {pages} {138}}\BibitemShut {NoStop}%
\bibitem [{\citenamefont {Bazavov}\ \emph {et~al.}(2012)\citenamefont
  {Bazavov}, \citenamefont {Bhattacharya}, \citenamefont {DeTar}, \citenamefont
  {Ding}, \citenamefont {Gottlieb}, \citenamefont {Gupta}, \citenamefont
  {Hegde}, \citenamefont {Heller}, \citenamefont {Karsch}, \citenamefont
  {Laermann}, \citenamefont {Levkova}, \citenamefont {Mukherjee}, \citenamefont
  {Petreczky}, \citenamefont {Schmidt}, \citenamefont {Soltz}, \citenamefont
  {Soeldner}, \citenamefont {Sugar},\ and\ \citenamefont {Vranas}}]{IEoS28}%
  \BibitemOpen
  \bibfield  {author} {\bibinfo {author} {\bibfnamefont {A.}~\bibnamefont
  {Bazavov}}, \bibinfo {author} {\bibfnamefont {T.}~\bibnamefont
  {Bhattacharya}}, \bibinfo {author} {\bibfnamefont {C.~E.}\ \bibnamefont
  {DeTar}}, \bibinfo {author} {\bibfnamefont {H.-T.}\ \bibnamefont {Ding}},
  \bibinfo {author} {\bibfnamefont {S.}~\bibnamefont {Gottlieb}}, \bibinfo
  {author} {\bibfnamefont {R.}~\bibnamefont {Gupta}}, \bibinfo {author}
  {\bibfnamefont {P.}~\bibnamefont {Hegde}}, \bibinfo {author} {\bibfnamefont
  {U.~M.}\ \bibnamefont {Heller}}, \bibinfo {author} {\bibfnamefont
  {F.}~\bibnamefont {Karsch}}, \bibinfo {author} {\bibfnamefont
  {E.}~\bibnamefont {Laermann}}, \bibinfo {author} {\bibfnamefont
  {L.}~\bibnamefont {Levkova}}, \bibinfo {author} {\bibfnamefont
  {S.}~\bibnamefont {Mukherjee}}, \bibinfo {author} {\bibfnamefont
  {P.}~\bibnamefont {Petreczky}}, \bibinfo {author} {\bibfnamefont
  {C.}~\bibnamefont {Schmidt}}, \bibinfo {author} {\bibfnamefont {R.~A.}\
  \bibnamefont {Soltz}}, \bibinfo {author} {\bibfnamefont {W.}~\bibnamefont
  {Soeldner}}, \bibinfo {author} {\bibfnamefont {R.}~\bibnamefont {Sugar}},\
  and\ \bibinfo {author} {\bibfnamefont {P.~M.}\ \bibnamefont {Vranas}}
  (\bibinfo {collaboration} {HotQCD Collaboration}),\ }\href
  {https://doi.org/10.1103/PhysRevD.86.034509} {\bibfield  {journal} {\bibinfo
  {journal} {Phys. Rev. D}\ }\textbf {\bibinfo {volume} {86}},\ \bibinfo
  {pages} {034509} (\bibinfo {year} {2012})}\BibitemShut {NoStop}%
\bibitem [{\citenamefont {Bellwied}\ \emph {et~al.}(2015)\citenamefont
  {Bellwied}, \citenamefont {Bors\'anyi}, \citenamefont {Fodor}, \citenamefont
  {Katz}, \citenamefont {P\'asztor}, \citenamefont {Ratti},\ and\ \citenamefont
  {Szab\'o}}]{IEoS29}%
  \BibitemOpen
  \bibfield  {author} {\bibinfo {author} {\bibfnamefont {R.}~\bibnamefont
  {Bellwied}}, \bibinfo {author} {\bibfnamefont {S.}~\bibnamefont
  {Bors\'anyi}}, \bibinfo {author} {\bibfnamefont {Z.}~\bibnamefont {Fodor}},
  \bibinfo {author} {\bibfnamefont {S.~D.}\ \bibnamefont {Katz}}, \bibinfo
  {author} {\bibfnamefont {A.}~\bibnamefont {P\'asztor}}, \bibinfo {author}
  {\bibfnamefont {C.}~\bibnamefont {Ratti}},\ and\ \bibinfo {author}
  {\bibfnamefont {K.~K.}\ \bibnamefont {Szab\'o}},\ }\href
  {https://doi.org/10.1103/PhysRevD.92.114505} {\bibfield  {journal} {\bibinfo
  {journal} {Phys. Rev. D}\ }\textbf {\bibinfo {volume} {92}},\ \bibinfo
  {pages} {114505} (\bibinfo {year} {2015})}\BibitemShut {NoStop}%
\bibitem [{\citenamefont {Bellwied}\ \emph {et~al.}(2013)\citenamefont
  {Bellwied}, \citenamefont {Borsanyi}, \citenamefont {Fodor}, \citenamefont
  {Katz},\ and\ \citenamefont {Ratti}}]{IEoS30}%
  \BibitemOpen
  \bibfield  {author} {\bibinfo {author} {\bibfnamefont {R.}~\bibnamefont
  {Bellwied}}, \bibinfo {author} {\bibfnamefont {S.}~\bibnamefont {Borsanyi}},
  \bibinfo {author} {\bibfnamefont {Z.}~\bibnamefont {Fodor}}, \bibinfo
  {author} {\bibfnamefont {S.~D.}\ \bibnamefont {Katz}},\ and\ \bibinfo
  {author} {\bibfnamefont {C.}~\bibnamefont {Ratti}},\ }\href
  {https://doi.org/10.1103/PhysRevLett.111.202302} {\bibfield  {journal}
  {\bibinfo  {journal} {Phys. Rev. Lett.}\ }\textbf {\bibinfo {volume} {111}},\
  \bibinfo {pages} {202302} (\bibinfo {year} {2013})}\BibitemShut {NoStop}%
\bibitem [{\citenamefont {Rischke}\ \emph
  {et~al.}(1991{\natexlab{a}})\citenamefont {Rischke}, \citenamefont
  {Gorenstein}, \citenamefont {Stöcker},\ and\ \citenamefont
  {Greiner}}]{IEoS31}%
  \BibitemOpen
  \bibfield  {author} {\bibinfo {author} {\bibfnamefont {D.~H.}\ \bibnamefont
  {Rischke}}, \bibinfo {author} {\bibfnamefont {M.}~\bibnamefont {Gorenstein}},
  \bibinfo {author} {\bibfnamefont {H.}~\bibnamefont {Stöcker}},\ and\
  \bibinfo {author} {\bibfnamefont {W.}~\bibnamefont {Greiner}},\ }\href
  {https://doi.org/10.1007/BF01548574} {\bibfield  {journal} {\bibinfo
  {journal} {Z. Phys. C}\ }\textbf {\bibinfo {volume} {51}},\ \bibinfo {pages}
  {485} (\bibinfo {year} {1991}{\natexlab{a}})}\BibitemShut {NoStop}%
\bibitem [{\citenamefont {Yen}\ \emph {et~al.}(1997)\citenamefont {Yen},
  \citenamefont {Gorenstein}, \citenamefont {Greiner},\ and\ \citenamefont
  {Yang}}]{IEoS32}%
  \BibitemOpen
  \bibfield  {author} {\bibinfo {author} {\bibfnamefont {G.~D.}\ \bibnamefont
  {Yen}}, \bibinfo {author} {\bibfnamefont {M.~I.}\ \bibnamefont {Gorenstein}},
  \bibinfo {author} {\bibfnamefont {W.}~\bibnamefont {Greiner}},\ and\ \bibinfo
  {author} {\bibfnamefont {S.~N.}\ \bibnamefont {Yang}},\ }\href
  {https://doi.org/10.1103/PhysRevC.56.2210} {\bibfield  {journal} {\bibinfo
  {journal} {Phys. Rev. C}\ }\textbf {\bibinfo {volume} {56}},\ \bibinfo
  {pages} {2210} (\bibinfo {year} {1997})}\BibitemShut {NoStop}%
\bibitem [{\citenamefont {Yen}\ and\ \citenamefont
  {Gorenstein}(1999)}]{IEoS33}%
  \BibitemOpen
  \bibfield  {author} {\bibinfo {author} {\bibfnamefont {G.~D.}\ \bibnamefont
  {Yen}}\ and\ \bibinfo {author} {\bibfnamefont {M.~I.}\ \bibnamefont
  {Gorenstein}},\ }\href {https://doi.org/10.1103/PhysRevC.59.2788} {\bibfield
  {journal} {\bibinfo  {journal} {Phys. Rev. C}\ }\textbf {\bibinfo {volume}
  {59}},\ \bibinfo {pages} {2788} (\bibinfo {year} {1999})}\BibitemShut
  {NoStop}%
\bibitem [{\citenamefont {Satarov}\ \emph {et~al.}(2017)\citenamefont
  {Satarov}, \citenamefont {Vovchenko}, \citenamefont {Alba}, \citenamefont
  {Gorenstein},\ and\ \citenamefont {Stoecker}}]{IEoS34}%
  \BibitemOpen
  \bibfield  {author} {\bibinfo {author} {\bibfnamefont {L.~M.}\ \bibnamefont
  {Satarov}}, \bibinfo {author} {\bibfnamefont {V.}~\bibnamefont {Vovchenko}},
  \bibinfo {author} {\bibfnamefont {P.}~\bibnamefont {Alba}}, \bibinfo {author}
  {\bibfnamefont {M.~I.}\ \bibnamefont {Gorenstein}},\ and\ \bibinfo {author}
  {\bibfnamefont {H.}~\bibnamefont {Stoecker}},\ }\href
  {https://doi.org/10.1103/PhysRevC.95.024902} {\bibfield  {journal} {\bibinfo
  {journal} {Phys. Rev. C}\ }\textbf {\bibinfo {volume} {95}},\ \bibinfo
  {pages} {024902} (\bibinfo {year} {2017})}\BibitemShut {NoStop}%
\bibitem [{\citenamefont {Cleymans}\ and\ \citenamefont
  {Satz}(1993{\natexlab{a}})}]{cleymanssatz}%
  \BibitemOpen
  \bibfield  {author} {\bibinfo {author} {\bibfnamefont {J.}~\bibnamefont
  {Cleymans}}\ and\ \bibinfo {author} {\bibfnamefont {H.}~\bibnamefont
  {Satz}},\ }\href {https://doi.org/10.1007/BF01555746} {\bibfield  {journal}
  {\bibinfo  {journal} {Z. Phys. C}\ }\textbf {\bibinfo {volume} {57}},\
  \bibinfo {pages} {135} (\bibinfo {year} {1993}{\natexlab{a}})}\BibitemShut
  {NoStop}%
\bibitem [{\citenamefont {Cleymans}\ \emph {et~al.}(1993)\citenamefont
  {Cleymans}, \citenamefont {Redlich}, \citenamefont {Satz},\ and\
  \citenamefont {Suhonen}}]{cleymansredlich}%
  \BibitemOpen
  \bibfield  {author} {\bibinfo {author} {\bibfnamefont {J.}~\bibnamefont
  {Cleymans}}, \bibinfo {author} {\bibfnamefont {K.}~\bibnamefont {Redlich}},
  \bibinfo {author} {\bibfnamefont {H.}~\bibnamefont {Satz}},\ and\ \bibinfo
  {author} {\bibfnamefont {E.}~\bibnamefont {Suhonen}},\ }\href
  {https://doi.org/10.1007/BF01560356} {\bibfield  {journal} {\bibinfo
  {journal} {Z. Phys. C}\ }\textbf {\bibinfo {volume} {58}},\ \bibinfo {pages}
  {347} (\bibinfo {year} {1993})}\BibitemShut {NoStop}%
\bibitem [{\citenamefont {Braun-Munzinger}\ \emph {et~al.}(1995)\citenamefont
  {Braun-Munzinger}, \citenamefont {Stachel}, \citenamefont {Wessels},\ and\
  \citenamefont {Xu}}]{MunzingerXu}%
  \BibitemOpen
  \bibfield  {author} {\bibinfo {author} {\bibfnamefont {P.}~\bibnamefont
  {Braun-Munzinger}}, \bibinfo {author} {\bibfnamefont {J.}~\bibnamefont
  {Stachel}}, \bibinfo {author} {\bibfnamefont {J.~P.}\ \bibnamefont
  {Wessels}},\ and\ \bibinfo {author} {\bibfnamefont {N.}~\bibnamefont {Xu}},\
  }\href {https://doi.org/10.1016/0370-2693(94)01534-J} {\bibfield  {journal}
  {\bibinfo  {journal} {Phys. lett. B}\ }\textbf {\bibinfo {volume} {344}},\
  \bibinfo {pages} {43} (\bibinfo {year} {1995})}\BibitemShut {NoStop}%
\bibitem [{\citenamefont {Braun-Munzinger}\ \emph {et~al.}(1996)\citenamefont
  {Braun-Munzinger}, \citenamefont {Stachel}, \citenamefont {Wessels},\ and\
  \citenamefont {Xu}}]{IEoS39}%
  \BibitemOpen
  \bibfield  {author} {\bibinfo {author} {\bibfnamefont {P.}~\bibnamefont
  {Braun-Munzinger}}, \bibinfo {author} {\bibfnamefont {J.}~\bibnamefont
  {Stachel}}, \bibinfo {author} {\bibfnamefont {J.}~\bibnamefont {Wessels}},\
  and\ \bibinfo {author} {\bibfnamefont {N.}~\bibnamefont {Xu}},\ }\href
  {https://doi.org/https://doi.org/10.1016/0370-2693(95)01258-3} {\bibfield
  {journal} {\bibinfo  {journal} {Phys. Lett. B}\ }\textbf {\bibinfo {volume}
  {365}},\ \bibinfo {pages} {1} (\bibinfo {year} {1996})}\BibitemShut {NoStop}%
\bibitem [{\citenamefont {Ritchie}\ \emph {et~al.}(2014)\citenamefont
  {Ritchie}, \citenamefont {Gorenstein},\ and\ \citenamefont
  {Miller}}]{IEoS40}%
  \BibitemOpen
  \bibfield  {author} {\bibinfo {author} {\bibfnamefont {R.~A.}\ \bibnamefont
  {Ritchie}}, \bibinfo {author} {\bibfnamefont {M.~I.}\ \bibnamefont
  {Gorenstein}},\ and\ \bibinfo {author} {\bibfnamefont {H.~G.}\ \bibnamefont
  {Miller}},\ }\href {https://doi.org/10.1007/s002880050497} {\bibfield
  {journal} {\bibinfo  {journal} {Z. Phys. C}\ }\textbf {\bibinfo {volume}
  {75}},\ \bibinfo {pages} {535} (\bibinfo {year} {2014})}\BibitemShut
  {NoStop}%
\bibitem [{\citenamefont {Singh}\ \emph {et~al.}(2011)\citenamefont {Singh},
  \citenamefont {Srivastava},\ and\ \citenamefont {Tiwari}}]{Bag}%
  \BibitemOpen
  \bibfield  {author} {\bibinfo {author} {\bibfnamefont {C.~P.}\ \bibnamefont
  {Singh}}, \bibinfo {author} {\bibfnamefont {P.~K.}\ \bibnamefont
  {Srivastava}},\ and\ \bibinfo {author} {\bibfnamefont {S.~K.}\ \bibnamefont
  {Tiwari}},\ }\href {https://doi.org/10.1103/PhysRevD.83.039904} {\bibfield
  {journal} {\bibinfo  {journal} {Phys. Rev. D}\ }\textbf {\bibinfo {volume}
  {83}},\ \bibinfo {pages} {039904} (\bibinfo {year} {2011})}\BibitemShut
  {NoStop}%
\bibitem [{\citenamefont {Cleymans}\ and\ \citenamefont
  {Suhonen}(1987)}]{cleymans-Suhonen}%
  \BibitemOpen
  \bibfield  {author} {\bibinfo {author} {\bibfnamefont {J.}~\bibnamefont
  {Cleymans}}\ and\ \bibinfo {author} {\bibfnamefont {E.}~\bibnamefont
  {Suhonen}},\ }\href {https://doi.org/10.1007/BF01442067} {\bibfield
  {journal} {\bibinfo  {journal} {Z. Phys. C}\ }\textbf {\bibinfo {volume}
  {37}},\ \bibinfo {pages} {51} (\bibinfo {year} {1987})}\BibitemShut {NoStop}%
\bibitem [{\citenamefont {Hagedorn}\ and\ \citenamefont
  {Rafelski}(1980{\natexlab{a}})}]{IEoS52}%
  \BibitemOpen
  \bibfield  {author} {\bibinfo {author} {\bibfnamefont {R.}~\bibnamefont
  {Hagedorn}}\ and\ \bibinfo {author} {\bibfnamefont {J.}~\bibnamefont
  {Rafelski}},\ }\href
  {https://doi.org/https://doi.org/10.1016/0370-2693(80)90566-3} {\bibfield
  {journal} {\bibinfo  {journal} {Phys. Lett. B}\ }\textbf {\bibinfo {volume}
  {97}},\ \bibinfo {pages} {136} (\bibinfo {year}
  {1980}{\natexlab{a}})}\BibitemShut {NoStop}%
\bibitem [{\citenamefont {Hagedorn}(1983)}]{hagedorn}%
  \BibitemOpen
  \bibfield  {author} {\bibinfo {author} {\bibfnamefont {R.}~\bibnamefont
  {Hagedorn}},\ }\href {https://doi.org/10.1007/BF01578153} {\bibfield
  {journal} {\bibinfo  {journal} {Z. Phys. C}\ }\textbf {\bibinfo {volume}
  {17}},\ \bibinfo {pages} {265} (\bibinfo {year} {1983})}\BibitemShut
  {NoStop}%
\bibitem [{\citenamefont {Kuono}\ and\ \citenamefont
  {Takagi}(1989)}]{kuonotakagi}%
  \BibitemOpen
  \bibfield  {author} {\bibinfo {author} {\bibfnamefont {H.}~\bibnamefont
  {Kuono}}\ and\ \bibinfo {author} {\bibfnamefont {F.}~\bibnamefont {Takagi}},\
  }\href {https://doi.org/10.1007/BF01555858} {\bibfield  {journal} {\bibinfo
  {journal} {Z. Phys. C}\ }\textbf {\bibinfo {volume} {42}},\ \bibinfo {pages}
  {209} (\bibinfo {year} {1989})}\BibitemShut {NoStop}%
\bibitem [{\citenamefont {Rischke}\ \emph
  {et~al.}(1991{\natexlab{b}})\citenamefont {Rischke}, \citenamefont
  {Gorenstein}, \citenamefont {Stöcker},\ and\ \citenamefont
  {Greiner}}]{rischke}%
  \BibitemOpen
  \bibfield  {author} {\bibinfo {author} {\bibfnamefont {D.}~\bibnamefont
  {Rischke}}, \bibinfo {author} {\bibfnamefont {M.}~\bibnamefont {Gorenstein}},
  \bibinfo {author} {\bibfnamefont {H.}~\bibnamefont {Stöcker}},\ and\
  \bibinfo {author} {\bibfnamefont {W.}~\bibnamefont {Greiner}},\ }\href
  {https://doi.org/10.1007/BF01548574} {\bibfield  {journal} {\bibinfo
  {journal} {Z. Phys. C}\ }\textbf {\bibinfo {volume} {51}},\ \bibinfo {pages}
  {485} (\bibinfo {year} {1991}{\natexlab{b}})}\BibitemShut {NoStop}%
\bibitem [{\citenamefont {Kapusta}(1981)}]{IEoS55}%
  \BibitemOpen
  \bibfield  {author} {\bibinfo {author} {\bibfnamefont {J.~I.}\ \bibnamefont
  {Kapusta}},\ }\href {https://doi.org/10.1103/PhysRevD.23.2444} {\bibfield
  {journal} {\bibinfo  {journal} {Phys. Rev. D}\ }\textbf {\bibinfo {volume}
  {23}},\ \bibinfo {pages} {2444} (\bibinfo {year} {1981})}\BibitemShut
  {NoStop}%
\bibitem [{\citenamefont {Kapusta}\ and\ \citenamefont {Olive}(1988)}]{IEoS56}%
  \BibitemOpen
  \bibfield  {author} {\bibinfo {author} {\bibfnamefont {J.~I.}\ \bibnamefont
  {Kapusta}}\ and\ \bibinfo {author} {\bibfnamefont {K.~A.}\ \bibnamefont
  {Olive}},\ }\href
  {https://doi.org/https://doi.org/10.1016/0370-2693(88)90949-5} {\bibfield
  {journal} {\bibinfo  {journal} {Phys. Lett. B}\ }\textbf {\bibinfo {volume}
  {209}},\ \bibinfo {pages} {295} (\bibinfo {year} {1988})}\BibitemShut
  {NoStop}%
\bibitem [{\citenamefont {Zhang}(1995)}]{IEoS58}%
  \BibitemOpen
  \bibfield  {author} {\bibinfo {author} {\bibfnamefont {Q.~R.}\ \bibnamefont
  {Zhang}},\ }\href {https://doi.org/10.1007/BF01292790} {\bibfield  {journal}
  {\bibinfo  {journal} {Z. Phys. C}\ }\textbf {\bibinfo {volume} {351}},\
  \bibinfo {pages} {89} (\bibinfo {year} {1995})}\BibitemShut {NoStop}%
\bibitem [{\citenamefont {Ma}\ \emph {et~al.}(1993)\citenamefont {Ma},
  \citenamefont {Zhang}, \citenamefont {Rischke},\ and\ \citenamefont
  {Greiner}}]{IEoS59}%
  \BibitemOpen
  \bibfield  {author} {\bibinfo {author} {\bibfnamefont {B.-Q.}\ \bibnamefont
  {Ma}}, \bibinfo {author} {\bibfnamefont {Q.-R.}\ \bibnamefont {Zhang}},
  \bibinfo {author} {\bibfnamefont {D.}~\bibnamefont {Rischke}},\ and\ \bibinfo
  {author} {\bibfnamefont {W.}~\bibnamefont {Greiner}},\ }\href
  {https://doi.org/https://doi.org/10.1016/0370-2693(93)90153-9} {\bibfield
  {journal} {\bibinfo  {journal} {Phys. Lett. B}\ }\textbf {\bibinfo {volume}
  {315}},\ \bibinfo {pages} {29} (\bibinfo {year} {1993})}\BibitemShut
  {NoStop}%
\bibitem [{\citenamefont {Uddin}\ and\ \citenamefont
  {Singh}(1993)}]{uddinsingh}%
  \BibitemOpen
  \bibfield  {author} {\bibinfo {author} {\bibfnamefont {S.}~\bibnamefont
  {Uddin}}\ and\ \bibinfo {author} {\bibfnamefont {C.~P.}\ \bibnamefont
  {Singh}},\ }\href {https://doi.org/10.1007/BF01577554} {\bibfield  {journal}
  {\bibinfo  {journal} {Z. Phys. C}\ }\textbf {\bibinfo {volume} {63}},\
  \bibinfo {pages} {147} (\bibinfo {year} {1993})}\BibitemShut {NoStop}%
\bibitem [{\citenamefont {Uddin}(1995)}]{uddinplb}%
  \BibitemOpen
  \bibfield  {author} {\bibinfo {author} {\bibfnamefont {S.}~\bibnamefont
  {Uddin}},\ }\href {https://doi.org/10.1016/0370-2693(95)80015-P} {\bibfield
  {journal} {\bibinfo  {journal} {Phys. lett. B}\ }\textbf {\bibinfo {volume}
  {341}},\ \bibinfo {pages} {361} (\bibinfo {year} {1995})}\BibitemShut
  {NoStop}%
\bibitem [{\citenamefont {Mir}\ \emph {et~al.}(2025)\citenamefont {Mir},
  \citenamefont {Rather}, \citenamefont {Mohi Ud~Din},\ and\ \citenamefont
  {Uddin}}]{Sam:2025}%
  \BibitemOpen
  \bibfield  {author} {\bibinfo {author} {\bibfnamefont {S.~A.}\ \bibnamefont
  {Mir}}, \bibinfo {author} {\bibfnamefont {N.~A.}\ \bibnamefont {Rather}},
  \bibinfo {author} {\bibfnamefont {I.}~\bibnamefont {Mohi Ud~Din}},\ and\
  \bibinfo {author} {\bibfnamefont {S.}~\bibnamefont {Uddin}},\ }\href
  {https://doi.org/10.1088/1361-6471/adb6c2} {\bibfield  {journal} {\bibinfo
  {journal} {J. Phys. G: Nucl. Part. Phys.}\ }\textbf {\bibinfo {volume}
  {52}},\ \bibinfo {pages} {035003} (\bibinfo {year} {2025})},\ \Eprint
  {https://arxiv.org/abs/2312.13079} {arXiv:2312.13079 [hep-ph]} \BibitemShut
  {NoStop}%
\bibitem [{\citenamefont {Bugaev}\ \emph {et~al.}(2000)\citenamefont {Bugaev},
  \citenamefont {Gorenstein}, \citenamefont {Stöcker},\ and\ \citenamefont
  {Greiner}}]{Ref4/3}%
  \BibitemOpen
  \bibfield  {author} {\bibinfo {author} {\bibfnamefont {K.}~\bibnamefont
  {Bugaev}}, \bibinfo {author} {\bibfnamefont {M.}~\bibnamefont {Gorenstein}},
  \bibinfo {author} {\bibfnamefont {H.}~\bibnamefont {Stöcker}},\ and\
  \bibinfo {author} {\bibfnamefont {W.}~\bibnamefont {Greiner}},\ }\href
  {https://doi.org/https://doi.org/10.1016/S0370-2693(00)00690-0} {\bibfield
  {journal} {\bibinfo  {journal} {Phys. Lett. B}\ }\textbf {\bibinfo {volume}
  {485}},\ \bibinfo {pages} {121} (\bibinfo {year} {2000})}\BibitemShut
  {NoStop}%
\bibitem [{\citenamefont {Tiwari}\ \emph {et~al.}(2012)\citenamefont {Tiwari},
  \citenamefont {Srivastava},\ and\ \citenamefont {Singh}}]{SKT}%
  \BibitemOpen
  \bibfield  {author} {\bibinfo {author} {\bibfnamefont {S.~K.}\ \bibnamefont
  {Tiwari}}, \bibinfo {author} {\bibfnamefont {P.~K.}\ \bibnamefont
  {Srivastava}},\ and\ \bibinfo {author} {\bibfnamefont {C.~P.}\ \bibnamefont
  {Singh}},\ }\href {https://doi.org/10.1103/PhysRevC.85.014908} {\bibfield
  {journal} {\bibinfo  {journal} {Phys. Rev. C}\ }\textbf {\bibinfo {volume}
  {85}},\ \bibinfo {pages} {014908} (\bibinfo {year} {2012})}\BibitemShut
  {NoStop}%
\bibitem [{\citenamefont {Wong}(2000)}]{Baryonstopping}%
  \BibitemOpen
  \bibfield  {author} {\bibinfo {author} {\bibfnamefont {S.}~\bibnamefont
  {Wong}},\ }\href
  {https://doi.org/https://doi.org/10.1016/S0370-2693(00)00408-1} {\bibfield
  {journal} {\bibinfo  {journal} {Phys. Lett. B}\ }\textbf {\bibinfo {volume}
  {480}},\ \bibinfo {pages} {65} (\bibinfo {year} {2000})}\BibitemShut
  {NoStop}%
\bibitem [{\citenamefont {Ding}\ \emph {et~al.}(2001)\citenamefont {Ding},
  \citenamefont {Glässel},\ and\ \citenamefont
  {Hüfner}}]{Nucleartransparency}%
  \BibitemOpen
  \bibfield  {author} {\bibinfo {author} {\bibfnamefont {H.}~\bibnamefont
  {Ding}}, \bibinfo {author} {\bibfnamefont {P.}~\bibnamefont {Glässel}},\
  and\ \bibinfo {author} {\bibfnamefont {J.}~\bibnamefont {Hüfner}},\ }\href
  {https://doi.org/https://doi.org/10.1016/S0375-9474(01)00657-1} {\bibfield
  {journal} {\bibinfo  {journal} {Nucl. Phys. A}\ }\textbf {\bibinfo {volume}
  {692}},\ \bibinfo {pages} {549} (\bibinfo {year} {2001})}\BibitemShut
  {NoStop}%
\bibitem [{\citenamefont {Bearden~$et\: al.$}(2004)}]{Transparency}%
  \BibitemOpen
  \bibfield  {author} {\bibinfo {author} {\bibfnamefont {I.~G.}\ \bibnamefont
  {Bearden~$et\: al.$}} (\bibinfo {collaboration} {BRAHMS Collaboration}),\
  }\href {https://doi.org/10.1103/PhysRevLett.93.102301} {\bibfield  {journal}
  {\bibinfo  {journal} {Phys. Rev. Lett.}\ }\textbf {\bibinfo {volume} {93}},\
  \bibinfo {pages} {102301} (\bibinfo {year} {2004})}\BibitemShut {NoStop}%
\bibitem [{\citenamefont {Cleymans}\ \emph {et~al.}(2008)\citenamefont
  {Cleymans}, \citenamefont {Str\"umpfer},\ and\ \citenamefont
  {Turko}}]{Transparency1}%
  \BibitemOpen
  \bibfield  {author} {\bibinfo {author} {\bibfnamefont {J.}~\bibnamefont
  {Cleymans}}, \bibinfo {author} {\bibfnamefont {J.}~\bibnamefont
  {Str\"umpfer}},\ and\ \bibinfo {author} {\bibfnamefont {L.}~\bibnamefont
  {Turko}},\ }\href {https://doi.org/10.1103/PhysRevC.78.017901} {\bibfield
  {journal} {\bibinfo  {journal} {Phys. Rev. C}\ }\textbf {\bibinfo {volume}
  {78}},\ \bibinfo {pages} {017901} (\bibinfo {year} {2008})}\BibitemShut
  {NoStop}%
\bibitem [{\citenamefont {Buzzatti}\ and\ \citenamefont
  {Gyulassy}(2013)}]{Buzzatti}%
  \BibitemOpen
  \bibfield  {author} {\bibinfo {author} {\bibfnamefont {A.}~\bibnamefont
  {Buzzatti}}\ and\ \bibinfo {author} {\bibfnamefont {M.}~\bibnamefont
  {Gyulassy}},\ }\href
  {https://doi.org/https://doi.org/10.1016/j.nuclphysa.2013.02.133} {\bibfield
  {journal} {\bibinfo  {journal} {Nucl. Phys. A}\ }\textbf {\bibinfo {volume}
  {904-905}},\ \bibinfo {pages} {779c} (\bibinfo {year} {2013})},\ \bibinfo
  {note} {the Quark Matter 2012}\BibitemShut {NoStop}%
\bibitem [{\citenamefont {Andronic}\ \emph {et~al.}(2006)\citenamefont
  {Andronic}, \citenamefont {Braun-Munzinger},\ and\ \citenamefont
  {Stachel}}]{Andronic}%
  \BibitemOpen
  \bibfield  {author} {\bibinfo {author} {\bibfnamefont {A.}~\bibnamefont
  {Andronic}}, \bibinfo {author} {\bibfnamefont {P.}~\bibnamefont
  {Braun-Munzinger}},\ and\ \bibinfo {author} {\bibfnamefont {J.}~\bibnamefont
  {Stachel}},\ }\href {https://doi.org/10.1016/j.nuclphysa.2006.03.012}
  {\bibfield  {journal} {\bibinfo  {journal} {Nucl. Phys. A}\ }\textbf
  {\bibinfo {volume} {772}},\ \bibinfo {pages} {167} (\bibinfo {year}
  {2006})}\BibitemShut {NoStop}%
\bibitem [{\citenamefont {Wheaton}\ \emph {et~al.}(2009)\citenamefont
  {Wheaton}, \citenamefont {Cleymans},\ and\ \citenamefont {Hauer}}]{thermus}%
  \BibitemOpen
  \bibfield  {author} {\bibinfo {author} {\bibfnamefont {S.}~\bibnamefont
  {Wheaton}}, \bibinfo {author} {\bibfnamefont {J.}~\bibnamefont {Cleymans}},\
  and\ \bibinfo {author} {\bibfnamefont {M.}~\bibnamefont {Hauer}},\ }\href
  {https://doi.org/https://doi.org/10.1016/j.cpc.2008.08.001} {\bibfield
  {journal} {\bibinfo  {journal} {Comput. Phys. Commun.}\ }\textbf {\bibinfo
  {volume} {180}},\ \bibinfo {pages} {84} (\bibinfo {year} {2009})}\BibitemShut
  {NoStop}%
\bibitem [{\citenamefont {Vovchenko}\ and\ \citenamefont
  {Stoecker}(2019)}]{fist}%
  \BibitemOpen
  \bibfield  {author} {\bibinfo {author} {\bibfnamefont {V.}~\bibnamefont
  {Vovchenko}}\ and\ \bibinfo {author} {\bibfnamefont {H.}~\bibnamefont
  {Stoecker}},\ }\href
  {https://doi.org/https://doi.org/10.1016/j.cpc.2019.06.024} {\bibfield
  {journal} {\bibinfo  {journal} {Comput. Phys. Commun.}\ }\textbf {\bibinfo
  {volume} {244}},\ \bibinfo {pages} {295} (\bibinfo {year}
  {2019})}\BibitemShut {NoStop}%
\bibitem [{\citenamefont {Kisiel}\ \emph {et~al.}(2006)\citenamefont {Kisiel},
  \citenamefont {Tałuć}, \citenamefont {Broniowski},\ and\ \citenamefont
  {Florkowski}}]{therminator}%
  \BibitemOpen
  \bibfield  {author} {\bibinfo {author} {\bibfnamefont {A.}~\bibnamefont
  {Kisiel}}, \bibinfo {author} {\bibfnamefont {T.}~\bibnamefont {Tałuć}},
  \bibinfo {author} {\bibfnamefont {W.}~\bibnamefont {Broniowski}},\ and\
  \bibinfo {author} {\bibfnamefont {W.}~\bibnamefont {Florkowski}},\ }\href
  {https://doi.org/https://doi.org/10.1016/j.cpc.2005.11.010} {\bibfield
  {journal} {\bibinfo  {journal} {Comput. Phys. Commun.}\ }\textbf {\bibinfo
  {volume} {174}},\ \bibinfo {pages} {669} (\bibinfo {year}
  {2006})}\BibitemShut {NoStop}%
\bibitem [{\citenamefont {Gorenstein}\ \emph {et~al.}(1981)\citenamefont
  {Gorenstein}, \citenamefont {Petrov},\ and\ \citenamefont
  {Zinovjev}}]{GorensteinPLB}%
  \BibitemOpen
  \bibfield  {author} {\bibinfo {author} {\bibfnamefont {M.}~\bibnamefont
  {Gorenstein}}, \bibinfo {author} {\bibfnamefont {V.}~\bibnamefont {Petrov}},\
  and\ \bibinfo {author} {\bibfnamefont {G.}~\bibnamefont {Zinovjev}},\ }\href
  {https://doi.org/https://doi.org/10.1016/0370-2693(81)90546-3} {\bibfield
  {journal} {\bibinfo  {journal} {Phys. Lett. B}\ }\textbf {\bibinfo {volume}
  {106}},\ \bibinfo {pages} {327} (\bibinfo {year} {1981})}\BibitemShut
  {NoStop}%
\bibitem [{\citenamefont {Hagedorn}\ and\ \citenamefont
  {Rafelski}(1980{\natexlab{b}})}]{hagedorn-rafelski}%
  \BibitemOpen
  \bibfield  {author} {\bibinfo {author} {\bibfnamefont {R.}~\bibnamefont
  {Hagedorn}}\ and\ \bibinfo {author} {\bibfnamefont {J.}~\bibnamefont
  {Rafelski}},\ }\href {https://doi.org/10.1016/0370-2693(80)90566-3}
  {\bibfield  {journal} {\bibinfo  {journal} {Phys. Lett. B}\ }\textbf
  {\bibinfo {volume} {97}},\ \bibinfo {pages} {136} (\bibinfo {year}
  {1980}{\natexlab{b}})}\BibitemShut {NoStop}%
\bibitem [{\citenamefont {Vovchenko}\ \emph {et~al.}(2017)\citenamefont
  {Vovchenko}, \citenamefont {Gorenstein},\ and\ \citenamefont
  {Stoecker}}]{vovchenkoprl}%
  \BibitemOpen
  \bibfield  {author} {\bibinfo {author} {\bibfnamefont {V.}~\bibnamefont
  {Vovchenko}}, \bibinfo {author} {\bibfnamefont {M.~I.}\ \bibnamefont
  {Gorenstein}},\ and\ \bibinfo {author} {\bibfnamefont {H.}~\bibnamefont
  {Stoecker}},\ }\href {https://doi.org/10.1103/PhysRevLett.118.182301}
  {\bibfield  {journal} {\bibinfo  {journal} {Phys. Rev. Lett.}\ }\textbf
  {\bibinfo {volume} {118}},\ \bibinfo {pages} {182301} (\bibinfo {year}
  {2017})}\BibitemShut {NoStop}%
\bibitem [{\citenamefont {Vovchenko}\ \emph
  {et~al.}(2015{\natexlab{a}})\citenamefont {Vovchenko}, \citenamefont
  {Anchishkin},\ and\ \citenamefont {Gorenstein}}]{vovchenkoprc}%
  \BibitemOpen
  \bibfield  {author} {\bibinfo {author} {\bibfnamefont {V.}~\bibnamefont
  {Vovchenko}}, \bibinfo {author} {\bibfnamefont {D.~V.}\ \bibnamefont
  {Anchishkin}},\ and\ \bibinfo {author} {\bibfnamefont {M.~I.}\ \bibnamefont
  {Gorenstein}},\ }\href {https://doi.org/10.1103/PhysRevC.91.064314}
  {\bibfield  {journal} {\bibinfo  {journal} {Phys. Rev. C}\ }\textbf {\bibinfo
  {volume} {91}},\ \bibinfo {pages} {064314} (\bibinfo {year}
  {2015}{\natexlab{a}})}\BibitemShut {NoStop}%
\bibitem [{\citenamefont {Vovchenko}\ \emph
  {et~al.}(2015{\natexlab{b}})\citenamefont {Vovchenko}, \citenamefont
  {Anchishkin}, \citenamefont {Gorenstein},\ and\ \citenamefont
  {Poberezhnyuk}}]{vovchenkoprc1}%
  \BibitemOpen
  \bibfield  {author} {\bibinfo {author} {\bibfnamefont {V.}~\bibnamefont
  {Vovchenko}}, \bibinfo {author} {\bibfnamefont {D.}~\bibnamefont
  {Anchishkin}}, \bibinfo {author} {\bibfnamefont {M.~I.}\ \bibnamefont
  {Gorenstein}},\ and\ \bibinfo {author} {\bibfnamefont {R.~V.}\ \bibnamefont
  {Poberezhnyuk}},\ }\href {https://doi.org/10.1103/PhysRevC.92.054901}
  {\bibfield  {journal} {\bibinfo  {journal} {Phys. Rev. C}\ }\textbf {\bibinfo
  {volume} {92}},\ \bibinfo {pages} {054901} (\bibinfo {year}
  {2015}{\natexlab{b}})}\BibitemShut {NoStop}%
\bibitem [{\citenamefont {Samanta}\ and\ \citenamefont
  {Mohanty}(2018)}]{samanthamohanty}%
  \BibitemOpen
  \bibfield  {author} {\bibinfo {author} {\bibfnamefont {S.}~\bibnamefont
  {Samanta}}\ and\ \bibinfo {author} {\bibfnamefont {B.}~\bibnamefont
  {Mohanty}},\ }\href {https://doi.org/10.1103/PhysRevC.97.015201} {\bibfield
  {journal} {\bibinfo  {journal} {Phys. Rev. C}\ }\textbf {\bibinfo {volume}
  {97}},\ \bibinfo {pages} {015201} (\bibinfo {year} {2018})}\BibitemShut
  {NoStop}%
\bibitem [{\citenamefont {Andronic}\ \emph {et~al.}(2012)\citenamefont
  {Andronic}, \citenamefont {Braun-Munzinger}, \citenamefont {Stachel},\ and\
  \citenamefont {Winn}}]{andronicplb}%
  \BibitemOpen
  \bibfield  {author} {\bibinfo {author} {\bibfnamefont {A.}~\bibnamefont
  {Andronic}}, \bibinfo {author} {\bibfnamefont {P.}~\bibnamefont
  {Braun-Munzinger}}, \bibinfo {author} {\bibfnamefont {J.}~\bibnamefont
  {Stachel}},\ and\ \bibinfo {author} {\bibfnamefont {M.}~\bibnamefont
  {Winn}},\ }\href {https://doi.org/10.1016/j.physletb.2012.10.001} {\bibfield
  {journal} {\bibinfo  {journal} {Phys. lett. B}\ }\textbf {\bibinfo {volume}
  {718}},\ \bibinfo {pages} {80} (\bibinfo {year} {2012})}\BibitemShut
  {NoStop}%
\bibitem [{\citenamefont {Granddon}\ \emph {et~al.}(1997)\citenamefont
  {Granddon}, \citenamefont {Gorenstein}, \citenamefont {Greiner},\ and\
  \citenamefont {Nan~Yang}}]{granddon}%
  \BibitemOpen
  \bibfield  {author} {\bibinfo {author} {\bibfnamefont {D.}~\bibnamefont
  {Granddon}}, \bibinfo {author} {\bibfnamefont {M.~I.}\ \bibnamefont
  {Gorenstein}}, \bibinfo {author} {\bibfnamefont {W.}~\bibnamefont
  {Greiner}},\ and\ \bibinfo {author} {\bibfnamefont {S.}~\bibnamefont
  {Nan~Yang}},\ }\href {https://doi.org/10.1103/PhysRevC.56.2210} {\bibfield
  {journal} {\bibinfo  {journal} {Phys. Rev. C}\ }\textbf {\bibinfo {volume}
  {56}},\ \bibinfo {pages} {2210} (\bibinfo {year} {1997})}\BibitemShut
  {NoStop}%
\bibitem [{\citenamefont {Pradhan}\ \emph {et~al.}(2023)\citenamefont
  {Pradhan}, \citenamefont {Sahu}, \citenamefont {Scaria},\ and\ \citenamefont
  {Sahoo}}]{kkpradhan}%
  \BibitemOpen
  \bibfield  {author} {\bibinfo {author} {\bibfnamefont {K.~K.}\ \bibnamefont
  {Pradhan}}, \bibinfo {author} {\bibfnamefont {D.}~\bibnamefont {Sahu}},
  \bibinfo {author} {\bibfnamefont {R.}~\bibnamefont {Scaria}},\ and\ \bibinfo
  {author} {\bibfnamefont {R.}~\bibnamefont {Sahoo}},\ }\href
  {https://doi.org/10.1103/PhysRevC.107.014910} {\bibfield  {journal} {\bibinfo
   {journal} {Phys. Rev. C}\ }\textbf {\bibinfo {volume} {107}},\ \bibinfo
  {pages} {014910} (\bibinfo {year} {2023})}\BibitemShut {NoStop}%
\bibitem [{\citenamefont {Gorenstein}(2012)}]{GorensteinSingle}%
  \BibitemOpen
  \bibfield  {author} {\bibinfo {author} {\bibfnamefont {M.~I.}\ \bibnamefont
  {Gorenstein}},\ }\href {https://doi.org/10.1103/PhysRevC.86.044907}
  {\bibfield  {journal} {\bibinfo  {journal} {Phys. Rev. C}\ }\textbf {\bibinfo
  {volume} {86}},\ \bibinfo {pages} {044907} (\bibinfo {year}
  {2012})}\BibitemShut {NoStop}%
\bibitem [{\citenamefont {Redlich}\ \emph {et~al.}(1994)\citenamefont
  {Redlich}, \citenamefont {Cleymans}, \citenamefont {Satz},\ and\
  \citenamefont {Suhonen}}]{redlich}%
  \BibitemOpen
  \bibfield  {author} {\bibinfo {author} {\bibfnamefont {K.}~\bibnamefont
  {Redlich}}, \bibinfo {author} {\bibfnamefont {J.}~\bibnamefont {Cleymans}},
  \bibinfo {author} {\bibfnamefont {H.}~\bibnamefont {Satz}},\ and\ \bibinfo
  {author} {\bibfnamefont {E.}~\bibnamefont {Suhonen}},\ }\href
  {https://doi.org/10.1016/0375-9474(94)90652-1} {\bibfield  {journal}
  {\bibinfo  {journal} {Nucl. Phys. A}\ }\textbf {\bibinfo {volume} {566}},\
  \bibinfo {pages} {391} (\bibinfo {year} {1994})}\BibitemShut {NoStop}%
\bibitem [{\citenamefont {Chodos}\ \emph {et~al.}(1974)\citenamefont {Chodos},
  \citenamefont {Jaffe}, \citenamefont {Johnson}, \citenamefont {Thorn},\ and\
  \citenamefont {Weisskopf}}]{ChodosBag}%
  \BibitemOpen
  \bibfield  {author} {\bibinfo {author} {\bibfnamefont {A.}~\bibnamefont
  {Chodos}}, \bibinfo {author} {\bibfnamefont {R.~L.}\ \bibnamefont {Jaffe}},
  \bibinfo {author} {\bibfnamefont {K.}~\bibnamefont {Johnson}}, \bibinfo
  {author} {\bibfnamefont {C.~B.}\ \bibnamefont {Thorn}},\ and\ \bibinfo
  {author} {\bibfnamefont {V.~F.}\ \bibnamefont {Weisskopf}},\ }\href
  {https://doi.org/10.1103/PhysRevD.9.3471} {\bibfield  {journal} {\bibinfo
  {journal} {Phys. Rev. D}\ }\textbf {\bibinfo {volume} {9}},\ \bibinfo {pages}
  {3471} (\bibinfo {year} {1974})}\BibitemShut {NoStop}%
\bibitem [{\citenamefont {DeGrand}\ \emph {et~al.}(1975)\citenamefont
  {DeGrand}, \citenamefont {Jaffe}, \citenamefont {Johnson},\ and\
  \citenamefont {Kiskis}}]{DeGrandBag}%
  \BibitemOpen
  \bibfield  {author} {\bibinfo {author} {\bibfnamefont {T.}~\bibnamefont
  {DeGrand}}, \bibinfo {author} {\bibfnamefont {R.~L.}\ \bibnamefont {Jaffe}},
  \bibinfo {author} {\bibfnamefont {K.}~\bibnamefont {Johnson}},\ and\ \bibinfo
  {author} {\bibfnamefont {J.}~\bibnamefont {Kiskis}},\ }\href
  {https://doi.org/10.1103/PhysRevD.12.2060} {\bibfield  {journal} {\bibinfo
  {journal} {Phys. Rev. D}\ }\textbf {\bibinfo {volume} {12}},\ \bibinfo
  {pages} {2060} (\bibinfo {year} {1975})}\BibitemShut {NoStop}%
\bibitem [{\citenamefont {Cleymans}\ \emph {et~al.}(1986)\citenamefont
  {Cleymans}, \citenamefont {Gavai},\ and\ \citenamefont
  {Suhonen}}]{CleymansBag}%
  \BibitemOpen
  \bibfield  {author} {\bibinfo {author} {\bibfnamefont {J.}~\bibnamefont
  {Cleymans}}, \bibinfo {author} {\bibfnamefont {R.}~\bibnamefont {Gavai}},\
  and\ \bibinfo {author} {\bibfnamefont {E.}~\bibnamefont {Suhonen}},\ }\href
  {https://doi.org/https://doi.org/10.1016/0370-1573(86)90169-9} {\bibfield
  {journal} {\bibinfo  {journal} {Phys. Rep.}\ }\textbf {\bibinfo {volume}
  {130}},\ \bibinfo {pages} {217} (\bibinfo {year} {1986})}\BibitemShut
  {NoStop}%
\bibitem [{\citenamefont {Kadam}\ and\ \citenamefont {Mishra}(2016)}]{Guru1}%
  \BibitemOpen
  \bibfield  {author} {\bibinfo {author} {\bibfnamefont {G.~P.}\ \bibnamefont
  {Kadam}}\ and\ \bibinfo {author} {\bibfnamefont {H.}~\bibnamefont {Mishra}},\
  }\href {https://doi.org/10.1103/PhysRevC.93.025205} {\bibfield  {journal}
  {\bibinfo  {journal} {Phys. Rev. C}\ }\textbf {\bibinfo {volume} {93}},\
  \bibinfo {pages} {025205} (\bibinfo {year} {2016})}\BibitemShut {NoStop}%
\bibitem [{\citenamefont {Koch}\ \emph {et~al.}(1986)\citenamefont {Koch},
  \citenamefont {Müller},\ and\ \citenamefont {Rafelski}}]{IEoS16}%
  \BibitemOpen
  \bibfield  {author} {\bibinfo {author} {\bibfnamefont {P.}~\bibnamefont
  {Koch}}, \bibinfo {author} {\bibfnamefont {B.}~\bibnamefont {Müller}},\ and\
  \bibinfo {author} {\bibfnamefont {J.}~\bibnamefont {Rafelski}},\ }\href
  {https://doi.org/https://doi.org/10.1016/0370-1573(86)90096-7} {\bibfield
  {journal} {\bibinfo  {journal} {Phys. Rep.}\ }\textbf {\bibinfo {volume}
  {142}},\ \bibinfo {pages} {167} (\bibinfo {year} {1986})}\BibitemShut
  {NoStop}%
\bibitem [{\citenamefont {Alba}\ \emph {et~al.}(2014)\citenamefont {Alba},
  \citenamefont {Alberico}, \citenamefont {Bellwied}, \citenamefont {Bluhm},
  \citenamefont {Mantovani~Sarti}, \citenamefont {Nahrgang},\ and\
  \citenamefont {Ratti}}]{Albactob}%
  \BibitemOpen
  \bibfield  {author} {\bibinfo {author} {\bibfnamefont {P.}~\bibnamefont
  {Alba}}, \bibinfo {author} {\bibfnamefont {W.}~\bibnamefont {Alberico}},
  \bibinfo {author} {\bibfnamefont {R.}~\bibnamefont {Bellwied}}, \bibinfo
  {author} {\bibfnamefont {M.}~\bibnamefont {Bluhm}}, \bibinfo {author}
  {\bibfnamefont {V.}~\bibnamefont {Mantovani~Sarti}}, \bibinfo {author}
  {\bibfnamefont {M.}~\bibnamefont {Nahrgang}},\ and\ \bibinfo {author}
  {\bibfnamefont {C.}~\bibnamefont {Ratti}},\ }\href
  {https://doi.org/10.1016/j.physletb.2014.09.052} {\bibfield  {journal}
  {\bibinfo  {journal} {Phys. lett. B}\ }\textbf {\bibinfo {volume} {738}},\
  \bibinfo {pages} {305} (\bibinfo {year} {2014})}\BibitemShut {NoStop}%
\bibitem [{\citenamefont {Bhattacharyya}\ \emph {et~al.}(2020)\citenamefont
  {Bhattacharyya}, \citenamefont {Biswas}, \citenamefont {K.~Ghosh},
  \citenamefont {Ray},\ and\ \citenamefont {Singha}}]{Sumana}%
  \BibitemOpen
  \bibfield  {author} {\bibinfo {author} {\bibfnamefont {S.}~\bibnamefont
  {Bhattacharyya}}, \bibinfo {author} {\bibfnamefont {D.}~\bibnamefont
  {Biswas}}, \bibinfo {author} {\bibfnamefont {S.}~\bibnamefont {K.~Ghosh}},
  \bibinfo {author} {\bibfnamefont {R.}~\bibnamefont {Ray}},\ and\ \bibinfo
  {author} {\bibfnamefont {P.}~\bibnamefont {Singha}},\ }\href
  {https://doi.org/10.1103/PhysRevD.101.054002} {\bibfield  {journal} {\bibinfo
   {journal} {Phys. Rev. D}\ }\textbf {\bibinfo {volume} {101}},\ \bibinfo
  {pages} {054002} (\bibinfo {year} {2020})}\BibitemShut {NoStop}%
\bibitem [{\citenamefont {Bashir}\ \emph {et~al.}(2018)\citenamefont {Bashir},
  \citenamefont {Parra}, \citenamefont {Nanda},\ and\ \citenamefont
  {Uddin}}]{ansatztempchepot1}%
  \BibitemOpen
  \bibfield  {author} {\bibinfo {author} {\bibfnamefont {I.}~\bibnamefont
  {Bashir}}, \bibinfo {author} {\bibfnamefont {R.~A.}\ \bibnamefont {Parra}},
  \bibinfo {author} {\bibfnamefont {H.}~\bibnamefont {Nanda}},\ and\ \bibinfo
  {author} {\bibfnamefont {S.}~\bibnamefont {Uddin}},\ }\href
  {https://doi.org/10.1155/2018/9285759} {\bibfield  {journal} {\bibinfo
  {journal} {Adv. High Energy Phys.}\ }\textbf {\bibinfo {volume} {9285759}},\
  \bibinfo {pages} {1} (\bibinfo {year} {2018})}\BibitemShut {NoStop}%
\bibitem [{\citenamefont {Cleymans}\ \emph
  {et~al.}(2006{\natexlab{a}})\citenamefont {Cleymans}, \citenamefont
  {Oeschler}, \citenamefont {Redlich},\ and\ \citenamefont
  {Wheaton}}]{ansatztempchepot2}%
  \BibitemOpen
  \bibfield  {author} {\bibinfo {author} {\bibfnamefont {J.}~\bibnamefont
  {Cleymans}}, \bibinfo {author} {\bibfnamefont {H.}~\bibnamefont {Oeschler}},
  \bibinfo {author} {\bibfnamefont {K.}~\bibnamefont {Redlich}},\ and\ \bibinfo
  {author} {\bibfnamefont {S.}~\bibnamefont {Wheaton}},\ }\href
  {https://doi.org/10.1103/PhysRevC.73.034905} {\bibfield  {journal} {\bibinfo
  {journal} {Phys. Rev. C}\ }\textbf {\bibinfo {volume} {73}},\ \bibinfo
  {pages} {034905} (\bibinfo {year} {2006}{\natexlab{a}})}\BibitemShut
  {NoStop}%
\bibitem [{\citenamefont {Mishra}\ and\ \citenamefont
  {Singh}(2008)}]{ansatztempchepot3}%
  \BibitemOpen
  \bibfield  {author} {\bibinfo {author} {\bibfnamefont {M.}~\bibnamefont
  {Mishra}}\ and\ \bibinfo {author} {\bibfnamefont {C.~P.}\ \bibnamefont
  {Singh}},\ }\href {https://doi.org/10.1103/PhysRevC.78.024910} {\bibfield
  {journal} {\bibinfo  {journal} {Phys. Rev. C}\ }\textbf {\bibinfo {volume}
  {78}},\ \bibinfo {pages} {024910} (\bibinfo {year} {2008})}\BibitemShut
  {NoStop}%
\bibitem [{\citenamefont {Cleymans}\ \emph
  {et~al.}(2006{\natexlab{b}})\citenamefont {Cleymans}, \citenamefont
  {Oeschler}, \citenamefont {Redlich},\ and\ \citenamefont
  {Wheaton}}]{ansatztempchepot4}%
  \BibitemOpen
  \bibfield  {author} {\bibinfo {author} {\bibfnamefont {J.}~\bibnamefont
  {Cleymans}}, \bibinfo {author} {\bibfnamefont {H.}~\bibnamefont {Oeschler}},
  \bibinfo {author} {\bibfnamefont {K.}~\bibnamefont {Redlich}},\ and\ \bibinfo
  {author} {\bibfnamefont {S.}~\bibnamefont {Wheaton}},\ }\href
  {https://doi.org/10.1088/0954-3899/32/12/S21} {\bibfield  {journal} {\bibinfo
   {journal} {J. Phys. G: Nucl. Part. Phys.}\ }\textbf {\bibinfo {volume}
  {32}},\ \bibinfo {pages} {S165} (\bibinfo {year}
  {2006}{\natexlab{b}})}\BibitemShut {NoStop}%
\bibitem [{\citenamefont {Tiwari}\ and\ \citenamefont
  {Singh}(2013)}]{ansatztempchepot5}%
  \BibitemOpen
  \bibfield  {author} {\bibinfo {author} {\bibfnamefont {S.}~\bibnamefont
  {Tiwari}}\ and\ \bibinfo {author} {\bibfnamefont {C.~P.}\ \bibnamefont
  {Singh}},\ }\href {https://doi.org/10.1155/2013/805413} {\bibfield  {journal}
  {\bibinfo  {journal} {Adv. High Energy Phys.}\ }\textbf {\bibinfo {volume}
  {805413}},\ \bibinfo {pages} {1} (\bibinfo {year} {2013})}\BibitemShut
  {NoStop}%
\bibitem [{\citenamefont {Bashir}\ \emph {et~al.}(2016)\citenamefont {Bashir},
  \citenamefont {Nanda},\ and\ \citenamefont {Uddin}}]{ansatztempchepot6}%
  \BibitemOpen
  \bibfield  {author} {\bibinfo {author} {\bibfnamefont {I.}~\bibnamefont
  {Bashir}}, \bibinfo {author} {\bibfnamefont {H.}~\bibnamefont {Nanda}},\ and\
  \bibinfo {author} {\bibfnamefont {S.}~\bibnamefont {Uddin}},\ }\href
  {https://doi.org/10.1134/S1063776116050022} {\bibfield  {journal} {\bibinfo
  {journal} {J. Exp. Theor. Phys.}\ }\textbf {\bibinfo {volume} {122}},\
  \bibinfo {pages} {1032} (\bibinfo {year} {2016})}\BibitemShut {NoStop}%
\bibitem [{\citenamefont {Kadam}\ and\ \citenamefont {Mishra}(2019)}]{Guru}%
  \BibitemOpen
  \bibfield  {author} {\bibinfo {author} {\bibfnamefont {G.~P.}\ \bibnamefont
  {Kadam}}\ and\ \bibinfo {author} {\bibfnamefont {H.}~\bibnamefont {Mishra}},\
  }\href {https://doi.org/10.1103/PhysRevD.100.074015} {\bibfield  {journal}
  {\bibinfo  {journal} {Phys. Rev. D}\ }\textbf {\bibinfo {volume} {100}},\
  \bibinfo {pages} {074015} (\bibinfo {year} {2019})}\BibitemShut {NoStop}%
\bibitem [{\citenamefont {Becattini}\ \emph {et~al.}(2006)\citenamefont
  {Becattini}, \citenamefont {Manninen},\ and\ \citenamefont
  {Ga\ifmmode~\acute{z}\else \'{z}\fi{}dzicki}}]{Becattini}%
  \BibitemOpen
  \bibfield  {author} {\bibinfo {author} {\bibfnamefont {F.}~\bibnamefont
  {Becattini}}, \bibinfo {author} {\bibfnamefont {J.}~\bibnamefont
  {Manninen}},\ and\ \bibinfo {author} {\bibfnamefont {M.}~\bibnamefont
  {Ga\ifmmode~\acute{z}\else \'{z}\fi{}dzicki}},\ }\href
  {https://doi.org/10.1103/PhysRevC.73.044905} {\bibfield  {journal} {\bibinfo
  {journal} {Phys. Rev. C}\ }\textbf {\bibinfo {volume} {73}},\ \bibinfo
  {pages} {044905} (\bibinfo {year} {2006})}\BibitemShut {NoStop}%
\bibitem [{\citenamefont {Cleymans}\ and\ \citenamefont
  {Satz}(1993{\natexlab{b}})}]{hp2GeV1}%
  \BibitemOpen
  \bibfield  {author} {\bibinfo {author} {\bibfnamefont {J.}~\bibnamefont
  {Cleymans}}\ and\ \bibinfo {author} {\bibfnamefont {H.}~\bibnamefont
  {Satz}},\ }\href {https://doi.org/10.1007/BF01555746} {\bibfield  {journal}
  {\bibinfo  {journal} {Z. Phys. C}\ }\textbf {\bibinfo {volume} {57}},\
  \bibinfo {pages} {135} (\bibinfo {year} {1993}{\natexlab{b}})}\BibitemShut
  {NoStop}%
\bibitem [{\citenamefont {Khuntia}\ \emph {et~al.}(2019)\citenamefont
  {Khuntia}, \citenamefont {Tiwari}, \citenamefont {Sharma}, \citenamefont
  {Sahoo},\ and\ \citenamefont {Nayak}}]{hp2GeV2}%
  \BibitemOpen
  \bibfield  {author} {\bibinfo {author} {\bibfnamefont {A.}~\bibnamefont
  {Khuntia}}, \bibinfo {author} {\bibfnamefont {S.~K.}\ \bibnamefont {Tiwari}},
  \bibinfo {author} {\bibfnamefont {P.}~\bibnamefont {Sharma}}, \bibinfo
  {author} {\bibfnamefont {R.}~\bibnamefont {Sahoo}},\ and\ \bibinfo {author}
  {\bibfnamefont {T.~K.}\ \bibnamefont {Nayak}},\ }\href
  {https://doi.org/10.1103/PhysRevC.100.014910} {\bibfield  {journal} {\bibinfo
   {journal} {Phys. Rev. C}\ }\textbf {\bibinfo {volume} {100}},\ \bibinfo
  {pages} {014910} (\bibinfo {year} {2019})}\BibitemShut {NoStop}%
\bibitem [{\citenamefont {Beringer~$et\: al.$}(2012)}]{pdg}%
  \BibitemOpen
  \bibfield  {author} {\bibinfo {author} {\bibfnamefont {J.}~\bibnamefont
  {Beringer~$et\: al.$}},\ }\href {https://doi.org/10.1103/PhysRevD.86.010001}
  {\bibfield  {journal} {\bibinfo  {journal} {Phys. Rev. D}\ }\textbf {\bibinfo
  {volume} {86}},\ \bibinfo {pages} {010001} (\bibinfo {year}
  {2012})}\BibitemShut {NoStop}%
\bibitem [{\citenamefont {Adamczyk~$et\: al.$}(2017)}]{Adamczyk2017}%
  \BibitemOpen
  \bibfield  {author} {\bibinfo {author} {\bibfnamefont {L.}~\bibnamefont
  {Adamczyk~$et\: al.$}} (\bibinfo {collaboration} {STAR Collaboration}),\
  }\href {https://doi.org/10.1103/PhysRevC.96.044904} {\bibfield  {journal}
  {\bibinfo  {journal} {Phys. Rev. C}\ }\textbf {\bibinfo {volume} {96}},\
  \bibinfo {pages} {044904} (\bibinfo {year} {2017})}\BibitemShut {NoStop}%
\bibitem [{\citenamefont {Rossi}\ \emph {et~al.}(1975)\citenamefont {Rossi},
  \citenamefont {Vannini}, \citenamefont {Bussière}, \citenamefont {Albini},
  \citenamefont {D'Alessandro},\ and\ \citenamefont {Giacomelli}}]{Rossi1975}%
  \BibitemOpen
  \bibfield  {author} {\bibinfo {author} {\bibfnamefont {A.}~\bibnamefont
  {Rossi}}, \bibinfo {author} {\bibfnamefont {G.}~\bibnamefont {Vannini}},
  \bibinfo {author} {\bibfnamefont {A.}~\bibnamefont {Bussière}}, \bibinfo
  {author} {\bibfnamefont {E.}~\bibnamefont {Albini}}, \bibinfo {author}
  {\bibfnamefont {D.}~\bibnamefont {D'Alessandro}},\ and\ \bibinfo {author}
  {\bibfnamefont {G.}~\bibnamefont {Giacomelli}},\ }\href
  {https://doi.org/https://doi.org/10.1016/0550-3213(75)90307-7} {\bibfield
  {journal} {\bibinfo  {journal} {Nucl. Phys. B}\ }\textbf {\bibinfo {volume}
  {84}},\ \bibinfo {pages} {269} (\bibinfo {year} {1975})}\BibitemShut
  {NoStop}%
\bibitem [{\citenamefont {Aggarwal~$et\: al.$}(2011)}]{Aggarwal2023}%
  \BibitemOpen
  \bibfield  {author} {\bibinfo {author} {\bibfnamefont {M.~M.}\ \bibnamefont
  {Aggarwal~$et\: al.$}} (\bibinfo {collaboration} {STAR Collaboration}),\
  }\href {https://doi.org/10.1103/PhysRevC.83.024901} {\bibfield  {journal}
  {\bibinfo  {journal} {Phys. Rev. C}\ }\textbf {\bibinfo {volume} {83}},\
  \bibinfo {pages} {024901} (\bibinfo {year} {2011})}\BibitemShut {NoStop}%
\bibitem [{\citenamefont {Adams~$et\: al.$}(2003{\natexlab{a}})}]{Adams2003}%
  \BibitemOpen
  \bibfield  {author} {\bibinfo {author} {\bibfnamefont {J.}~\bibnamefont
  {Adams~$et\: al.$}} (\bibinfo {collaboration} {STAR Collaboration}),\ }\href
  {https://doi.org/https://doi.org/10.1016/j.physletb.2003.06.039} {\bibfield
  {journal} {\bibinfo  {journal} {Phys. Lett. B}\ }\textbf {\bibinfo {volume}
  {567}},\ \bibinfo {pages} {167} (\bibinfo {year}
  {2003}{\natexlab{a}})}\BibitemShut {NoStop}%
\bibitem [{\citenamefont {Abbas $et\:~al.$ (The
  ALICE~Collaboration)}(2013)}]{Abbas2013}%
  \BibitemOpen
  \bibfield  {author} {\bibinfo {author} {\bibfnamefont {E.}~\bibnamefont
  {Abbas $et\:~al.$ (The ALICE~Collaboration)}},\ }\href
  {https://doi.org/10.1140/epjc/s10052-013-2496-5} {\bibfield  {journal}
  {\bibinfo  {journal} {Eur. Phys. J. C}\ }\textbf {\bibinfo {volume} {73}},\
  \bibinfo {pages} {2496} (\bibinfo {year} {2013})}\BibitemShut {NoStop}%
\bibitem [{\citenamefont {Ahle~$et\: al.$}(1999)}]{Ahle1999}%
  \BibitemOpen
  \bibfield  {author} {\bibinfo {author} {\bibfnamefont {L.}~\bibnamefont
  {Ahle~$et\: al.$}} (\bibinfo {collaboration} {E802 Collaboration}),\ }\href
  {https://doi.org/10.1103/PhysRevC.60.064901} {\bibfield  {journal} {\bibinfo
  {journal} {Phys. Rev. C}\ }\textbf {\bibinfo {volume} {60}},\ \bibinfo
  {pages} {064901} (\bibinfo {year} {1999})}\BibitemShut {NoStop}%
\bibitem [{\citenamefont {Abelev~$et\: al.$}(2010{\natexlab{a}})}]{Abelev2010}%
  \BibitemOpen
  \bibfield  {author} {\bibinfo {author} {\bibfnamefont {B.~I.}\ \bibnamefont
  {Abelev~$et\: al.$}} (\bibinfo {collaboration} {STAR Collaboration}),\ }\href
  {https://doi.org/10.1103/PhysRevC.81.024911} {\bibfield  {journal} {\bibinfo
  {journal} {Phys. Rev. C}\ }\textbf {\bibinfo {volume} {81}},\ \bibinfo
  {pages} {024911} (\bibinfo {year} {2010}{\natexlab{a}})}\BibitemShut
  {NoStop}%
\bibitem [{\citenamefont {Adam~$et\: al.$}(2020)}]{Adam2020}%
  \BibitemOpen
  \bibfield  {author} {\bibinfo {author} {\bibfnamefont {J.}~\bibnamefont
  {Adam~$et\: al.$}} (\bibinfo {collaboration} {STAR Collaboration}),\ }\href
  {https://doi.org/10.1103/PhysRevC.102.034909} {\bibfield  {journal} {\bibinfo
   {journal} {Phys. Rev. C}\ }\textbf {\bibinfo {volume} {102}},\ \bibinfo
  {pages} {034909} (\bibinfo {year} {2020})}\BibitemShut {NoStop}%
\bibitem [{\citenamefont {Mischke~$et\: al.$}(2002)}]{Mischke2002}%
  \BibitemOpen
  \bibfield  {author} {\bibinfo {author} {\bibfnamefont {A.}~\bibnamefont
  {Mischke~$et\: al.$}} (\bibinfo {collaboration} {NA49 Collaboration}),\
  }\href {https://doi.org/10.1088/0954-3899/28/7/330} {\bibfield  {journal}
  {\bibinfo  {journal} {J. Phys. G: Nucl. Part. Phys.}\ }\textbf {\bibinfo
  {volume} {28}},\ \bibinfo {pages} {1761} (\bibinfo {year}
  {2002})}\BibitemShut {NoStop}%
\bibitem [{\citenamefont {Adler $et\:~al.$
  (STAR~Collaboration)}(2001)}]{decayprotonsprl}%
  \BibitemOpen
  \bibfield  {author} {\bibinfo {author} {\bibfnamefont {C.}~\bibnamefont
  {Adler $et\:~al.$ (STAR~Collaboration)}},\ }\href
  {https://doi.org/10.1103/PhysRevLett.87.262302} {\bibfield  {journal}
  {\bibinfo  {journal} {Phys. Rev. Lett.}\ }\textbf {\bibinfo {volume} {87}},\
  \bibinfo {pages} {262302} (\bibinfo {year} {2001})}\BibitemShut {NoStop}%
\bibitem [{\citenamefont {Broniowski}\ and\ \citenamefont
  {Florkowski}(2000)}]{Broniowski2}%
  \BibitemOpen
  \bibfield  {author} {\bibinfo {author} {\bibfnamefont {W.}~\bibnamefont
  {Broniowski}}\ and\ \bibinfo {author} {\bibfnamefont {W.}~\bibnamefont
  {Florkowski}},\ }\href {https://doi.org/10.1016/S0370-2693(00)00992-8}
  {\bibfield  {journal} {\bibinfo  {journal} {Phys. lett. B}\ }\textbf
  {\bibinfo {volume} {490}},\ \bibinfo {pages} {223} (\bibinfo {year}
  {2000})}\BibitemShut {NoStop}%
\bibitem [{\citenamefont {Broniowski}\ \emph {et~al.}(2004)\citenamefont
  {Broniowski}, \citenamefont {Florkowski},\ and\ \citenamefont
  {Ya.~Glozman}}]{Broniowski1}%
  \BibitemOpen
  \bibfield  {author} {\bibinfo {author} {\bibfnamefont {W.}~\bibnamefont
  {Broniowski}}, \bibinfo {author} {\bibfnamefont {W.}~\bibnamefont
  {Florkowski}},\ and\ \bibinfo {author} {\bibfnamefont {L.}~\bibnamefont
  {Ya.~Glozman}},\ }\href {https://doi.org/10.1103/PhysRevD.70.117503}
  {\bibfield  {journal} {\bibinfo  {journal} {Phys. Rev. D}\ }\textbf {\bibinfo
  {volume} {70}},\ \bibinfo {pages} {117503} (\bibinfo {year}
  {2004})}\BibitemShut {NoStop}%
\bibitem [{\citenamefont {Abelev~$et\: al.$}(2013)}]{Abelev2013}%
  \BibitemOpen
  \bibfield  {author} {\bibinfo {author} {\bibfnamefont {B.}~\bibnamefont
  {Abelev~$et\: al.$}} (\bibinfo {collaboration} {ALICE Collaboration}),\
  }\href {https://doi.org/10.1103/PhysRevC.88.044910} {\bibfield  {journal}
  {\bibinfo  {journal} {Phys. Rev. C}\ }\textbf {\bibinfo {volume} {88}},\
  \bibinfo {pages} {044910} (\bibinfo {year} {2013})}\BibitemShut {NoStop}%
\bibitem [{\citenamefont {Adams~$et\: al.$}(2003{\natexlab{b}})}]{IEoS46}%
  \BibitemOpen
  \bibfield  {author} {\bibinfo {author} {\bibfnamefont {J.}~\bibnamefont
  {Adams~$et\: al.$}} (\bibinfo {collaboration} {STAR Collaboration}),\ }\href
  {https://doi.org/10.1103/PhysRevLett.90.172301} {\bibfield  {journal}
  {\bibinfo  {journal} {Phys. Rev. Lett.}\ }\textbf {\bibinfo {volume} {90}},\
  \bibinfo {pages} {172301} (\bibinfo {year} {2003}{\natexlab{b}})}\BibitemShut
  {NoStop}%
\bibitem [{\citenamefont {Pratt}\ and\ \citenamefont {Plumberg}(2021)}]{Pratt}%
  \BibitemOpen
  \bibfield  {author} {\bibinfo {author} {\bibfnamefont {S.}~\bibnamefont
  {Pratt}}\ and\ \bibinfo {author} {\bibfnamefont {C.}~\bibnamefont
  {Plumberg}},\ }\href {https://doi.org/10.1103/PhysRevC.104.014906} {\bibfield
   {journal} {\bibinfo  {journal} {Phys. Rev. C}\ }\textbf {\bibinfo {volume}
  {104}},\ \bibinfo {pages} {014906} (\bibinfo {year} {2021})}\BibitemShut
  {NoStop}%
\bibitem [{\citenamefont {Abelev~$et\:
  al.$}(2010{\natexlab{b}})}]{ChargePratt4}%
  \BibitemOpen
  \bibfield  {author} {\bibinfo {author} {\bibfnamefont {B.~I.}\ \bibnamefont
  {Abelev~$et\: al.$}} (\bibinfo {collaboration} {STAR Collaboration}),\ }\href
  {https://doi.org/https://doi.org/10.1016/j.physletb.2010.05.028} {\bibfield
  {journal} {\bibinfo  {journal} {Phys. Lett. B}\ }\textbf {\bibinfo {volume}
  {690}},\ \bibinfo {pages} {239} (\bibinfo {year}
  {2010}{\natexlab{b}})}\BibitemShut {NoStop}%
\bibitem [{\citenamefont {Rafelski}\ and\ \citenamefont
  {M\"uller}(1982)}]{RafelskiMuller}%
  \BibitemOpen
  \bibfield  {author} {\bibinfo {author} {\bibfnamefont {J.}~\bibnamefont
  {Rafelski}}\ and\ \bibinfo {author} {\bibfnamefont {B.}~\bibnamefont
  {M\"uller}},\ }\href {https://doi.org/10.1103/PhysRevLett.48.1066} {\bibfield
   {journal} {\bibinfo  {journal} {Phys. Rev. Lett.}\ }\textbf {\bibinfo
  {volume} {48}},\ \bibinfo {pages} {1066} (\bibinfo {year}
  {1982})}\BibitemShut {NoStop}%
\bibitem [{\citenamefont {Pal}\ \emph {et~al.}(2001)\citenamefont {Pal},
  \citenamefont {Ko},\ and\ \citenamefont {Lin}}]{Pal}%
  \BibitemOpen
  \bibfield  {author} {\bibinfo {author} {\bibfnamefont {S.}~\bibnamefont
  {Pal}}, \bibinfo {author} {\bibfnamefont {C.~M.}\ \bibnamefont {Ko}},\ and\
  \bibinfo {author} {\bibfnamefont {Z.-w.}\ \bibnamefont {Lin}},\ }\href
  {https://doi.org/10.1103/PhysRevC.64.042201} {\bibfield  {journal} {\bibinfo
  {journal} {Phys. Rev. C}\ }\textbf {\bibinfo {volume} {64}},\ \bibinfo
  {pages} {042201} (\bibinfo {year} {2001})}\BibitemShut {NoStop}%
\bibitem [{\citenamefont {Bass}\ \emph {et~al.}(1999)\citenamefont {Bass},
  \citenamefont {Dumitru}, \citenamefont {Bleicher}, \citenamefont {Bravina},
  \citenamefont {Zabrodin}, \citenamefont {St\"ocker},\ and\ \citenamefont
  {Greiner}}]{Bass}%
  \BibitemOpen
  \bibfield  {author} {\bibinfo {author} {\bibfnamefont {S.~A.}\ \bibnamefont
  {Bass}}, \bibinfo {author} {\bibfnamefont {A.}~\bibnamefont {Dumitru}},
  \bibinfo {author} {\bibfnamefont {M.}~\bibnamefont {Bleicher}}, \bibinfo
  {author} {\bibfnamefont {L.}~\bibnamefont {Bravina}}, \bibinfo {author}
  {\bibfnamefont {E.}~\bibnamefont {Zabrodin}}, \bibinfo {author}
  {\bibfnamefont {H.}~\bibnamefont {St\"ocker}},\ and\ \bibinfo {author}
  {\bibfnamefont {W.}~\bibnamefont {Greiner}},\ }\href
  {https://doi.org/10.1103/PhysRevC.60.021902} {\bibfield  {journal} {\bibinfo
  {journal} {Phys. Rev. C}\ }\textbf {\bibinfo {volume} {60}},\ \bibinfo
  {pages} {021902} (\bibinfo {year} {1999})}\BibitemShut {NoStop}%
\bibitem [{\citenamefont {Alt $et\:~al.$
  (NA49~Collaboration)}(2008)}]{Adamczyk64}%
  \BibitemOpen
  \bibfield  {author} {\bibinfo {author} {\bibfnamefont {C.}~\bibnamefont {Alt
  $et\:~al.$ (NA49~Collaboration)}},\ }\href
  {https://doi.org/10.1103/PhysRevC.77.024903} {\bibfield  {journal} {\bibinfo
  {journal} {Phys. Rev. C}\ }\textbf {\bibinfo {volume} {77}},\ \bibinfo
  {pages} {024903} (\bibinfo {year} {2008})}\BibitemShut {NoStop}%
\bibitem [{\citenamefont {Wang}\ \emph {et~al.}(2000)\citenamefont {Wang},
  \citenamefont {Liu}, \citenamefont {Sorge}, \citenamefont {Xu},\ and\
  \citenamefont {Yang}}]{WangLiu}%
  \BibitemOpen
  \bibfield  {author} {\bibinfo {author} {\bibfnamefont {F.}~\bibnamefont
  {Wang}}, \bibinfo {author} {\bibfnamefont {H.}~\bibnamefont {Liu}}, \bibinfo
  {author} {\bibfnamefont {H.}~\bibnamefont {Sorge}}, \bibinfo {author}
  {\bibfnamefont {N.}~\bibnamefont {Xu}},\ and\ \bibinfo {author}
  {\bibfnamefont {J.}~\bibnamefont {Yang}},\ }\href
  {https://doi.org/10.1103/PhysRevC.61.064904} {\bibfield  {journal} {\bibinfo
  {journal} {Phys. Rev. C}\ }\textbf {\bibinfo {volume} {61}},\ \bibinfo
  {pages} {064904} (\bibinfo {year} {2000})}\BibitemShut {NoStop}%
\bibitem [{\citenamefont {Kapusta}\ and\ \citenamefont
  {Mekjian}(1986)}]{KapustaMekjian}%
  \BibitemOpen
  \bibfield  {author} {\bibinfo {author} {\bibfnamefont {J.}~\bibnamefont
  {Kapusta}}\ and\ \bibinfo {author} {\bibfnamefont {A.}~\bibnamefont
  {Mekjian}},\ }\href {https://doi.org/10.1103/PhysRevD.33.1304} {\bibfield
  {journal} {\bibinfo  {journal} {Phys. Rev. D}\ }\textbf {\bibinfo {volume}
  {33}},\ \bibinfo {pages} {1304} (\bibinfo {year} {1986})}\BibitemShut
  {NoStop}%
\bibitem [{\citenamefont {S.~Lee}\ \emph {et~al.}(1988)\citenamefont {S.~Lee},
  \citenamefont {Rhoades-Brown},\ and\ \citenamefont {Heinz}}]{LeeRhodesHeinz}%
  \BibitemOpen
  \bibfield  {author} {\bibinfo {author} {\bibfnamefont {K.}~\bibnamefont
  {S.~Lee}}, \bibinfo {author} {\bibfnamefont {M.~J.}\ \bibnamefont
  {Rhoades-Brown}},\ and\ \bibinfo {author} {\bibfnamefont {U.}~\bibnamefont
  {Heinz}},\ }\href {https://doi.org/10.1103/PhysRevC.37.1452} {\bibfield
  {journal} {\bibinfo  {journal} {Phys. Rev. C}\ }\textbf {\bibinfo {volume}
  {37}},\ \bibinfo {pages} {1452} (\bibinfo {year} {1988})}\BibitemShut
  {NoStop}%
\bibitem [{\citenamefont {Baym}(1988)}]{Baym}%
  \BibitemOpen
  \bibfield  {author} {\bibinfo {author} {\bibfnamefont {G.}~\bibnamefont
  {Baym}},\ }\href {https://doi.org/10.1016/0375-9474(88)90427-7} {\bibfield
  {journal} {\bibinfo  {journal} {Nucl. Phys. A}\ }\textbf {\bibinfo {volume}
  {479}},\ \bibinfo {pages} {27} (\bibinfo {year} {1988})}\BibitemShut
  {NoStop}%
\bibitem [{\citenamefont {McLarren}(1986)}]{McLarren}%
  \BibitemOpen
  \bibfield  {author} {\bibinfo {author} {\bibfnamefont {L.}~\bibnamefont
  {McLarren}},\ }\href {https://doi.org/10.1103/RevModPhys.58.1021} {\bibfield
  {journal} {\bibinfo  {journal} {Rev. Mod. Phys.}\ }\textbf {\bibinfo {volume}
  {58}},\ \bibinfo {pages} {1021} (\bibinfo {year} {1986})}\BibitemShut
  {NoStop}%
\bibitem [{\citenamefont {Tawfik}\ \emph {et~al.}(2021)\citenamefont {Tawfik},
  \citenamefont {Abou~Elyazeed},\ and\ \citenamefont
  {Yassin}}]{Tawfik:2021txc}%
  \BibitemOpen
  \bibfield  {author} {\bibinfo {author} {\bibfnamefont {A.~N.}\ \bibnamefont
  {Tawfik}}, \bibinfo {author} {\bibfnamefont {E.~R.}\ \bibnamefont
  {Abou~Elyazeed}},\ and\ \bibinfo {author} {\bibfnamefont {H.}~\bibnamefont
  {Yassin}},\ }\href@noop {} {\bibfield  {journal} {\bibinfo  {journal} {SSRN
  Electronic Journal}\ } (\bibinfo {year} {2021})},\ \Eprint
  {https://arxiv.org/abs/2108.10320} {arXiv:2108.10320 [hep-ph]} \BibitemShut
  {NoStop}%
\bibitem [{\citenamefont {Biswas}(2021)}]{Biswas:2020dsc}%
  \BibitemOpen
  \bibfield  {author} {\bibinfo {author} {\bibfnamefont {D.}~\bibnamefont
  {Biswas}},\ }\href {https://doi.org/10.1155/2021/6611394} {\bibfield
  {journal} {\bibinfo  {journal} {Adv. High Energy Phys.}\ }\textbf {\bibinfo
  {volume} {2021}},\ \bibinfo {pages} {6611394} (\bibinfo {year} {2021})},\
  \Eprint {https://arxiv.org/abs/2003.10425} {arXiv:2003.10425 [hep-ph]}
  \BibitemShut {NoStop}%
\bibitem [{\citenamefont {Gaździcki}(1996)}]{Gazdzicki}%
  \BibitemOpen
  \bibfield  {author} {\bibinfo {author} {\bibfnamefont {M.}~\bibnamefont
  {Gaździcki}},\ }\href {https://doi.org/10.1007/BF03155597} {\bibfield
  {journal} {\bibinfo  {journal} {APH N.S. Heavy Ion Physics}\ }\textbf
  {\bibinfo {volume} {4}},\ \bibinfo {pages} {33} (\bibinfo {year}
  {1996})}\BibitemShut {NoStop}%
\bibitem [{\citenamefont {Gaździcki}(1881)}]{GazdzickiJ.Ph}%
  \BibitemOpen
  \bibfield  {author} {\bibinfo {author} {\bibfnamefont {M.}~\bibnamefont
  {Gaździcki}},\ }\href {https://doi.org/10.1088/0954-3899/23/12/012}
  {\bibfield  {journal} {\bibinfo  {journal} {J. Phys. G: Nucl. Part. Phys.}\
  }\textbf {\bibinfo {volume} {23}},\ \bibinfo {pages} {12} (\bibinfo {year}
  {1881})}\BibitemShut {NoStop}%
\bibitem [{\citenamefont {Bugaev}\ \emph {et~al.}(2013)\citenamefont {Bugaev},
  \citenamefont {Oliinychenko}, \citenamefont {Cleymans}, \citenamefont
  {Ivanytskyi}, \citenamefont {Mishustin}, \citenamefont {Nikonov},\ and\
  \citenamefont {Sagun}}]{Bugaev:2013sfa}%
  \BibitemOpen
  \bibfield  {author} {\bibinfo {author} {\bibfnamefont {K.~A.}\ \bibnamefont
  {Bugaev}}, \bibinfo {author} {\bibfnamefont {D.~R.}\ \bibnamefont
  {Oliinychenko}}, \bibinfo {author} {\bibfnamefont {J.}~\bibnamefont
  {Cleymans}}, \bibinfo {author} {\bibfnamefont {A.~I.}\ \bibnamefont
  {Ivanytskyi}}, \bibinfo {author} {\bibfnamefont {I.~N.}\ \bibnamefont
  {Mishustin}}, \bibinfo {author} {\bibfnamefont {E.~G.}\ \bibnamefont
  {Nikonov}},\ and\ \bibinfo {author} {\bibfnamefont {V.~V.}\ \bibnamefont
  {Sagun}},\ }\href {https://doi.org/10.1209/0295-5075/104/22002} {\bibfield
  {journal} {\bibinfo  {journal} {EPL}\ }\textbf {\bibinfo {volume} {104}},\
  \bibinfo {pages} {22002} (\bibinfo {year} {2013})},\ \Eprint
  {https://arxiv.org/abs/1308.3594} {arXiv:1308.3594 [hep-ph]} \BibitemShut
  {NoStop}%
\bibitem [{\citenamefont {Sagun}(2014)}]{Sagun:2014mka}%
  \BibitemOpen
  \bibfield  {author} {\bibinfo {author} {\bibfnamefont {V.~V.}\ \bibnamefont
  {Sagun}},\ }\href@noop {} {\bibfield  {journal} {\bibinfo  {journal} {Ukr. J.
  Phys.}\ }\textbf {\bibinfo {volume} {59}},\ \bibinfo {pages} {755} (\bibinfo
  {year} {2014})},\ \Eprint {https://arxiv.org/abs/1408.6110} {arXiv:1408.6110
  [hep-ph]} \BibitemShut {NoStop}%
\bibitem [{\citenamefont {Rafelski}\ \emph {et~al.}(2002)\citenamefont
  {Rafelski}, \citenamefont {Letessier},\ and\ \citenamefont
  {Torrieri}}]{Torrieri}%
  \BibitemOpen
  \bibfield  {author} {\bibinfo {author} {\bibfnamefont {J.}~\bibnamefont
  {Rafelski}}, \bibinfo {author} {\bibfnamefont {J.}~\bibnamefont
  {Letessier}},\ and\ \bibinfo {author} {\bibfnamefont {G.}~\bibnamefont
  {Torrieri}},\ }\href {https://doi.org/10.1103/PhysRevC.65.069902} {\bibfield
  {journal} {\bibinfo  {journal} {Phys. Rev. C}\ }\textbf {\bibinfo {volume}
  {65}},\ \bibinfo {pages} {069902} (\bibinfo {year} {2002})}\BibitemShut
  {NoStop}%
\bibitem [{\citenamefont {Blume}\ and\ \citenamefont {Markert}(2011)}]{Blume}%
  \BibitemOpen
  \bibfield  {author} {\bibinfo {author} {\bibfnamefont {C.}~\bibnamefont
  {Blume}}\ and\ \bibinfo {author} {\bibfnamefont {C.}~\bibnamefont
  {Markert}},\ }\href
  {https://doi.org/https://doi.org/10.1016/j.ppnp.2011.05.001} {\bibfield
  {journal} {\bibinfo  {journal} {Prog. Part. Nucl. Phys.}\ }\textbf {\bibinfo
  {volume} {66}},\ \bibinfo {pages} {834} (\bibinfo {year} {2011})}\BibitemShut
  {NoStop}%
\bibitem [{\citenamefont {Tiwari}\ and\ \citenamefont {Singh}(1998)}]{B421}%
  \BibitemOpen
  \bibfield  {author} {\bibinfo {author} {\bibfnamefont {V.~K.}\ \bibnamefont
  {Tiwari}}\ and\ \bibinfo {author} {\bibfnamefont {C.~P.}\ \bibnamefont
  {Singh}},\ }\href {https://doi.org/10.1016/S0370-2693(97)01594-3} {\bibfield
  {journal} {\bibinfo  {journal} {Phys. Lett. B}\ }\textbf {\bibinfo {volume}
  {421}},\ \bibinfo {pages} {363} (\bibinfo {year} {1998})}\BibitemShut
  {NoStop}%
\bibitem [{\citenamefont {Nayak}\ \emph {et~al.}(2010)\citenamefont {Nayak},
  \citenamefont {Banik},\ and\ \citenamefont {Alam}}]{Jajati}%
  \BibitemOpen
  \bibfield  {author} {\bibinfo {author} {\bibfnamefont {J.~K.}\ \bibnamefont
  {Nayak}}, \bibinfo {author} {\bibfnamefont {S.}~\bibnamefont {Banik}},\ and\
  \bibinfo {author} {\bibfnamefont {J.-e.}\ \bibnamefont {Alam}},\ }\href
  {https://doi.org/10.1103/PhysRevC.82.024914} {\bibfield  {journal} {\bibinfo
  {journal} {Phys. Rev. C}\ }\textbf {\bibinfo {volume} {82}},\ \bibinfo
  {pages} {024914} (\bibinfo {year} {2010})}\BibitemShut {NoStop}%
\bibitem [{\citenamefont {Cleymans}\ \emph
  {et~al.}(2006{\natexlab{c}})\citenamefont {Cleymans}, \citenamefont
  {Oeschler}, \citenamefont {Redlich},\ and\ \citenamefont
  {Wheaton}}]{Cleymans1}%
  \BibitemOpen
  \bibfield  {author} {\bibinfo {author} {\bibfnamefont {J.}~\bibnamefont
  {Cleymans}}, \bibinfo {author} {\bibfnamefont {H.}~\bibnamefont {Oeschler}},
  \bibinfo {author} {\bibfnamefont {K.}~\bibnamefont {Redlich}},\ and\ \bibinfo
  {author} {\bibfnamefont {S.}~\bibnamefont {Wheaton}},\ }\href
  {https://doi.org/10.1140/epja/i2005-10309-6} {\bibfield  {journal} {\bibinfo
  {journal} {Eur. Phys. J. A}\ }\textbf {\bibinfo {volume} {29}},\ \bibinfo
  {pages} {119} (\bibinfo {year} {2006}{\natexlab{c}})}\BibitemShut {NoStop}%
\bibitem [{\citenamefont {Busza}\ and\ \citenamefont
  {Goldhaber}(1984)}]{Busza}%
  \BibitemOpen
  \bibfield  {author} {\bibinfo {author} {\bibfnamefont {W.}~\bibnamefont
  {Busza}}\ and\ \bibinfo {author} {\bibfnamefont {A.~S.}\ \bibnamefont
  {Goldhaber}},\ }\href {https://doi.org/10.1016/0370-2693(84)91070-0}
  {\bibfield  {journal} {\bibinfo  {journal} {Phys. lett. B}\ }\textbf
  {\bibinfo {volume} {139}},\ \bibinfo {pages} {235} (\bibinfo {year}
  {1984})}\BibitemShut {NoStop}%
\bibitem [{\citenamefont {Wang}(2000)}]{WangPLB}%
  \BibitemOpen
  \bibfield  {author} {\bibinfo {author} {\bibfnamefont {F.}~\bibnamefont
  {Wang}},\ }\href {https://doi.org/10.1016/S0370-2693(00)00975-8} {\bibfield
  {journal} {\bibinfo  {journal} {Phys. lett. B}\ }\textbf {\bibinfo {volume}
  {489}},\ \bibinfo {pages} {273} (\bibinfo {year} {2000})}\BibitemShut
  {NoStop}%
\bibitem [{\citenamefont {Wang}\ and\ \citenamefont {Xu}(2000)}]{WangXu}%
  \BibitemOpen
  \bibfield  {author} {\bibinfo {author} {\bibfnamefont {F.}~\bibnamefont
  {Wang}}\ and\ \bibinfo {author} {\bibfnamefont {N.}~\bibnamefont {Xu}},\
  }\href {https://doi.org/10.1103/PhysRevC.61.021904} {\bibfield  {journal}
  {\bibinfo  {journal} {Phys. Rev. C}\ }\textbf {\bibinfo {volume} {61}},\
  \bibinfo {pages} {021904(R)} (\bibinfo {year} {2000})}\BibitemShut {NoStop}%
\bibitem [{\citenamefont {Abelev}\ \emph {et~al.}(2009)\citenamefont {Abelev}
  \emph {et~al.}}]{STAR:2008bgi}%
  \BibitemOpen
  \bibfield  {author} {\bibinfo {author} {\bibfnamefont {B.~I.}\ \bibnamefont
  {Abelev}} \emph {et~al.} (\bibinfo {collaboration} {STAR}),\ }\href
  {https://doi.org/10.1103/PhysRevC.79.064903} {\bibfield  {journal} {\bibinfo
  {journal} {Phys. Rev. C}\ }\textbf {\bibinfo {volume} {79}},\ \bibinfo
  {pages} {064903} (\bibinfo {year} {2009})},\ \Eprint
  {https://arxiv.org/abs/0809.4737} {arXiv:0809.4737 [nucl-ex]} \BibitemShut
  {NoStop}%
\end{thebibliography}%
\end{document}